\documentclass[a4paper,onecolumn,superscriptaddress,10pt,accepted=2018-05-17]{quantumarticle}
\pdfoutput=1
\usepackage[utf8]{inputenc}
\usepackage[english]{babel}
\usepackage[T1]{fontenc}
\usepackage{amsmath}
\usepackage[unicode]{hyperref}
\usepackage[numbers,sort&compress]{natbib}

\usepackage{amsmath}
\usepackage{amssymb}
\usepackage{graphicx}
\usepackage{dcolumn}
\usepackage{bm}
\usepackage{bbold}
\usepackage{braket}
\usepackage{mathtools}

\begin{document}

\title{Scattering into one-dimensional waveguides from a coherently-driven quantum-optical system}

\author{Kevin A. Fischer}\email{kevinf@stanford.edu}
\affiliation{E. L. Ginzton Laboratory, Stanford University, Stanford CA 94305, USA}
\author{Rahul Trivedi}
\affiliation{E. L. Ginzton Laboratory, Stanford University, Stanford CA 94305, USA}
\author{Vinay Ramasesh}
\affiliation{Department of Physics, University of California, Berkeley CA 94720, USA}
\author{Irfan Siddiqi}
\affiliation{Department of Physics, University of California, Berkeley CA 94720, USA}
\author{Jelena Vu\v{c}kovi\'{c}}
\affiliation{E. L. Ginzton Laboratory, Stanford University, Stanford CA 94305, USA}

\begin{abstract}
We develop a new computational tool and framework for characterizing the scattering of photons by energy-nonconserving Hamiltonians into unidirectional (chiral) waveguides, for example, with coherent pulsed excitation. The temporal waveguide modes are a natural basis for characterizing scattering in quantum optics, and afford a powerful technique based on a coarse discretization of time. This overcomes limitations imposed by singularities in the waveguide-system coupling.  Moreover, the integrated discretized equations can be faithfully converted to a continuous-time result by taking the appropriate limit. This approach provides a complete solution to the scattered photon field in the waveguide, and can also be used to track system-waveguide entanglement during evolution. We further develop a direct connection between quantum measurement theory and evolution of the scattered field, demonstrating the correspondence between quantum trajectories and the scattered photon state. Our method is most applicable when the number of photons scattered is known to be small, i.e. for a single-photon or photon-pair source. We illustrate two examples: analytical solutions for short laser pulses scattering off a two-level system and numerically exact solutions for short laser pulses scattering off a spontaneous parametric downconversion (SPDC) or spontaneous four-wave mixing (SFWM) source. Finally, we note that our technique can easily be extended to systems with multiple ground states and generalized scattering problems with both finite photon number input \textit{and} coherent state drive, potentially enhancing the understanding of, e.g., light-matter entanglement and photon phase gates.
\end{abstract}

\noindent \textsf{Numerical package in collaboration with Ben Bartlett (Stanford University), implemented in \href{http://qutip.org/}{QuTiP: The Quantum Toolbox in Python} \cite{johansson2012qutip} with \href{http://nbviewer.jupyter.org/github/qutip/qutip-notebooks/blob/master/examples/temporal-photon-scattering.ipynb}{tutorial notebook} numerically reproducing results shown in this paper.}

\newpage
\tableofcontents


\newpage
\section{INTRODUCTION}

A central object of study in quantum optics is a finite-dimensional quantum system (e.g. an atom, quantum dot, superconducting circuit) coupled to a bath with an infinite number of degrees of freedom.  In this work, we consider the bath to be a unidirectional waveguide with photonic modes which exchange energy with the quantum system. Historically, the infinite dimensionality of this problem led many to believe that general solutions for the evolution operator of the composite system were intractable.

Thus, quantum optical methods initially focused on the dynamics of the system, i.e. by tracing out the state of the waveguide, often using assumptions that are violated in practice. For example, the canonical Lindblad master equation governing these reduced dynamics \cite{carmichael2009open} was originally derived under the assumption that the state of the bath and the system factorizes at all times.  This separability is obviously not valid for the plethora of spin-photon entanglement schemes in which the photonic bath is maximally entangled with a low-dimensional system, see e.g. Refs. \cite{gao2012observation,yao2005theory}. Mollow, ever prescient, made strides in solving for the emission dynamics (including the scattered field) of a driven two-level system in 1975 \cite{mollow1975pure}, but this solution remains largely unrecognized, potentially due to the lack of generality in his technique. Instead, this paper was remembered as the first `unravelling' of the density matrix evolution for a quantum stochastic master equation (SME).

Next, Gardiner and Collett made an important contribution to the understanding of the output photon field in 1985 \cite{gardiner1985input}, showing that the entire photonic state could be determined by computing all the correlation functions \cite{glauber1963quantum,glauber1963coherent} associated with the \emph{system's} coupling operators. For example, if the system operator $a$ linearly couples to a waveguide, then the total state of the field could be extracted from the entire family of correlations
\begin{equation}
G(t_1,\dots,t_n,t_1',\dots,t_m')=\braket{a^\dagger(t_1)\cdots a^\dagger(t_n)a(t_1')\cdots a(t_m')}.\label{eq:allcorr}
\end{equation}
This technique has been used, to much success, in understanding fields in quantum-optical problems. However, it has two major drawbacks: first, if the field is in a pure state, many correlations contain redundant information; second, an arbitrarily large number of correlations may be needed to determine the $N$-photon state of the field.

More recently, theorists have made rapid progress in exploring alternative ways to arrive at the state of the field in the waveguide. We briefly review a few of the many excellent techniques developed.
\begin{itemize}
\item Calculations of the $N$-photon scattering probabilities from a two-level system based on Heisenberg picture approaches \cite{domokos2002quantum} and on an operational translation of the fact that a two-level system can only contain one excitation~\cite{roulet2016solving}.
\item Various types of diagrammatic summations and scattering matrix formalisms for energy-nonconserving systems \cite{pletyukhov2015quantum,pletyukhov2012scattering,chang2016deterministic}.
\item Scattering matrices, Lehmann-Symanzik-Zimmermann reductions, Dyson series, and Green's functions based on photon transport in energy conserving scenarios \cite{shi2009lehmann,see2017diagrammatic,liao2016photon,roy2016strongly,zheng2010waveguide,xu2017input,xu2015input2,fan2010input,caneva2015quantum,schneider2016green,hurst2017analytic,rahul}.
\item Theory of quantum integrable systems \cite{bombardelli2016s,yudson2008multiphoton,yudson1985dynamics,yudson1988dynamics,rupasov1982complete,bassi1999one} which leads to Bethe anzatz approaches \cite{rupasov1984exact,liao2016photon,roy2016strongly,leclair1997qed,konik1998scattering}.
\item Analogy to the Kondo problem \cite{leclair1997maxwell,leclair1999eigenstates}.
\item Pure-state wavefunction approaches \cite{mollow1975pure,konyk2016quantum,nysteen2015scattering}.
\item Stochastic master equation approaches that include the state of a temporally coarse-grained field vector \cite{caneva2015quantum,baragiola2017quantum,pan2016exact}.
\item Generalized master equations, that allow for few-photon inputs to drive the reduced system, or to potentially calculate the output field \cite{caneva2015quantum,shi2015multiphoton,baragiola2012n}.
\item Connection between Matrix Product State (MPS) or continuous Matrix Product State (cMPS) techniques in quantum field theories and the $(0+1)$-dimensional field theories of quantum optics \cite{verstraete2010continuous,osborne2010holographic,guimond2017delayed,pichler2016photonic,cuevas2017continuum}.
\end{itemize}
Quite remarkably, most of these results were discovered within the last ten years.  As an example of the types of calculations enabled by these new methods, we point the interested reader to recent works calculating field states in quantum feedback problems \cite{guimond2017delayed,pichler2016photonic,whalen2017open}.

\begin{figure}
  \centering
  \includegraphics[scale=0.9]{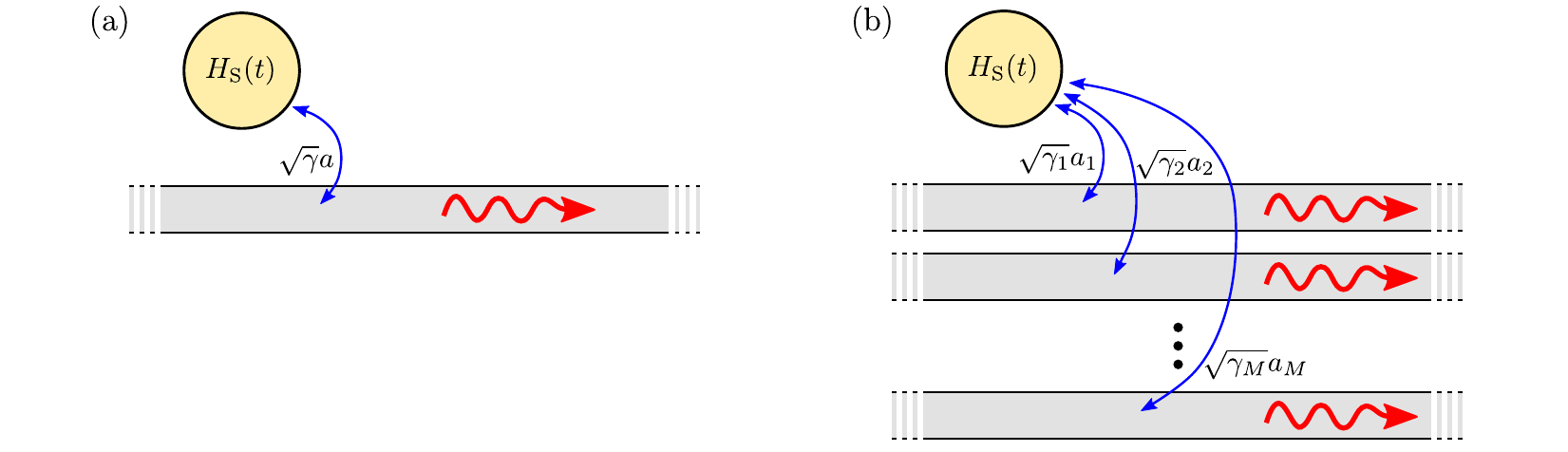}
  \caption{The general problem we solve in this article is to compute the field scattered into unidirectional (chiral) waveguide(s) from an energy-nonconserving system Hamiltonian with a unique ground state. This class of Hamiltonian is often used to represent coherent laser pulses scattering off quantum-optical systems such as a two-level system, Jaynes-Cummings system, or entangled photon pair source. First, we discuss just a single waveguide (a) and later extend to multiple waveguides (b).}
  \label{figure:1}
\end{figure}

In this work, we take a different approach to solve for the scattered fields, presenting a general technique for directly integrating the total state vector for the combined waveguide(s) and quantum-optical system.  We are primarily interested in the scattered field here, and will mostly ignore the state vector during the emission process. We show how to overcome the singularities posed by the infinite dimensionality of the baths to find a general solution for the composite evolution operator of the bath(s) and system. We illustrate our technique first for a single waveguide coupled to a low-dimensional system (Fig. \ref{figure:1}a) and then extend the formalism to handle multiple waveguides (Fig. \ref{figure:1}b). This technique, to our knowledge, is the first to provide a general method of obtaining the scattered state vector in the case where coherent laser pulses are incident on a system and source the energy of the scattered photons. Our interest in this problem is to understand the dynamics of few-photon sources as potential state generators for quantum communication or computation applications \cite{o2009photonic}.

To this end, we detail how our technique based on coarse-graining of time \cite{lee2017effective,guimond2017delayed,pichler2016photonic,whalen2017open} can be used to show how a coherently driven two-level system acts as a single- or two-photon source and spontaneous parametric downconversion or four-wave mixing act as photon pair sources. Notably, we arrive at analytic expressions for the $N$-photon scattered states when a two-level system is undergoing Rabi oscillations. For photon pair sources, our formalism allows us to distill the fundamental physics of the photon pair emission and provide the first numerically exact solutions (all previous models were perturbative, e.g. \cite{yang2008spontaneous,liscidini2012asymptotic,helt2015spontaneous,dezfouli2014heisenberg}).

Here, our main contributions are to:
\begin{enumerate}
\item Present a complete introduction to the scattering problem between a local system and a set of waveguides. 
\item Show a general treatment of using temporal modes to describe the waveguide field and the system's coupling to the waveguides.
\item Allow the local system to be driven over some time interval $t\in(0,T_P)$, and hence the system does not conserve energy there (so the problem is not quantum integrable).
\end{enumerate}
Our pedagogical choice is to present the introductory material without the use of quantum noise approaches, stochastic calculus, or input-output theory, and we direct the reader to, e.g. Refs. \cite{xu2015input2,pan2016exact,gardiner2004quantum,baragiola2012n,wiseman1994quantum} and references therein to learn about these connections.

The paper is structured as follows. In Sec. II, we build up a general framework for scattering theory between temporal waveguide modes, including when the system is driven over the interval $t\in(0,T_P)$ and hence briefly does not conserve energy. Then in Sec. III, we discuss the derivation of our general solution for a low-dimensional quantum system scattering photons into waveguide(s)---there we arrive at our central result. Finally, in Sec. IV we discuss two prototypical examples of single- and two-photon sources, based on two-level systems and spontaneous parametric downconversion or four-wave mixing.


\section{PROBLEM DEFINITION}

We consider a system described (in the Schr{\"o}dinger picture) by the time-dependent Hamiltonian $H_\textrm{S}(t)$, coupled to a bath of modes described by the static Hamiltonian $H_{0\text{B}}$ via the coupling operator $V$. We first discuss the three components of the total Hamiltonian and their properties separately.


\subsection{System Hamiltonian}

The system Hamiltonians we will consider take the form
\begin{equation}
H_\textrm{S}(t) =
\begin{cases}
H_\textrm{0S}+ H_\textrm{1S}(t) & \text{if } 0<t<T_P\\
H_\textrm{0S} & \text{otherwise}.\label{eq:ham}
\end{cases}
\end{equation}
While the above form of Hamiltonian is applicable to general systems which are driven from time $t=0$ to $t = T_P$, we place four additional restrictions on the system:
\begin{enumerate}
\item There exists an operator $N_\textrm{S}$ counting the total number of excitations in the system, which is conserved in the absence of the drive, i.e. $[N_\textrm{S}, H_\textrm{0S}]=0$, which is satisfied as long as $H_\textrm{0S}$ is Hermitian and time-independent.
\item No internal phase evolution occurs when the system is in its state with zero excitations. Practically, this can almost always be arranged by applying the appropriate transformation to $H_\textrm{0S}$.
\item The spacing between consecutive eigenfrequencies of $H_\textrm{0S}$ either be close to 0 or be clustered near some optical frequency $\omega_0$.
\item The magnitude of $H_\textrm{S}(t)$ is small compared with $\omega_0$, a standard assumption in quantum optics.
\end{enumerate}
Notably, the time-\textit{dependent} part of the Hamiltonian, while Hermitian, is not required to conserve excitation number, i.e. in general $[N_\textrm{S}, H_\textrm{1S}(t)]\neq 0$.  
As a result, we can think of $H_\textrm{S}(t)$ as conserving excitation number except for a brief period when some external interaction causes the system's Hamiltonian to acquire a time-dependence---we emphasize that $H_\textrm{S}(t)$ acts only on the system of interest, not the bath. 

A natural basis for the Hilbert space of the system $\mathcal{H}_\textrm{S}$ is a number basis, with vectors
\begin{equation}
\vec{\bm{n}}\equiv\ket{n_1,\dots, n_p}\quad\textrm{ and }\quad 
N_\textrm{S}\ket{\vec{\bm{n}}}=\sum_q n_q \ket{\vec{\bm{n}}}
\end{equation}
where $n_q$ is the number of excitations in the $q^\textrm{th}$ degree of freedom, and we assume the number of degrees $p$ is finite. For example, these degrees may represent: a cavity and atom in the case of a Jaynes-Cummings system, multiple cavity modes in spontaneous parametric downconversion (SPDC) or spontaneous four-wave mixing (SFWM), many atomic degrees in multi-emitter cavity systems, or just a single cavity or two-level system. Including the state with zero excitations
\begin{equation}
\ket{\mathbf{0}_\textrm{S}}\equiv\ket{0_1,\dots, 0_p}\quad\textrm{ where }\quad N_\textrm{S}\ket{\mathbf{0}_\textrm{S}}=0,
\end{equation}
the number states form a complete orthonormal basis
\begin{equation}
\braket{\vec{\bm{n}}|\vec{\bm{n}}'}=\delta_{\vec{\bm{n}}\vec{\bm{n}}'},
\end{equation}
with $\delta_{\vec{\bm{n}}\vec{\bm{n}}'}$ as the Kronecker-delta function. Then, the system's wavefunction is expressed as
\begin{equation}
\ket{\psi_\textrm{S}} = \sum_{\vec{\bm{n}}} \braket{\vec{\bm{n}}|\psi_\textrm{S}}\ket{\vec{\bm{n}}}.
\end{equation}


\subsection{Bath Hamiltonian}

Consider the bath to represent a single chiral channel of a waveguide, i.e. a waveguide with a single transverse spatial profile that carries energy only along one direction \cite{fan2010input}. This type of bath forms the basis for more complicated waveguide geometries and is easily extensible to multiple channels. Such a channel is described by the Hamiltonian (with $\hbar=1$)
\begin{equation}
H_{0\text{B}} = \int_0^\infty \mathop{\textrm{d}\beta} \omega(\beta) \,b_\beta^\dagger b_\beta,\label{eq:hb0}
\end{equation}
where $\omega(\beta)$ is the waveguide's dispersion relation and $b_\beta$ is the annihilation operator for a delta-normalized plane-wave excitation with wavevector $\beta$. The $b_\beta$ obey the commutation relations $[b_\beta, b^\dagger_{\beta'}]=\delta(\beta-\beta')$ and $[b_\beta, b_{\beta'}]=0$.

To transition to the standard annihilation operators $b_\omega$, i.e. for modes labeled by their frequency, we need the relationship $b_\beta = b_{\omega(\beta)}/\sqrt{\mathop{\textrm{d}\omega}/\mathop{\textrm{d}\beta}}$. The normalization is by the group velocity $v_\textrm{g}=\mathop{\textrm{d}\omega}/\mathop{\textrm{d}\beta}$ (which can either be written with a wavevector or frequency dependence). This results in
\begin{equation}
H_{0\textrm{B}} = \int_{0}^{\infty} \mathop{\textrm{d}\omega} \omega \,b^\dagger_\omega b_\omega,\label{eq:hbw0}
\end{equation}
where the mode operators similarly obey
\begin{equation}
[b_\omega, b^\dagger_{\omega'}]=\delta(\omega-\omega')\quad\textrm{ and }\quad[b_\omega, b_{\omega'}]=0.\label{commute0}
\end{equation}
We comment Eq. \ref{eq:hbw0} is exact for a complete basis of waveguide modes when $\omega(\beta)$ is one-to-one.


\subsubsection{\textit{Frequency mode basis}}

Because the bath's excitation number operator
\begin{equation}
N_\textrm{B}=\int_{0}^{\infty} \mathop{\textrm{d}\omega} b^\dagger_\omega b_\omega\textrm{ commutes as }\left[ N_\textrm{B}, H_{0\textrm{B}} \right]=0,\label{eq:numop}
\end{equation}
the waveguide's state with zero excitations is defined by $N_\textrm{B}\ket{\mathbf{0}_\textrm{B}}=0$, and a natural basis for $\mathcal{H}_\textrm{B}$ is then
\begin{equation}\label{eq:bath_states}
\ket{\vec{\bm{\omega}}^{(m)}}\equiv b^\dagger_{\omega_1} \cdots b^\dagger_{\omega_m}\ket{\mathbf{0}_\textrm{B}}/\sqrt{m!}\quad\textrm{ with }\quad N_\textrm{B}\ket{\vec{\bm{\omega}}^{(m)}}=m\ket{\vec{\bm{\omega}}^{(m)}}.
\end{equation}
We define the $m$-dimensional vector that parameterizes the infinite-dimensional states in $\mathcal{H}_\textrm{B}$ as
\begin{equation}
\vec{\bm{\omega}}^{(m)}=\{\omega_1,\dots,\omega_m\}\quad\textrm{ with }\quad 0\leq\omega_1,\dots,\omega_m.\label{eq:order}
\end{equation}
Our notation is meant to include the vectors $\vec{\bm{\omega}}^{(1)}\equiv\{\omega_1\}$ and $\vec{\bm{\omega}}^{(0)}\equiv\{\emptyset\}$, and hence the states
\begin{equation}
\ket{\{\omega_1\}}\equiv b^\dagger_{\omega_1}\ket{\mathbf{0}_\textrm{B}}\quad\textrm{ and }\quad\ket{\{\emptyset\}}\equiv\ket{\mathbf{0}_\textrm{B}}
\end{equation}
for completeness. 

Also from applying Eq. \ref{commute0}, these states form a complete orthogonal basis that is delta normalized
\begin{equation}
\braket{\vec{\bm{\omega}}^{(m)}|\vec{\bm{\omega}}'^{(m')}}=\delta_{mm'}\delta(\vec{\bm{\omega}}^{(m)}-\vec{\bm{\omega}}'^{(m)}).\label{eq:deltanorm}
\end{equation}
We can then write the bath's wavefunction as a projection onto all $\ket{\vec{\bm{\omega}}^{(m)}}$
\begin{equation}
\ket{\psi_\textrm{B}} = \sum_{m=0}^\infty \int\mathop{\textrm{d}\vec{\bm{\omega}}^{(m)}} \braket{\vec{\bm{\omega}}^{(m)}|\psi_\textrm{B}}\ket{\vec{\bm{\omega}}^{(m)}}\label{eq:basis}
\end{equation}
where
\begin{equation}
\int\mathop{\textrm{d}\vec{\bm{\omega}}^{(m)}} \equiv 
\int_0^{\infty}\mathop{\textrm{d}\omega_1} \int_{0}^{\infty}\mathop{\textrm{d}\omega_2} \cdots \int_{0}^{\infty}\mathop{\textrm{d}\omega_m}.\label{eq:omsum}
\end{equation}
To clarify two cases
\begin{equation}
\int\mathop{\textrm{d}\{\omega_1\}}\equiv\int_0^{\infty}\mathop{\textrm{d}\omega_1}\quad\textrm{ and }\quad\int\mathop{\textrm{d}\{\emptyset\}\equiv 1}.\label{eq:w1sum}
\end{equation}
Finally, we note the normalization of the wavefunction requires
\begin{equation}
\sum_{m=0}^\infty\int\mathop{\textrm{d}\vec{\bm{\omega}}^{(m)}} \left|\braket{\vec{\bm{\omega}}^{(m)}|\psi_\textrm{B}}\right|^2=1.\label{eq:norm}
\end{equation}


\subsubsection{\textit{Temporal mode basis}\label{sec:tempmode}}

At this point, we make a standard approximation in quantum optics that only modes near the characteristic frequencies of the system $\omega_0$ will be occupied \cite{fan2010input}. Thus, we are free to define fictitious waveguide modes of negative frequency and extend the limit of $\omega$ from $0\rightarrow-\infty$ in Eqs. \ref{eq:hbw0}-\ref{eq:w1sum}. This extension conveniently allows for convergence of integrals over all frequencies, and hence it is possible to define the Fourier transformed operator of~$b_\omega$
\begin{equation}
b_\tau\equiv\int_{-\infty}^\infty\frac{\mathop{\textrm{d}\omega}}{\sqrt{2\pi}}\textrm{e}^{-\textrm{\scriptsize i}\omega \tau}b_\omega ,
\end{equation}
whose the inverse transform is also well-defined. Then,
\begin{equation}
[b_\tau, b^\dagger_{\tau'}]=\delta(\tau-\tau')\quad\textrm{ and }\quad [b_\tau, b_{\tau'}]=0\label{eq:commuttebt0}
\end{equation}
so we can interpret $b_\tau$ as the annihilation operator for a `temporal' mode \cite{loudon2000quantum,brecht2015photon}. Notably, $\tau$ is an index of the operators $b_\tau$ not an actual time evolution (see Sec. \ref{sec:freeevo}). Because $\ket{\vec{\bm{\omega}}^{(m)}}$ form a complete basis for $\mathcal{H}_\textrm{B}$ and the Fourier transform is unitary, 
\begin{equation}
\ket{\vec{\bm{\tau}}^{(m)}}\equiv b^\dagger_{\tau_1} \cdots b^\dagger_{\tau_m}\ket{\mathbf{0}_\textrm{B}}\quad\textrm{ with }\quad N_\textrm{B}\ket{\vec{\bm{\tau}}^{(m)}}=m\ket{\vec{\bm{\tau}}^{(m)}},
\end{equation}
also form a complete, orthogonal, and delta-normalized basis in $\mathcal{H}_\textrm{B}$. Although the $\tau$'s can take on any real values, for convenience \textit{we will only consider waveguide states for which they are positive and ordered}
\begin{equation}
\vec{\bm{\tau}}^{(m)}=\{\tau_1,\dots,\tau_m\}\quad\textrm{ with }\quad 0<\tau_1<\cdots<\tau_m.\label{eq:subset}
\end{equation}
We are free to impose the order without loss of generality due to Eq. \ref{eq:commuttebt0}, though the ordered and non-ordered basis states differ by a normalization of $\sqrt{m!}$. Although this basis excludes states having multiple photons with the same time index, the subset of the full Hilbert space described by Eq. \ref{eq:subset} forms a complete basis for the scattered states given a linear system-waveguide interaction Hamiltonian (described in Sec. \ref{sec:freeevo}). As we will later show, for linear interactions with the waveguide, the amplitude of scattering two photons precisely at the same time instant vanishes because the choice of system-bath interaction involves only single excitation exchange.

The bath wavefunction may thus alternatively be written as a projection onto all $\ket{\vec{\bm{\tau}}^{(m)}}$ as
\begin{equation}
\ket{\psi_\textrm{B}} = \sum_{m=0}^\infty \int\mathop{\textrm{d}\vec{\bm{\tau}}^{(m)}} \braket{\vec{\bm{\tau}}^{(m)}|\psi_\textrm{B}}\ket{\vec{\bm{\tau}}^{(m)}}
\end{equation}
where
\begin{equation}
\int\mathop{\textrm{d}\vec{\bm{\tau}}^{(m)}} \equiv 
\int_0^{\infty}\mathop{\textrm{d}\tau_1} \int_{\tau_1}^{\infty}\mathop{\textrm{d}\tau_2} \cdots \int_{\tau_{m-1}}^{\infty}\mathop{\textrm{d}\tau_m}.
\end{equation}
To again clarify two cases
\begin{equation}
\int\mathop{\textrm{d}\{\tau_1\}}\equiv\int_0^{\infty}\mathop{\textrm{d}\tau_1}\quad\textrm{ and }\quad\int\mathop{\textrm{d}\{\emptyset\}\equiv 1}.
\end{equation}
Similar as before, the normalization of the wavefunction requires
\begin{equation}
\sum_{m=0}^\infty\int\mathop{\textrm{d}\vec{\bm{\tau}}^{(m)}} \left|\braket{\vec{\bm{\tau}}^{(m)}|\psi_\textrm{B}}\right|^2=1.
\end{equation}

The temporal mode weighting functions $\braket{\vec{\bm{\omega}}^{(m)}|\psi_\textrm{B}}$ and $\braket{\vec{\bm{\tau}}^{(m)}|\psi_\textrm{B}}$ are related via $m$-dimensional symmetric Fourier transforms, with the trivial relation $\braket{\vec{\bm{\omega}}^{(0)}|\psi_\textrm{B}}=\braket{\vec{\bm{\tau}}^{(0)}|\psi_\textrm{B}}=\braket{\mathbf{0}_\textrm{B}|\psi_\textrm{B}}$. When the bath is in a number eigenstate other than the ground state, it is said to be in an $m$-photon Fock state if $\braket{(\cdot)|\psi_\textrm{B}}$ is separable into like functions, i.e. when $\braket{\vec{\bm{\omega}}^{(m)}|\psi_\textrm{B}}=\phi(\omega_1)\cdots\phi(\omega_m)$ or equivalently $\braket{\vec{\bm{\tau}}^{(m)}|\psi_\textrm{B}}=\phi(\tau_1)\cdots\phi(\tau_m)$. Notably, in the continuous case an $m$-photon Fock state is not unique.

Please see Appendix A for a precise discussion on the origin of the term `temporal mode'.


\subsubsection{\textit{Free waveguide evolution}\label{sec:freeevo}}

Under a unitary transformation that removes the free evolution of the bath, the frequency-indexed annihilation operators acquire phase based on
\begin{subequations}
\begin{eqnarray}
b_\omega(t) &=& \textrm{e}^{\textrm{i}H_{0\textrm{B}}t} b_\omega \textrm{e}^{-\textrm{i}H_{0\textrm{B}}t}\\
&=&\textrm{e}^{-\textrm{i}\omega t}b_\omega.
\end{eqnarray}
\end{subequations}
As a result, the free evolution causes the temporal mode operators to translate in time
\begin{subequations}
\begin{eqnarray}
b_\tau(t) &=& \textrm{e}^{\textrm{\scriptsize i}H_{0\textrm{B}}t} b_\tau \textrm{e}^{-\textrm{\scriptsize i}H_{0\textrm{B}}t}\label{eq:bt}\\
&=&b_{\tau+t}.
\end{eqnarray}
\end{subequations}
This transformation has the consequence that 
\begin{eqnarray}
b_{\tau=0}(t)\equiv b_0(t)=b_t,\label{eq:opequiv}
\end{eqnarray}
but it is important to keep in mind they are operators in different pictures. For instance, it's common to use a basis for $\ket{\psi_\textrm{B}}$ formed with the $b_\tau$ operators and a Hamiltonian from the $b_0(t)$ operators. Hence, the commutation relations between the two are important
\begin{equation}
\left[b_{0}(t), b^\dagger_{\tau}\right]=\delta(t-\tau)\;\textrm{ and }\;\left[b_{0}(t), b_{\tau}\right]=0.\label{eq:commute1}
\end{equation}


\subsection{Waveguide-system coupling}

Suppose the system couples to the waveguide via some operator 
\begin{equation}
\vec{A}=\vec{\lambda}a^\dagger+\vec{\lambda}^*a,
\end{equation}
acting on a sector of the system's Hilbert space $\mathcal{H_\textrm{S}}$. The vector $\vec{\lambda}$ has a value and interpretation determined by the specific realization of the system, e.g. it represents a two-level system's dipole moment or a cavity's electric field. We consider the case where $a$ is an annihilation operator that acts on the system's $1$-st degree of freedom, potentially truncated to some excitation number $n_\textrm{max}$. Thus, we write its action on the system's basis states as 
\begin{equation}
a \ket{\vec{\bm{n}}}=\sqrt{n_1}\ket{n_1-1}\braket{n_1|\vec{\bm{n}}}
\end{equation}
with $a \ket{0,n_2,  \cdots, n_p}=0$, and 
\begin{equation}
a^\dagger \ket{\vec{\bm{n}}}=\sqrt{n_1+1}\ket{n_1+1}\braket{n_1|\vec{\bm{n}}}
\end{equation}
with $(a^\dagger)^{n_\textrm{max}+1} \ket{0,n_2,  \cdots, n_p}=0$. Thus, it obeys the commutation
\begin{equation}
\left[a, a^\dagger\right]=\sum_{n_1=0}^{n_\textrm{max}-1} \ket{n_1}\bra{n_1} - n_\textrm{max}\ket{n_\textrm{max}}\bra{n_\textrm{max}},
\end{equation}
with an implicit identity operator in the remaining system degrees of freedom. For a cavity $n_\textrm{max}\rightarrow\infty$ and for a two-level system $n_\textrm{max}=1$.

The system, placed at position $\vec{r}=0$, is linearly coupled to the bath via the overlap between A and the waveguide's electric field operator $\vec{\textrm{E}}(\vec{r})$. The positive frequency part of $\vec{\textrm{E}}(\vec{r}=0)$ is given by
\begin{equation}
\vec{\mathcal{E}}(0)=\textrm{i} \int_0^\infty \mathop{\textrm{d}\beta} \sqrt{\omega(\beta)}\vec{u}_\beta(0) b_\beta,
\end{equation}
where $\vec{u}_\beta(\vec{r})$ represent the orthonormal spatial modes of the waveguide. Transforming again to a sum over modes indexed by their frequency,
\begin{equation}
\vec{\mathcal{E}}(0) = \textrm{i}\int_0^\infty \mathop{\textrm{d}\omega} \sqrt{\frac{\omega}{ v_\textrm{g}(\omega)}} \vec{u}_\omega(0) b_\omega.
\end{equation}
Therefore, we write the coupling Hamiltonian as
\begin{subequations}
\begin{eqnarray}
V &=& -\vec{A}\cdot\vec{\textrm{E}}(0)\\
&=&\int_0^{\infty} \mathop{\textrm{d}\omega} \kappa(\omega)\left(a +a^\dagger\right)\left(\textrm{i}b_\omega^\dagger - \textrm{i}b_\omega \right),
\end{eqnarray}
\end{subequations}
with the coupling rate $\kappa(\omega)=\vec{\lambda}\cdot\vec{u}(0)\sqrt{\frac{\omega}{ v_\textrm{g}(\omega)}}$, and we have chosen $\vec{\lambda}\cdot\vec{u}(0)$ as real-valued without loss of generality.

At this point, we note that the system will only excite waveguide modes within a narrow band near its natural frequency $\omega_0$, which requires $\kappa(\omega_0)\ll\omega_0$, and then we make the standard three approximations in one \cite{scully1999quantum,loudon2000quantum}:
\begin{enumerate}
\item Ignore terms in $V$ that do not conserve the composite system's total excitation number $N=N_\textrm{S}+N_\textrm{B}$ (in a rotating-wave approximation).
\item Make a Markovian approximation, whereby we assume a flat coupling constant between the system and the waveguide $\kappa(\omega_0)\rightarrow\sqrt{\frac{\gamma}{2\pi}}$.
\item Again, extend the lower limit of integration over frequency.
\end{enumerate}
\begin{subequations}
\begin{eqnarray}
V &\approx&\textrm{i}\sqrt{\gamma}\int_{-\infty}^{\infty} \frac{\mathop{\textrm{d}\omega}}{\sqrt{2\pi}}\left(b_\omega^\dagger a -b_\omega a^\dagger\right)\\
&=&\textrm{i}\sqrt{\gamma}\left(b_{\tau=0}^\dagger(0)a -b_{\tau=0}(0)a^\dagger\right).
\end{eqnarray}
\end{subequations}
Because $\left[N, V\right]=0$, the zero excitation state of the composite system in $\mathcal{H_\textrm{S}}\otimes\mathcal{H_\textrm{B}}$ is determined by $\left(N_\textrm{S}+N_\textrm{B}\right)\ket{\mathbf{0}_\textrm{S}}\otimes\ket{\mathbf{0}_\textrm{B}}=0$, which we define as
\begin{equation}
\ket{\mathbf{0}}\equiv\ket{\mathbf{0}_\textrm{S}}\otimes\ket{\mathbf{0}_\textrm{B}}.
\end{equation}
Both $a\ket{\mathbf{0}}=0$ and $b_\tau\ket{\mathbf{0}}=0$, and a natural set of basis vectors for the composite system is formed from
\begin{equation}
\ket{\vec{\bm{n}},\vec{\bm{\tau}}^{(m)}}\equiv\ket{\vec{\bm{n}}}\otimes\ket{\vec{\bm{\tau}}^{(m)}}\quad\textrm{ with }\quad\braket{\vec{\bm{n}},\vec{\bm{\tau}}^{(m)}|\vec{\bm{n}}',\vec{\bm{\tau}}'^{(m')}}=\delta_{mm'}\delta_{\vec{\bm{n}}\vec{\bm{n}}'}\delta(\vec{\bm{\tau}}^{(m)}-\vec{\bm{\tau}}'^{(m)}).
\end{equation}

We now present our final requirement in this paper, that the system have only one `ground' state in the parlance of typical quantum theory. In our context this means that the system-waveguide state factorizes as $t\rightarrow\infty$ and that the system's final state is unique, so we ask
\begin{equation}
\ket{\Psi(\infty)}=\ket{\psi_\textrm{S}(t\rightarrow\infty)}\otimes\ket{\psi_\textrm{B}(t\rightarrow\infty)}\quad\textrm{ with}\quad \ket{\psi_\textrm{S}(\infty)}=\ket{\mathbf{0}_\textrm{S}}.
\end{equation}
This will be true as long as any energy put in the system will leak out into the waveguide.


\subsection{Interaction-picture Hamiltonian}

The Schr{\"o}dinger-picture Hamiltonian is given by the sum of the bare Hamiltonians and their coupling Hamiltonian
\begin{equation}
H(t)=H_\textrm{S}(t) + V + H_{0\textrm{B}},
\end{equation}
with the time-evolution of the state vector
\begin{equation}
\textrm{i}\frac{\partial}{\partial t}\ket{\Psi(t)}=H\ket{\Psi(t)}.
\end{equation}
Our first step towards solving this equation is to transform to an interaction-picture with respect to the free waveguide evolution \cite{pichler2016photonic}.

The Hamiltonian transforms as
\begin{equation}
H_\textrm{I}(t)=H_\textrm{S}(t) + \textrm{e}^{\textrm{\scriptsize i}H_{0\textrm{B}} t}V\textrm{e}^{-\textrm{\scriptsize i}H_{0\textrm{B}} t},
\end{equation}
obeying the evolution equation 
\begin{equation}
\textrm{i}\frac{\partial}{\partial t}\ket{\Psi_\textrm{I}(t)}=H_\textrm{I}(t)\ket{\Psi_\textrm{I}(t)}.\label{eq:psi_int}
\end{equation}
The Schr{\"o}dinger- and interaction-picture state vectors are related via 
\begin{equation}
\ket{\Psi(t)}=\textrm{e}^{-\textrm{i}H_{0\textrm{B}}t}\ket{\Psi_\textrm{I}(t)}.
\end{equation}
We then find the transformed Hamiltonian
\begin{equation}
H_\textrm{I}(t)=H_\textrm{S}(t)+V(t)\label{eq:HI_t}
\end{equation}
with 
\begin{equation}
V(t)=\textrm{i}\sqrt{\gamma}\left(b_{\tau=0}^\dagger(t)a -b_{\tau=0}(t)a^\dagger\right),\label{eq:Vt}
\end{equation}
and making use of Eq. \ref{eq:bt}.

The Hamiltonian possesses a few notable properties. First, no phase evolution occurs in the ground state during the time-periods with energy conservation
\begin{equation}
H_\textrm{I}(t)\ket{\mathbf{0}}=\ket{\mathbf{0}}\,\textrm{ for }\{t\leq 0\textrm{, }t\geq T_P\}.\label{eq:Honzero}
\end{equation}
Second, due to the commutation relation in Eq. \ref{eq:commute1}
\begin{equation}
\left[b_\tau,H_\textrm{I}(t)\right]=\text{i}\sqrt{\gamma}a\,\delta(t-\tau),\label{eq:Hcommute}
\end{equation}
giving the interpretation that the Hamiltonian only interacts locally with one temporal mode at a time before visiting the next.

The formal solution to Eq. \ref{eq:psi_int} is given by 
\begin{equation}
\ket{\Psi_\textrm{I}(t_1)}=U_\textrm{I}(t_1,t_0)\ket{\Psi_\textrm{I}(t_0)}
\end{equation}
with the unitary time-evolution operator
\begin{equation}
U_\textrm{I}(t_1,t_0)=\mathcal{T}\textrm{e}^{-\textrm{i}\int_{t_0}^{t_1}\mathop{\textrm{d}t} H_\textrm{I}(t)}\label{eq:unitary}
\end{equation}
and $\mathcal{T}$ as the chronological operator. The time-evolution operator has the property that it can be partitioned into arbitrary time-intervals 
\begin{equation}
U_\textrm{I}(t_2,t_0)=U_\textrm{I}(t_2,t_1)U_\textrm{I}(t_1,t_0).
\end{equation}

We briefly note this solution can be transformed to the standard continuous Matrix Product State (cMPS) form \cite{verstraete2010continuous} of
\begin{equation}
\ket{\Psi_\textrm{cMPS}}=\textrm{Tr}_\textrm{anc}\left[\mathcal{P}\,\textrm{e}^{-\textrm{i}\int_{0}^{L}\mathop{\textrm{d}x}\, \left(Q(x)\otimes \mathbb{1}+\sum_\alpha R_\alpha\otimes \hat{\phi}^\dagger_\alpha (x)\right)}\right]\ket{\Omega},
\end{equation}
with the identifications of $\ket{\Omega}\rightarrow\ket{\mathbf{0}}$, $\mathcal{P}\rightarrow\mathcal{T}$, $x\rightarrow t$, $Q(x)\rightarrow H_\textrm{S}(t)-\text{i}\frac{1}{2}\gamma a^\dagger a$, $\hat{\phi}(x)\rightarrow b_0(t)$, and $R_\alpha\rightarrow \textrm{i} \sqrt{\gamma} a$.


\subsection{Scattering matrices\label{sec:scatteringmatrices}}

The scattering matrix is formally defined in the interaction picture \cite{lancaster2014quantum} as
\begin{subequations}
\begin{eqnarray}
\hat{S}&=&\lim_{\substack{t_0\rightarrow -\infty \\ t_1\rightarrow +\infty}}U_\textrm{I}(t_1,t_0)\\
&=&\mathcal{T}\textrm{e}^{-\textrm{i}\int_{-\infty}^{+\infty}\mathop{\textrm{d}t} H_\textrm{I}(t)}.
\end{eqnarray}
\end{subequations}
This matrix can be decomposed into the Moller wave operators as $\hat{S}=\hat{\Omega}_-^\dagger\hat{\Omega}_+$
with
\begin{equation}
\hat{\Omega}_+ =\lim_{t_0\rightarrow -\infty}U_\textrm{I}(0,t_0)\quad\textrm{and}\quad
\hat{\Omega}_- =\lim_{t_1\rightarrow +\infty}U_\textrm{I}(0,t_1).
\end{equation}
(We will drop further limit notation for compaction.)

Traditionally, the scattering matrix has been used to calculate the overlap between initial and final states like $\ket{\vec{\bm{\omega}}^{(m)}}\ket{\mathbf{0}_\textrm{S}}$ and also under the constraint that $H_\textrm{S}(t)$ conserves excitation number \cite{liao2016photon,roy2016strongly,zheng2010waveguide,xu2017input,xu2015input2,fan2010input,caneva2015quantum} (for our hypothetical system this would correspond to the case where $H_\textrm{S}(t)=H_{0\textrm{S}}$ at all times). This method precludes preparing the system in an excited state before scattering---here, we show how to calculate overlap between initial and final states with the operator $\hat{\Omega}_-^\dagger$ rather than~$\hat{S}$.

When excitation number is conserved, the initial and final states must have the same number of excitations, and hence matrix elements like 
\begin{equation}
\bra{\vec{\bm{\omega}}^{(m)}}\hat{\Omega}_-^\dagger \,_\textrm{conserving}^\textrm{energy}\ket{\mathbf{{0}}}=0\;\;\textrm{ for }m>0.
\end{equation}
However, in our work we explicitly are considering the case where $H_\textrm{I}(t)$ does not conserve energy while $0<t<T_P$, as would be the case for a coherent pulse driving, for example, a two-level system or photon pair source. Hence, for this type of Hamiltonian
\begin{equation}
\bra{\mathbf{0}_\textrm{S}}\bra{\vec{\bm{\omega}}^{(m)}}\hat{\Omega}_-^\dagger\ket{\mathbf{{0}}}\neq 0
\end{equation}
and we wish to calculate elements of this type.

Our next change relative to standard scattering theory is to use the temporal mode basis all the way through, from the beginning to end of our calculations \cite{pan2016exact}, and to compute the elements
\begin{equation}
\braket{\hat{\Omega}_-^\dagger}_{\vec{\bm{\tau}}^{(m)}}\equiv\bra{\mathbf{0}_\textrm{S}}\bra{\vec{\bm{\tau}}^{(m)}}\hat{\Omega}_-^\dagger\ket{\Psi(t=0)},\label{eq:scatterdef}
\end{equation}
with $\ket{\Psi(0)}=\ket{\psi_\textrm{S}(0)}\ket{\mathbf{0}_\textrm{B}}$. Specifically, in our calculations we assumed a system Hamiltonian with a unique ground state so that $\bra{\psi_\textrm{S}(t\rightarrow\infty)}=\bra{\mathbf{0}_\textrm{S}}$. Furthermore, we will only consider cases with the waveguide prepared in its zero-excitation state, but the system may be initially excited.

First, suppose every photon emission occurs during the energy-nonconserving phase, i.e. with $\tau_m< T_P$. By defining $\tau_+$ as the first time after $\tau$ where $[b_\tau,b_{\tau^+}^\dagger]=0$ and then using Eqs. \ref{eq:Hcommute} and \ref{eq:unitary}, we find
\begin{equation}
\left[b_\tau, U_\textrm{I}(+\infty,\tau^+)\right]=0\label{eq:commute2}.
\end{equation}
Using this commutation on Eq. \ref{eq:scatterdef} we may write
\begin{eqnarray}
\braket{\hat{\Omega}_-^\dagger}_{\vec{\bm{\tau}}^{(m)}}^{\tau_m<T_P}=\bra{\mathbf{0}}U_\textrm{I}(+\infty,T_P)\,b_{\tau_1} \cdots b_{\tau_m}U_\textrm{I}(T_P,0)\ket{\mathbf{0}_\textrm{B}}\ket{\psi_\textrm{S}(0)}.\label{eq:energy11}
\end{eqnarray}
This shows why we assumed positive values for all $\tau$'s---for any negative values, those mode operators would commute all the way to the right and annihilate the state. Because energy is conserved over $T_P<t<+\infty$,
\begin{equation}
U_\textrm{I}^\dagger(+\infty, T_P)\ket{\textbf{0}} = \ket{\textbf{0}}\quad\textrm{ and }\quad
\bra{\textbf{0}}U_\textrm{I}(+\infty, T_P) = \bra{\textbf{0}}.
\end{equation}
Substituting this result back into Eq. \ref{eq:energy11}, we have
\begin{equation}
\braket{\hat{\Omega}_-^\dagger}_{\vec{\bm{\tau}}^{(m)}}^{\tau_m<T_P}=\bra{\mathbf{0}}b_{\tau_1} \cdots b_{\tau_m}U_\textrm{I}(T_P,0)\ket{\mathbf{0}_\textrm{B}}\ket{\psi_\textrm{S}(0)}.\label{eq:energy3}
\end{equation}

Now suppose that at the end of the energy-nonconserving period instead, the system remains excited and thus can continue to emit photons. Then, we need to consider a final state with $\tau_1<\cdots<T_P<\cdots<\tau_m$ and
\begin{eqnarray}
\braket{\hat{\Omega}_-^\dagger}_{\vec{\bm{\tau}}^{(m)}}^{T_P<\tau_m}=\bra{\mathbf{0}}U_\textrm{I}(+\infty,\tau_m)\,b_{\tau_1} \cdots b_{\tau_m}U_\textrm{I}(\tau_m,0)\ket{\mathbf{0}_\textrm{B}}\ket{\psi_\textrm{S}(0)},
\end{eqnarray}
where we again made use of Eq. \ref{eq:commute2}. By the same arguments made in Eqs. \ref{eq:energy11}--\ref{eq:energy3},
\begin{equation}
\braket{\hat{\Omega}_-^\dagger}_{\vec{\bm{\tau}}^{(m)}}^{T_P<\tau_m}=\bra{\mathbf{0}}b_{\tau_1} \cdots b_{\tau_m}U_\textrm{I}(\tau_m,0)\ket{\mathbf{0}_\textrm{B}}\ket{\psi_\textrm{S}(0)}.
\end{equation}

Finally, we can combine these two cases
\begin{subequations}
\begin{eqnarray}
\braket{\hat{\Omega}_-^\dagger}_{\vec{\bm{\tau}}^{(m)}}&=&
\begin{cases}
\bra{\mathbf{0}_\textrm{S}}\bra{\vec{\bm{\tau}}^{(m)}}U_\textrm{I}(T_p,0)\ket{\mathbf{0}_\textrm{B}}\ket{\psi_\textrm{S}(0)} & \text{if } \tau_m<T_P\\
\bra{\mathbf{0}_\textrm{S}}\bra{\vec{\bm{\tau}}^{(m)}}U_\textrm{I}(\tau_m,0)\ket{\mathbf{0}_\textrm{B}}\ket{\psi_\textrm{S}(0)} & \text{if } T_P<\tau_m
\end{cases},\\
&=&\bra{\mathbf{0}_\textrm{S}}\bra{\vec{\bm{\tau}}^{(m)}}U_\textrm{I}(\tau_\text{max},0)\ket{\mathbf{0}_\textrm{B}}\ket{\psi_\textrm{S}(0)}\label{eq:scatter1}
\end{eqnarray}
\end{subequations}
and we find that only times before $\tau_\text{max}=\text{max}(T_P,\tau_m)$ matter!

Although we could simplify further by commuting the temporal mode operators towards the right, we will wait to perform these operations until after we have coarse-grained time. We note that if the Hamiltonian is energy-conserving for all time, i.e. $T_P=0$, then a solution to this scattering problem reduces to solving for the emitted photon wavepacket under spontaneous emission.


\section{DERIVATION FOR SCATTERING OF COHERENT PULSES}

Practically integrating Eq. \ref{eq:psi_int} is not straight forward because of the singularity at time $t$ in $H_\textrm{I}(t)$. Although realistically the singularity is regularized through the coupling rate $\kappa(\omega)$, we anyways want to obtain a result that is independent of the precise form of $\kappa(\omega)$. Thus, we look to coarse-grain the temporal dynamics at a scale of $\Delta t$ \cite{lee2017effective,guimond2017delayed,pichler2016photonic,whalen2017open}, in effect averaging over the singularity. We will solve the dynamics in a coarse-grained basis, using a technique that is conceptually similar to manually performing a path integral between our initial and final states---quantum-optical systems are simple enough that it is not necessary to use the formal machinery of path integrals! Then, we return to the continuous-mode basis by taking the limit as $\Delta t\rightarrow 0$.


\subsection{Coarse-graining of time}

We will look for a coarse-grained propagator $U[k+1,k]$ that maps the wavefunction from $t_k\rightarrow t_{k+1}$ (with $t_k=k \Delta t$), defined as
\begin{equation}
\ket{\Psi_\textrm{I}[k+1]} = U[k+1,k] \ket{\Psi_\textrm{I}[k]}
\end{equation}
and by extension
\begin{subequations}\label{eq:unitary33}
\begin{eqnarray}
U[n,m] &\equiv& U[n,n-1] \,U[n-1,n-2]\cdots U[m+1,m]\\
&= & \overleftarrow{\prod_{k=m}^{n-1}}U[k+1, k],
\end{eqnarray}
\end{subequations}
with the
\begin{equation}
\lim_{\Delta t\rightarrow0}U[\lfloor t_1/\Delta t\rfloor,\lfloor t_0/\Delta t\rfloor]=U_\textrm{I}(t_1,t_0).\label{eq:Ulim}
\end{equation}
As long as $U[k+1,k]$ is accurate to $\mathcal{O}(\Delta t)$ and the norm of the wavefunction is conserved, i.e. $\braket{\Psi_\textrm{I}(t)|\Psi_\textrm{I}(t)}=1$ for all time, this limit will hold. Note: we only define this map in the interaction picture and hence drop the labels $(\cdot)_\textrm{I}$ for all coarse-grained operators.


\subsubsection{\textit{Derivation of a dynamical map}}

To arrive at this map, consider the formal definition of Eq. \ref{eq:unitary} for $U(t_{k+1}, t_k)$, given by the Dyson series
\begin{subequations}
\begin{eqnarray}\label{eq:disc-evolution}
U(t_{k+1}, t_k) &=&U((k+1)\Delta t, k\Delta t)\\
&\equiv& 1 - \textrm{i}\int_{k\Delta t}^{(k+1)\Delta t} \mathop{\textrm{d}t} H_\textrm{I}(t) +\frac{(-\textrm{i})^2}{2!}\int_{k\Delta t}^{(k+1)\Delta t} \mathop{\textrm{d}t}\int_{k\Delta t}^{(k+1)\Delta t} \mathop{\textrm{d}t'} \mathcal{T} H_\textrm{I}(t) H_\textrm{I}(t')+\cdots\nonumber;\\
\end{eqnarray}
\end{subequations}
note the terms with more than one time index are chronologically ordered. To guarantee that we conserve the wavefunction's norm, we use the explicitly unitary map 
\begin{equation}
U[k+1,k] \approx \textrm{exp}\left[-\textrm{i}\int_{k\Delta t}^{(k+1)\Delta t}\mathop{\textrm{d}t} H_\textrm{I}(t)\right].
\end{equation}
We note that other fields within quantum optics approach this problem of preserving the norm differently---either through Pade approximations in density-matrix renormalization techniques \cite{garcia2006time} or explicit renormalization of the wavefunction at each time step in stochastic master equations~\cite{baragiola2017quantum}.

Our operator differs from the formally correct map via time ordering
\begin{subequations}
\begin{eqnarray}
\textrm{Error}&=&\textrm{exp}\left[-\textrm{i}\int_{k\Delta t}^{(k+1)\Delta t}\mathop{\textrm{d}t} H_\textrm{I}(t)\right] - \mathcal{T} \textrm{exp}\left[-\textrm{i}\int_{k\Delta t}^{(k+1)\Delta t}\mathop{\textrm{d}t} H_\textrm{I}(t)\right]\\
&=&\int_{k\Delta t}^{(k+1)\Delta t}\mathop{\textrm{d}t}\int_{t}^{(k+1)\Delta t}\mathop{\textrm{d}t'}\left[H_\textrm{I}(t),H_\textrm{I}(t')\right] + \cdots,\label{eq:erroranalysis}
\end{eqnarray}
\end{subequations}
where the limits of integration are only over the upper half of the coordinate plane for $t<t'$. Here, the operators $H_\textrm{I}(t)$ and $H_\textrm{I}(t')$ need to be reordered, which gives their commutator in Eq. \ref{eq:erroranalysis}. We can break up the commutation between the Hamiltonian at different times based on Eq. \ref{eq:HI_t}
\begin{subequations}
\begin{eqnarray}
\left[H_\textrm{I}(t),H_\textrm{I}(t')\right]&=&\left[H_\textrm{S}(t),H_\textrm{S}(t')\right]+\left[H_\textrm{S}(t),V(t')\right]+\left[V(t),H_\textrm{S}(t')\right]+\left[V(t),V(t')\right]\\
&=& \quad\quad\,\mathcal{O}(1)\quad\quad+\;\;\mathcal{O}(1/\sqrt{\Delta t})\;\,+\;\;\mathcal{O}(1/\sqrt{\Delta t})\;\:+\quad \quad 0.
\end{eqnarray}
\end{subequations}
The commutations between $H_\textrm{S}(t)$ and $V(t')$ are assigned an $\mathcal{O}(1/\sqrt{\Delta t})$ because the singular operators in $V(t)$ (Eq. \ref{eq:Vt}) obey the commutation
\begin{equation}
\left[\int_{k\Delta t}^{(k+1)\Delta t}\mathop{\textrm{d}t} b_0(t),\int_{k\Delta t}^{(k+1)\Delta t}\mathop{\textrm{d}t'} b_0^\dagger(t')\right]=\Delta t.
\end{equation}
Combining the orders of $\left[H_\textrm{I}(t),H_\textrm{I}(t')\right]$ in Eq. \ref{eq:erroranalysis}, we see that our choice of map is correct to $\mathcal{O}(\Delta t)$, with a leading error $\mathcal{O}(\Delta t^{3/2})$ from the commutation of $H_\textrm{S}(t)$ and $V(t')$. This approximation amounts to a coarse-graining of the system-bath-interaction dynamics to a timescale of $\Delta t$, which occur on a much faster timescale than the dynamics generated by the system evolution. Nevertheless, at the end we will take the limit of $\Delta t\rightarrow 0$ in Eq. \ref{eq:Ulim}, where the error vanishes and the map becomes exact.

For later convenience, we choose to work with the map decomposed into two exponential operators
\begin{eqnarray}
U[k+1,k] &\equiv& U_\textrm{S}[k+1,k]U_\textrm{swap}[k+1,k]\label{eq:decomposition22}
\end{eqnarray}
with
\begin{eqnarray}
U_\textrm{S}[k+1,k]&=&\textrm{exp}\left[-\textrm{i}\int_{k\Delta t}^{(k+1)\Delta t}\mathop{\textrm{d}t} H_\textrm{S}(t)\right]\label{eq:USitself}
\end{eqnarray}
and
\begin{eqnarray}
U_\textrm{swap}[k+1,k]&=&\textrm{exp}\left[-\textrm{i}\int_{k\Delta t}^{(k+1)\Delta t}\mathop{\textrm{d}t} V(t)\right],
\end{eqnarray}
which is still explicitly unitary. The leading error in this approximation is still $\mathcal{O}(\Delta t^{3/2})$ because the unitaries $U_\textrm{S}[k+1,k]$ and $U_\textrm{swap}[k+1,k]$ commute to order $\Delta t$ in the series expansion of Eq.~\ref{eq:decomposition22}.

Plugging in our expression for $V(t)$ from Eq. \ref{eq:Vt}, we can rewrite
\begin{subequations}
\begin{eqnarray}
U_\textrm{swap}[k+1,k] &=& \exp{\left[\sqrt{\gamma\Delta t}\int_{k\Delta t}^{(k+1)\Delta t} \mathop{\textrm{d}t} \Big(\frac{b^\dagger_0(t)}{\sqrt{\Delta t}} a - \frac{b_0(t)}{\sqrt{\Delta t}} a^\dagger\Big)\right]}\\
&\equiv&\exp{\left[\sqrt{\gamma\Delta t} \left(\Delta B_0^\dagger[k] a - \Delta B_0[k] a^\dagger\right)\right]},\label{eq:Uswp}
\end{eqnarray}
\end{subequations}
where we defined a coarse-grained interaction-picture operator
\begin{equation}
\Delta B_0[k]=\frac{1}{\sqrt{\Delta t}}\int_{k\Delta t}^{(k+1)\Delta t}\mathop{\textrm{d}t} b_0(t).
\end{equation}
We can similarly define a Schr{\"o}dinger-picture operator
\begin{equation}
\Delta B_j=\frac{1}{\sqrt{\Delta t}}\int_{j\Delta t}^{(j+1)\Delta t}\mathop{\textrm{d}\tau} b_\tau,
\end{equation}
and combined with Eq. \ref{eq:opequiv} we see that $\Delta B_j=\Delta B_0[j]$. Hence, these operators obey the commutation
\begin{equation}
\left[\Delta B_j,\Delta B_0^\dagger[k]\right]=\delta[j-k],\label{eq:commuteBjBk}
\end{equation}
where $\delta[l]$ is also the discrete Kronecker-delta function. Therefore, unlike the singular operator $b_\tau^\dagger$, our new operator $\Delta B_{j}^\dagger$ can be interpreted as creating a properly normalized excitation of a harmonic oscillator, which is indexed by the temporal-mode bin $j$.

Additionally, the waveguide mode operators commute with the dynamical map for nonequal bins
\begin{equation}
\left[\Delta B_j,U[k,j+1]\right]=0\quad\textrm{and}\quad
\left[\Delta B_j,U[j,l]\right]=0,\label{eq:commute4}
\end{equation}
where it is assumed that $k>j+1$ and $j>l$. The continuum temporal mode operators are recovered in the limit
\begin{equation}
\lim_{\Delta t\rightarrow 0}\frac{\Delta B_{\lfloor \tau/\Delta t\rfloor}}{\sqrt{\Delta t}} = b_\tau\label{eq:blim},
\end{equation}
and the subtleties of taking this limit have been discussed elsewhere \cite{verstraete2010continuous}. 


\subsubsection{\textit{Hilbert space}\label{sec:coarsehilbert}}

Consider the Hilbert space of the combined internal system $\mathcal{H}_\textrm{S}$ and the coarse-grained waveguide field $\mathcal{H}_\textrm{B}^\textrm{coarse}$
\begin{equation}
\mathcal{H}^\textrm{coarse}=\mathcal{H}_\textrm{S}\otimes\mathcal{H}_\textrm{B}^\textrm{coarse}.
\end{equation}
Our new coarse-grained wavefunction $\ket{\Psi[k]}$ lives in this space. Here, we are representing the Hilbert space of the waveguide as a combined space of each $n^\text{th}$ harmonic oscillator representing a coarse-grained temporal mode
\begin{equation}
\mathcal{H}_\textrm{B}^\textrm{coarse}=\bigotimes_{n=-\infty}^{+\infty} \mathcal{H}_n.
\end{equation}
The new number operator in $\mathcal{H}^\textrm{coarse}$
\begin{equation}
N_\textrm{C} = \sum_{j=-\infty}^{+\infty}\Delta B_j^\dagger \Delta B_j
\end{equation}
commutes with $U_\textrm{S}$. The total excitation number operator $N_\textrm{C}+N_\textrm{S}$ is conserved when $t<0$ and $t>T_P$ (or equivalently $k<0$ and $k>T_P/\Delta t$). Hence, the state $\ket{\mathbf{0}_\textrm{B}}$ now represents zero excitations in $\mathcal{H}_\textrm{B}^\textrm{coarse}$, and the global ground state is (again) the elementary direct product 
\begin{equation}
\ket{\mathbf{0}}\equiv\ket{\mathbf{0}_\textrm{S}}\otimes\ket{\mathbf{0}_\textrm{B}} \qquad \textrm{with}\qquad
\ket{\mathbf{0}_\textrm{B}}=\bigotimes_{n=-\infty}^{+\infty} \ket{0_n}.\label{eq:gndcoarse}
\end{equation}
The new orthonormal basis states for the field modes are
\begin{eqnarray}
\ket{\vec{\bm{j}}^{(m)}}&\equiv&\Delta B^\dagger_{j_1}\cdots\Delta B^\dagger_{j_m}\ket{\mathbf{0}_\textrm{B}}\label{eq:coarsebasis}\\
&=&\ket{1_{j_1}}\otimes\cdots\otimes\ket{1_{j_m}}\otimes\bigotimes_{n\neq \{j_1,\dots,j_m\}} \ket{ 0_n},
\end{eqnarray}
and $\vec{\bm{j}}^{(m)}=\{j_1,\dots,j_m\}$. Like for the $\tau$'s, we only consider the cases where $0<j_1<\cdots<j_m$. When these states are taken in a similar limit as Eq. \ref{eq:blim}, they form a complete basis for the waveguide state in the continuum.

\begin{figure}
  \centering
  \includegraphics[scale=0.9]{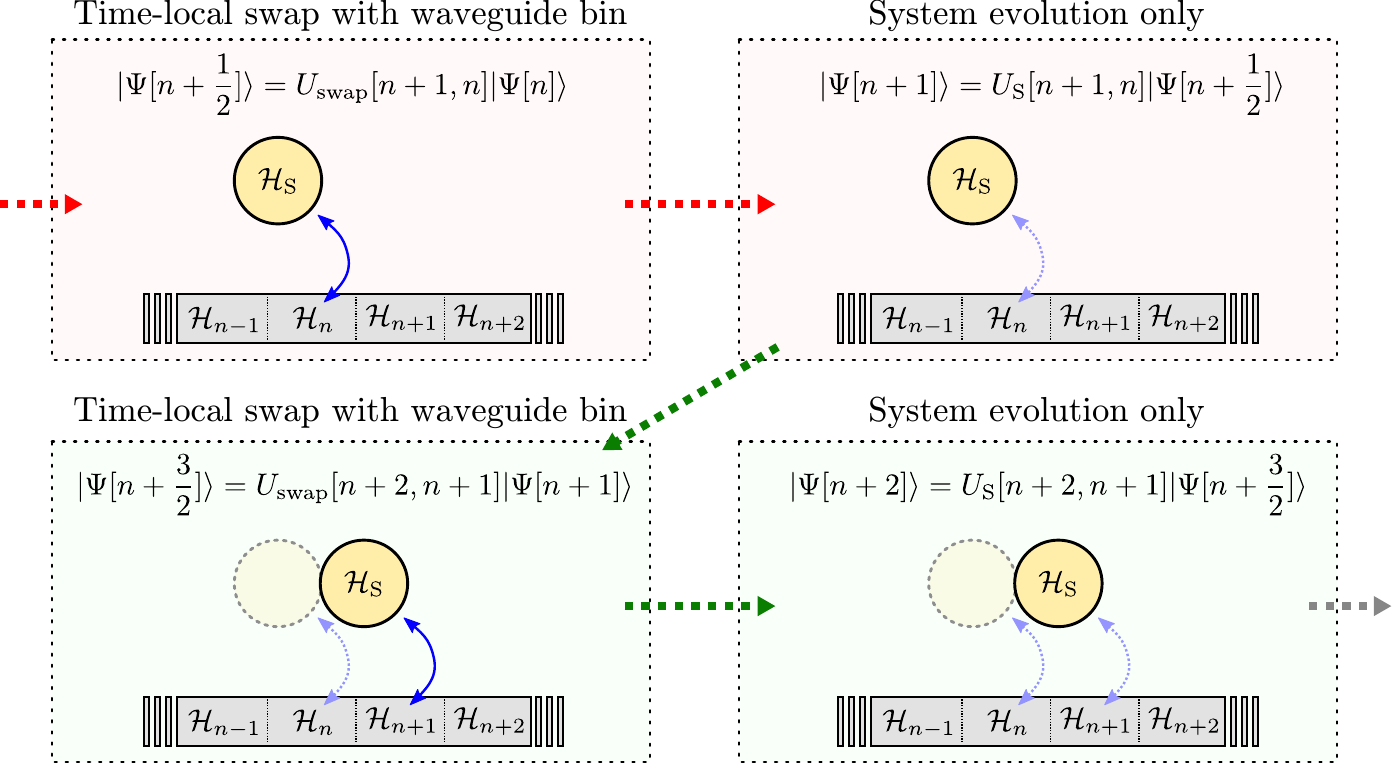}
  \caption{An intuitive picture of the coarse-grained dynamical map. Here, we use half indices to indicate the intermediate states after application of the swap but before the system evolution. This procedure is similar to the Trotterization in quantum-mechanical path integrals, although here the Hamiltonian is also varying in time. }
  \label{figure:2}
\end{figure}

At this point, it is clear that $U_\textrm{S}[k+1,k]$ acts on $\mathcal{H}_\textrm{S}$ and $U_\textrm{swap}[k+1,k]$ acts on $\mathcal{H}_\textrm{S}\otimes\mathcal{H}_{n=k}$. 
Furthermore, $U_\textrm{swap}[k+1,k]$ is the standard swap Hamiltonian between two modes. This leads to an intuitive picture (Fig. \ref{figure:2}):
\begin{enumerate}
\item At every time step the internal system performs a partial unitary swap operation with the $k^\text{th}$ waveguide bin, 
\item Following this swap, the system is briefly evolved in $\mathcal{H}_\textrm{S}$ for $\Delta t$,
\item The time index is shifted $k\rightarrow k+1$ and the process repeats.
\end{enumerate}


\subsection{A general solution\label{sec:gensol}}

Now, we define a coarse-grained scattering operator
\begin{equation}
\braket{\hat{\Omega}_-^\dagger}_{\vec{\bm{j}}^{(m)}}=
\begin{cases}
\bra{\mathbf{0}_\textrm{S}}\bra{\vec{\bm{j}}^{(m)}}U[P,0]\ket{\mathbf{0}_\textrm{B}}\ket{\psi_\textrm{S}(0)} & \text{if } j_m<P\\
\bra{\mathbf{0}_\textrm{S}}\bra{\vec{\bm{j}}^{(m)}}U[j_m+1,0]\ket{\mathbf{0}_\textrm{B}}\ket{\psi_\textrm{S}(0)} & \text{if } P<j_m.
\end{cases}
\end{equation}
From Eqs. \ref{eq:Ulim}, \ref{eq:blim}, and \ref{eq:coarsebasis}, we see that we can recover the scattering operator for continuum modes (Eq. \ref{eq:scatter1}) in the limit
\begin{equation}
\braket{\hat{\Omega}_-^\dagger}_{\vec{\bm{\tau}}^{(m)}}=
\lim_{\Delta t\rightarrow 0}\left(\frac{1}{\sqrt{\Delta t}}\right)^m\braket{\hat{\Omega}_-^\dagger}_{\vec{\bm{j}}^{(m)}}\label{eq:limfin}
\end{equation}
with $\vec{\bm{j}}^{(m)}\rightarrow\{\lfloor\tau_1/\Delta t\rfloor,\dots,\lfloor\tau_m/\Delta t\rfloor\}$ and $P\rightarrow \lfloor{T_P/\Delta t\rfloor}$. We now will begin performing manipulations on $\braket{\hat{\Omega}_-^\dagger}_{\vec{\bm{j}}^{(m)}}^{P<j_m}$ to arrive at this solution.

Our first step is to order all operators based on their action on the $j^\text{th}$ waveguide bins using Eq. \ref{eq:commute4}. Suppose that $j_1<\cdots <P<\cdots<j_m$, then
\begin{equation}
\braket{\hat{\Omega}_-^\dagger}_{\vec{\bm{j}}^{(m)}}^{P<j_m}=\bra{\mathbf{0}}\Delta B_{j_m} U[j_m+1, j_{m-1}+1]\cdots \Delta B_{j_2}U[j_2+1,j_1+1]\,\Delta B_{j_1}U[j_1+1,0]\ket{\mathbf{0}_\textrm{B}}\ket{\psi_\textrm{S}(0)}.\label{eq:orderedOm}
\end{equation}
Next, we make use of the commutation relation (obtained with the Baker-Haussdorff Lemma)
\begin{eqnarray}
\left[\Delta B_{j},U[j+1,j]\right]\approx U[j+1,j]\left(-\frac{\gamma \Delta t}{2}\left[a, a^\dagger\right]\Delta B_{j}  + \sqrt{\gamma \Delta t} \,a \right)+\mathcal{O}(\Delta t^2),
\label{eq:commute6}
\end{eqnarray}
which is normalized to $\mathcal{O}(\Delta t)$, to further commute $\Delta B_{j_1}$ to the right. From Eqs. \ref{eq:unitary33}, \ref{eq:commute4}, and \ref{eq:commute6},
\begin{subequations}
\begin{eqnarray}
&&\Delta B_{j_1}U[j_1+1,0]\ket{\mathbf{0}_\textrm{B}}\nonumber\\
&&\qquad\qquad=\Delta B_{j_1}U[j_1+1,j_1]U[j_1,0]\ket{\mathbf{0}_\textrm{B}}\label{eq:commuteBU0}\\
&&\qquad\qquad=U[j_1+1,j_1]\left(\Big(1-\frac{\gamma \Delta t}{2}\left[a, a^\dagger\right]\Big)\Delta B_{j_1}  + \sqrt{\gamma \Delta t} \,a \right)U[j_1,0]\ket{\mathbf{0}_\textrm{B}}\\
&&\qquad\qquad=\Big(1-\frac{\gamma \Delta t}{2}\left[a, a^\dagger\right]\Big)\,U[j_1+1,0]\Delta B_{j_1}\ket{\mathbf{0}_\textrm{B}}+U[j_1+1,j_1]\sqrt{\gamma \Delta t} \,a \,U[j_1,0]\ket{\mathbf{0}_\textrm{B}}\qquad\quad\\
&&\qquad\qquad=\sqrt{\Delta t}\,U[j_1+1,j_1] \,\sqrt{\gamma}a \,U[j_1,0]\ket{\mathbf{0}_\textrm{B}},\label{eq:commuteBU}
\end{eqnarray}
\end{subequations}
given that $\Delta B_{j}\ket{\mathbf{0}_\textrm{B}}=0$. (Note: we will keep the waveguide-coupling rate $\sqrt{\gamma}$ with the operators $a$ in preparation for the extension to multiple waveguides.) By sequentially commuting each $\Delta B_{j}$ in Eq. \ref{eq:orderedOm} towards the right (like with Eqs. \ref{eq:commuteBU0}--\ref{eq:commuteBU}), we are left with a result that contains only system operators and unitary evolutions
\begin{equation}
\braket{\hat{\Omega}_-^\dagger}_{\vec{\bm{j}}^{(m)}}^{P<j_m}=(\sqrt{\Delta t})^m\bra{\mathbf{0}_\textrm{S}}\bra{\mathbf{0}_\textrm{B}}\sqrt{\gamma}a\, U[j_m, j_{m-1}]\cdots \sqrt{\gamma}a\,U[j_2,j_1]\,\sqrt{\gamma}a\,U[j_1,0]\ket{\mathbf{0}_\textrm{B}}\ket{\psi_\textrm{S}(0)}.\label{eq:nowaveguideops}
\end{equation}

As discussed in Sec. \ref{sec:coarsehilbert}, $U[k+1,k]$ only acts on $\mathcal{H}_\textrm{S}\otimes\mathcal{H}_{n=k}$. With this fact and the help of Eqs. \ref{eq:unitary33} and \ref{eq:gndcoarse}, we may rewrite
\begin{subequations}
\begin{eqnarray}
\sqrt{\gamma}a\,U[j_1,0]\ket{\mathbf{0}_\textrm{B}}&=&\left(\prod_{n=-\infty}^{-1}\ket{{0}_n}\right)\sqrt{\gamma}a\,U[j_1,j_1-1]\cdots U[2,1]U[1,0]\prod_{n=0}^{+\infty}\ket{{0}_n}\qquad\\
&=&\left(\prod_{n=-\infty}^{-1}\ket{{0}_n}\right)\sqrt{\gamma}a\,U[j_1,j_1-1]\cdots U[2,1]\left(\prod_{n=1}^{+\infty} \ket{0_n}\right)U[1,0]\ket{{0}_0}\qquad\\
&=&\left(\prod_{n=-\infty}^{-1}\ket{{0}_n}\right)\sqrt{\gamma }a\,U[j_1,j_1-1]\cdots \left(\prod_{n=2}^{+\infty} \ket{0_n}\right)\overleftarrow{\prod_{k=0}^{1}} U[k+1,k]\ket{0_k}\\
&=&\left(\prod_{n=-\infty}^{-1}\ket{{0}_n}\right)\left(\prod_{n=j_1}^{+\infty}\ket{{0}_n}\right)\sqrt{\gamma}a\,\overleftarrow{\prod_{k=0}^{j_1-1}} U[k+1,k]\ket{0_k}
\end{eqnarray}
\end{subequations}
Using this procedure, we move each $\ket{0_n}$ element of $\ket{\mathbf{0}_\textrm{B}}$ in Eq. \ref{eq:nowaveguideops} as far left as possible, and we similarly move each $\bra{0_n}$ of $\bra{\mathbf{0}_\textrm{B}}$ as far right as possible. Upon doing this, each $k^\text{th}$ unitary map in Eq. \ref{eq:nowaveguideops} is evaluated for no absorption (due to $\ket{0_k}$) or emission (due to $\bra{0_k}$) of photons and
\begin{subequations}
\begin{eqnarray}
\braket{\hat{\Omega}_-^\dagger}_{\vec{\bm{j}}^{(m)}}^{P<j_m}&=&(\sqrt{\Delta t})^m \bra{\mathbf{0}}_\textrm{S}\bra{\mathbf{0}}_\textrm{B} \cdots \sqrt{\gamma}a\,U[j_q, j_{q-1}] \cdots \ket{\mathbf{0}}_\textrm{B}\ket{\psi_\textrm{S}(0)}\\
&=&(\sqrt{\Delta t})^m\bra{\mathbf{0}}_\textrm{S}\cdots \sqrt{\gamma }a\overleftarrow{\prod_{k=j_{q-1}}^{j_{q}-1}} \bra{0_k}U[k+1,k]\ket{0_k}\cdots\ket{\psi_\textrm{S}(0)}.
\end{eqnarray}
\end{subequations}
As a reminder, we approximated $U[k+1,k]=U_\textrm{S}[k+1,k] U_\textrm{swap}[k+1,k]$ as two sequential operations---$U_\textrm{S}[k+1,k]$ that acts on $\mathcal{H}_\textrm{S}$ and $U_\textrm{swap}[k+1,k]$ that acts on $\mathcal{H}_\textrm{S}\otimes\mathcal{H}_{n=k}$. Thus, we write
\begin{equation}
\left< 0_k \right|U[k+1,k]\left| 0_k \right>=U_\textrm{S}[k+1,k]\left< 0_k \right|U_\textrm{swap}[k+1,k]\left| 0_k \right>.
\end{equation}
From the definition of $U_\textrm{swap}[k+1,k]$ in Eq. \ref{eq:Uswp}, we evaluate
\begin{subequations}
\begin{eqnarray}
\bra{0_k}U_\textrm{swap}[k+1,k]\ket{0_k}&\approx& 1- \frac{\gamma \Delta t}{2} a^\dagger a +\mathcal{O}(\Delta t^2)  \\
&\approx&\exp{\left(-\frac{\gamma}{2}a^\dagger a \Delta t \right)}+\mathcal{O}(\Delta t^2)
\end{eqnarray}
\end{subequations}
for small $\Delta t$. We can combine $U_\textrm{S}[k+1,k]$ and $\left< 0_k \right|U[k+1,k]\left| 0_k \right>$ into one exponential, which is accurate to at least $\mathcal{O}(\Delta t)$ since $\left[U_\textrm{S}[k+1,k], \left< 0_k \right|U_\textrm{swap}[k+1,k]\left| 0_k \right>\right]\leq\mathcal{O}(\Delta t^2)$. Therefore, we define the effective non-unitary propagator as
\begin{subequations}
\begin{eqnarray}
U_\text{eff}[k+1,k]&\equiv&\bra{0_k}U[k+1,k]\ket{0_k}\\
&\approx& \textrm{exp}\left[-\textrm{i}\int_{k\Delta t}^{(k+1)\Delta t}\mathop{\textrm{d}t} H_\textrm{eff}(t)\right],
\end{eqnarray}
\end{subequations}
where $H_\textrm{eff}(t) = H_\textrm{S}(t)-\textrm{i}\frac{\gamma}{2} a^\dagger a$ is the non-Hermitian effective Hamiltonian. Importantly, $H_\textrm{eff}(t)$ has only system operators so $U_\text{eff}[k+1,k]$ acts only on $\mathcal{H}_\text{S}$. Specifically, because the field state is unchanged after application of this operator, it represents the evolution conditioned on no photon emission during the time interval $\Delta t$---this operator is the same as in quantum measurement theory, though we have derived it from explicit consideration of the waveguide state.

\begin{figure}
  \centering
  \includegraphics[scale=1]{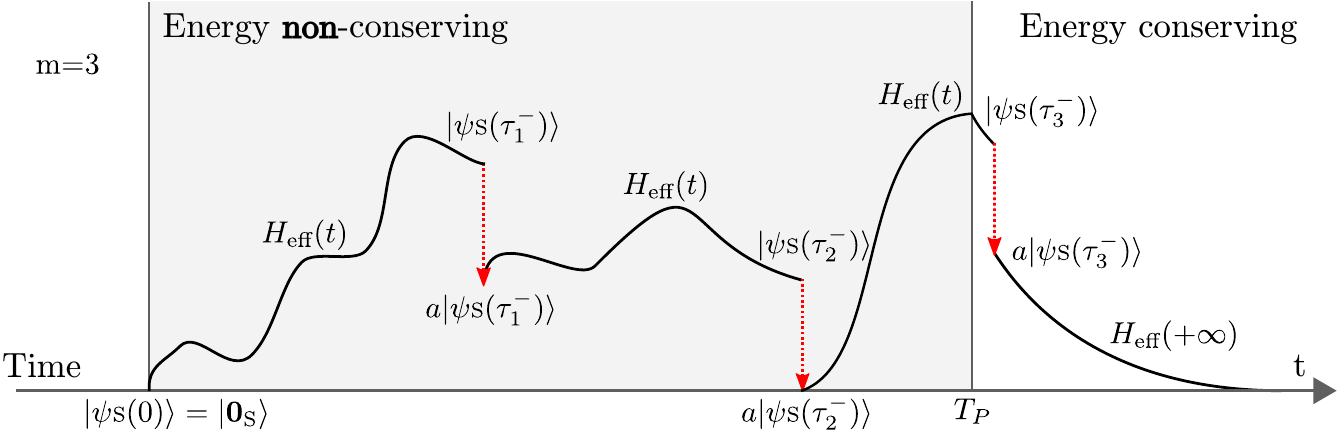}
  \caption{Sketch of a path formalized by Eqs. \ref{eq:path_formal1}--\ref{eq:projection}, specifically for the path amplitude with a realization of the photon emission times $\vec{\bm{\tau}}^{(3)}=\{\tau_1,\tau_2,\tau_3\}$ and $T_P<\tau_3$. Through a series of commutations, we reduced the problem of computing the scattered waveguide field to one of computing many system evolutions for different $\vec{\bm{\tau}}^{(m)}$, subject to $H_\textrm{eff}(t)$, and punctuated by state collapses at times $\vec{\bm{\tau}}^{(m)}$. This non-Hermitian Hamiltonian causes $\ket{\psi_\textrm{S}(t)}$ to leak probability and at the end of the evolution we calculate the overlap with the remaining amplitude in the ground state, i.e. with $\braket{\mathbf{0}_\textrm{S}|\psi_\textrm{S}(t)}$. But, since the system conserves energy after $T_P$, we can compute the remaining ground state amplitude just by following the evolution to the greater of $\tau_m$ or $T_P$, i.e. $\braket{\mathbf{0}_\textrm{S}|\psi_\textrm{S}(\tau_m^+)}$ or $\braket{\mathbf{0}_\textrm{S}|\psi_\textrm{S}(T_P)}$, respectively.}
  \label{figure:3}
\end{figure}

Then, we can combine periods of non-Hermitian evolution together
\begin{eqnarray}
U_\text{eff}[j_{q},j_{q-1}]&=&\overleftarrow{\prod_{k=j_{q-1}}^{j_{q}-1}} \bra{0_k}U[k+1,k]\ket{0_k}
\end{eqnarray}
to arrive at
\begin{equation}
\braket{\hat{\Omega}_-^\dagger}_{\vec{\bm{j}}^{(m)}}^{P<j_m}=(\sqrt{\Delta t})^m\bra{\mathbf{0}_\textrm{S}}\sqrt{\gamma}a\,U_\text{eff}[j_m,j_{m-1}]\cdots \sqrt{\gamma}a\,U_\text{eff}[j_2,j_1]\,\sqrt{\gamma}a\,U_\text{eff}[j_1,0]\ket{\psi_\textrm{S}(0)}.\label{eq:path_formal1}
\end{equation}
This integral has a very intuitive form, where $m$ evolution periods are governed by $U_\text{eff}$ and hence by the effective non-Hermitian Hamiltonian, which represent the amplitudes the system does not emit a photon in those periods (Fig. \ref{figure:3}). 
For compaction, we now define $j_0$=0 and write our result as
\begin{equation}
\braket{\hat{\Omega}_-^\dagger}_{\vec{\bm{j}}^{(m)}}^{P<j_m}=(\sqrt{\Delta t})^m\bra{\mathbf{0}_\textrm{S}}\overleftarrow{\prod_{q=1}^{m}}\sqrt{\gamma}a\,U_\text{eff}[j_q,j_{q-1}]\ket{\psi_\textrm{S}(0)},\label{eq:mleqPlast}
\end{equation}
where the product is ordered in $q$. The jumps in the state vector by $a$ at times $j_q$ correspond to photon emission events. We can carry out a similar procedure for the case where $j_m<P$ (as in Eqs. \ref{eq:orderedOm}--\ref{eq:mleqPlast}). Here, we obtain
\begin{eqnarray}
\braket{\hat{\Omega}_-^\dagger}_{\vec{\bm{j}}^{(m)}}^{j_m<P}=(\sqrt{\Delta t})^m\bra{\mathbf{0}_\textrm{S}}U_\text{eff}[P,j_m]\overleftarrow{\prod_{q=1}^{m}}\sqrt{\gamma }a\,U_\text{eff}[j_q,j_{q-1}]\ket{\psi_\textrm{S}(0)}.\label{eq:mleqPlast2}
\end{eqnarray}
Note the difference for this case: the trajectories end with a waiting period from $j_m$ to $P$ where no emissions occur.


\subsubsection{\textit{Result with interaction-picture operators}}

Taking the continuum limit in Eq. \ref{eq:limfin} from Eqs. \ref{eq:mleqPlast} and \ref{eq:mleqPlast2}, we obtain our solution to Eq. \ref{eq:psi_int} in the asymptotic limit as $t\rightarrow\infty$ and when the waveguide is initially prepared in the ground state. The solution in terms of a scattering operator is given by
\begin{subequations}
\begin{eqnarray}
\braket{\hat{\Omega}_-^\dagger}_{\vec{\bm{\tau}}^{(m)}}&=&
\begin{cases}
\bra{\mathbf{0}_\textrm{S}}U_\text{eff}(T_P, \tau_m)\,\overleftarrow{\prod}_{q=1}^{m}\sqrt{\gamma}a\,U_\text{eff}(\tau_q, \tau_{q-1})\ket{\psi_\textrm{S}(0)} & \text{if } \tau_m<T_P\\
\bra{\mathbf{0}_\textrm{S}}\overleftarrow{\prod}_{q=1}^{m}\sqrt{\gamma}a\,U_\text{eff}(\tau_q, \tau_{q-1})\ket{\psi_\textrm{S}(0)} & \text{if } T_P<\tau_m
\end{cases}\label{eq:scatter16}\\
&=&\bra{\mathbf{0}_\textrm{S}}U_\text{eff}(\tau_\text{max}, \tau_m)\,\overleftarrow{\prod_{q=1}^{m}}\sqrt{\gamma}a\,U_\text{eff}(\tau_q, \tau_{q-1})\ket{\psi_\textrm{S}(0)},\label{eq:scatter3}
\end{eqnarray}
\end{subequations}
where $\tau_0 = 0$, $\tau_\text{max}=\text{max}(T_P,\tau_m)$, and
\begin{subequations}
\begin{eqnarray}
U_\text{eff}(\tau_q, \tau_{q-1}) &=&\lim_{\Delta t\rightarrow 0}U_\text{eff}[\lfloor \tau_q/\Delta t \rfloor,\lfloor \tau_{q-1}/\Delta t \rfloor]\\
&=&\mathcal{T} \textrm{exp}\left[-\textrm{i} \int_{\tau_{q-1}}^{\tau_q} \mathop{\textrm{d}t} H_\text{eff}(t)\right]\label{eq:Ueff}
\end{eqnarray}
\end{subequations}
is the propagator corresponding to the effective Hamiltonian
\begin{equation}
H_\textrm{eff}(t) = H_\textrm{S}(t)-\textrm{i}\frac{\gamma}{2} a^\dagger a
\end{equation}
where $H_\textrm{S}(t)$ is again the system Hamiltonian from Eq. \ref{eq:ham}. It should be noted that $U_\text{eff}(\tau_q, \tau_{q-1})$ is not unitary since the effective Hamiltonian $H_\text{eff}$ is not Hermitian.

The $m$-photon temporal mode functions are given, for the interaction-picture wavefunction, by
\begin{equation}
\braket{\mathbf{0}_\textrm{S}, \vec{\bm{\tau}}^{(m)} | \Psi_\textrm{I}(t\rightarrow\infty)}=\braket{\hat{\Omega}_-^\dagger}_{\vec{\bm{\tau}}^{(m)}}.\label{eq:projection}
\end{equation}
Hence, our final result for the scattered field is
\begin{equation}
\ket{\psi_\textrm{B,I}(t\rightarrow\infty)}=\sum_{m=0}^\infty\int \mathop{\textrm{d}\vec{\bm{\tau}}^{(m)}}\braket{\hat{\Omega}_-^\dagger}_{\vec{\bm{\tau}}^{(m)}}\ket{\vec{\bm{\tau}}^{(m)}}.\label{eq:scatter11}
\end{equation}
We again emphasize that if the system Hamiltonian conserves energy for all times, i.e. $T_P=0$ so $H_\textrm{S}(t)=H_{0\textrm{S}}$, this result reduces to the solution for the system spontaneously emitting its energy into the waveguide after being prepared in $\ket{\psi_\textrm{S}(0)}$. Remarkably, the scattered field state after the system has finished decaying depends only on operators that act on the quantum-optical system's Hilbert space! 


\subsubsection{\textit{Result with Heisenberg-like operators}}
To arrive at an alternative form of expressing Eq.~\ref{eq:scatter3} that is more useful for computations with simple Heisenberg equations of motion, we define the `Heisenberg-like' system operator $\tilde{a}(\tau)$ via
\begin{equation}
\tilde{a}(\tau) = U_\text{eff}(0,\tau) \,a \, U_\text{eff}(\tau, 0),
\end{equation}
which satisfies the Heisenberg equation of motion
\begin{equation}
\textrm{i}\frac{\textrm{d} \tilde{a}(\tau)}{\textrm{d}\tau} = [\tilde{a}(\tau), \tilde{H}_\text{eff}(\tau)]
\end{equation}
where 
\begin{equation}
\tilde{H}_\text{eff}(\tau) = U_\text{eff}(0, \tau) H_\text{eff}(\tau) U_\text{eff}(\tau, 0).
\end{equation}
It can be noted that $\tilde{a}(\tau)$ is different from the Heisenberg-picture operator $a_H(t) = U(0,t)\, a\, U(t,0)$, where $U(t_2, t_1)$ is the (unitary) propagator for the total Hamiltonian (which includes the system, bath, and coupling Hamiltonians). Additionally, since $H_\text{eff}(t)$ is not Hermitian, the equation of motion for $a^\dagger$, i.e. $\tilde{a^\dagger}(\tau)$, is not the same as the adjoint of $a(\tau)$, i.e $\tilde{a}^\dagger(\tau)$. Using the property that $U_\text{eff}(\tau_q, 0)U_\text{eff}(0, \tau_{q-1})=U_\text{eff}(\tau_q, \tau_{q-1})$, Eq.~\ref{eq:scatter3} can be recast in terms of $\tilde{a}(\tau)$ 
\begin{equation}
\braket{\hat{\Omega}_-^\dagger}_{\vec{\bm{\tau}}^{(m)}}=
\bra{\mathbf{0}_\textrm{S}}U_\text{eff}(T_P, 0)\,\overleftarrow{\prod_{q=1}^{m}} \sqrt{\gamma}\,\tilde{a}(\tau_q)\ket{\psi_\textrm{S}(0)}.\label{eq:scatter_heisenberg}
\end{equation}
Here, we have used the fact that $\bra{\textbf{0}_S}U(\tau_m, T_P) = \bra{\textbf{0}_S}$ for $\tau_m > T_P$, since the decay rate of the system's ground state would be 0 outside the energy non-conserving interval. When the system Hamiltonian conserves energy for all times, i.e. $T_P=0$ so $H_\textrm{S}(t)=H_{0\textrm{S}}$, Eq. \ref{eq:scatter_heisenberg} reduces to $\bra{\mathbf{0}_\textrm{S}}\overleftarrow{\prod}_{q=1}^{m}\sqrt{\gamma}\,\tilde{a}(\tau_q)\ket{\psi_\textrm{S}(0)}$. This is the solution for the system spontaneously emitting its energy into the waveguide after being prepared in $\ket{\psi_\textrm{S}(0)}$, as we recently found by solving the scattering problem in the Heisenberg picture \cite{rahul}.


\subsection{Extension to computing the system-waveguide entangled state}

We could alternatively compute the entangled state of the system and waveguide \textit{during} the emission process from the elements
\begin{equation}
\braket{\vec{\bm{n}},\vec{\bm{\tau}}^{(m)}|U_{\textrm{I}}(t,0)|\Psi(t=0)}
\end{equation}
with the state then expressed as
\begin{equation}
\ket{\Psi_\textrm{I}(t)}=\sum_{\vec{\bm{n}}}\sum_{m=0}^\infty\int \mathop{\textrm{d}\vec{\bm{\tau}}^{(m)}}\braket{\vec{\bm{n}},\vec{\bm{\tau}}^{(m)}|U_{\textrm{I}}(t,0)|\Psi(t=0)}\ket{\vec{\bm{n}},\vec{\bm{\tau}}^{(m)}}.
\end{equation}
Here, the emission sequence is always terminated by a waiting period with no emission from $\tau_m$ until time $t$, i.e. with $\tau_m<t$, rather than being determined by $\tau_m$ relative to $T_P$. Following the computation of this element using the techniques in Sec. \ref{sec:gensol}, we make the replacements $\bra{\mathbf{0}_\textrm{S}}\rightarrow\bra{\vec{\bm{n}}}$ for the final system's state vector and $\tau_\text{max}\rightarrow t$ in Eq.~\ref{eq:scatter3}
\begin{eqnarray}
\hspace{0pt}\braket{\vec{\bm{n}},\vec{\bm{\tau}}^{(m)}|U_{\textrm{I}}(t,0)|\Psi(t=0)}=\bra{\vec{\bm{n}}}U_\text{eff}(t, \tau_m)\overleftarrow{\prod_{q=1}^{m}}\sqrt{\gamma}a\,U_\text{eff}(\tau_q, \tau_{q-1})\ket{\psi_\textrm{S}(0)}.
\end{eqnarray}
Hence, this is the full solution to Eq.~\ref{eq:psi_int} when the waveguide is initially prepared in the vacuum state at $t=0$. Often $\vec{\bm{n}}$ has only a few degrees of freedom, so it's very reasonable to calculate these elements as well. 

We also note the formal relation to the Lindblad master equation for the reduced system dynamics. This could be found by taking a partial trace over the bath degrees of freedom from a pure-state density operator $\rho_\textrm{S}(t)=\textrm{Tr}_\textrm{B}\{\ket{\Psi_\textrm{I}(t)}\bra{\Psi_\textrm{I}(t)}\}$, which is not too challenging for a two-level system coupled to a waveguide. However, there are much more direct ways to arrive at the reduced system dynamics using a coarse-grained picture \cite{fischer2017derivation,lee2017effective}.


\subsection{Extension to multiple output waveguides\label{sec:multiplewaveguides}}

The general framework outlined above can easily be extended to problems where a quantum-optical system couples to multiple ($M$) waveguides. The bath Hamiltonian and the bath-system interaction Hamiltonian for such a system is given by
\begin{equation}
H_{0\textrm{B}} = \sum_{i=1}^M \int \mathop{\textrm{d}\omega} \omega \,b_{i,\omega}^\dagger b_{i,\omega}  \quad\text{and}\quad V =  \text{i} \sum_{i=1}^M \sqrt{\gamma_i}\int \frac{\textrm{d}\omega}{\sqrt{2\pi}}(b_{i,\omega}^\dagger a_i - b_{i,\omega} a_i^\dagger),
\end{equation}
where $b_{i,\omega}$ are the annihilation operators for the delta-normalized plane-wave modes in the $i^\textrm{th}$ waveguide and $a_i$ are the system operators through which the quantum-optical system interacts with the $i^\text{th}$ waveguide. Note that $a_i$ need not be distinct operators, in which case they correspond to multiple waveguides coupling to the quantum-optical system through the same operator. This scenario is comparable to passing the emission of one waveguide through a multi-port beamsplitter.

A complete basis for the bath states can be constructed taking the tensor products of the bases for individual waveguides (Eq.~\ref{eq:bath_states})
\begin{equation}\label{eq:multiple_bath_states}
|\vec{\boldsymbol{\omega}}_1^{(m_1)},\vec{\boldsymbol{\omega}}_2^{(m_2)}, \cdots ,\vec{\boldsymbol{\omega}}_M^{(m_M)}\rangle = \prod_{i=1}^M b_{i,\omega_{i,1}}^\dagger b_{i,\omega_{i,2}}^\dagger\cdots b_{i,\omega_{i,m_i}}^\dagger |\mathbf{0}\rangle /\sqrt{m_1m_2\dots m_M},
\end{equation}
where $\boldsymbol{\omega}_i^{(m_i)} = \{\omega_{i,1}, \omega_{i,2},\dots, \omega_{i,m_i}\}$ parameterizes the state of the $i^\text{th}$ waveguide. A similar \textit{but ordered} temporal mode basis can be constructed for the bath states
\begin{equation}\label{eq:multiple_bath_states_temp}
|\vec{\boldsymbol{\tau}}_1^{(m_1)},\vec{\boldsymbol{\tau}}_2^{(m_2)},\dots, \vec{\boldsymbol{\tau}}_M^{(m_M)}\rangle = \prod_{i=1}^M b_{i,\tau_{i,1}}^\dagger b_{i,\tau_{i,2}}^\dagger\cdots b_{i,\tau_{i,m_i}}^\dagger |\mathbf{0}\rangle ,
\end{equation}
where $\boldsymbol{\tau}_i^{(m_i)} = \{\tau_{i,1}, \tau_{i,2},\dots, \tau_{i, m_i}\}$ parametrizes the state of the $i^\text{th}$ waveguide. One important point to note is regarding the time ordering of the indices. Since photons at the same time index \textit{in the same waveguide} are identical, it is possible to impose the ordering $\tau_{i,1}<\tau_{i,2}<\cdots<\tau_{i,m_i}$. However, photons in different waveguides are not identical and hence, it is not possible to impose an ordering of indices across different waveguides. 

Following a procedure similar to the problem of a single waveguide coupled to the local system, we consider a system initially in the state $|\psi_\text{S}(0)\rangle$ and the waveguides in the vacuum state. Then, we similarly assume the system has only one ground state at long times $|\textbf{0}_\text{S}\rangle$, so the waveguide state at $t\to\infty$ can be expressed as
\begin{equation}
|\psi_{\text{B,I}}(t\to\infty)\rangle = \sum_{m_1 =0}^\infty  \cdots \sum_{m_M=0}^\infty \int \mathop{\textrm{d}\vec{\boldsymbol{\tau}}_1^{(m_1)}} \cdots \int \mathop{\textrm{d}\vec{\boldsymbol{\tau}}_M^{(m_M)}} \braket{\hat{\Omega}_{-}^\dagger}_{\vec{\boldsymbol{\tau}}_1^{(m_1)},\dots, \vec{\boldsymbol{\tau}}_M^{(m_M)}} |\vec{\boldsymbol{\tau}}_1^{(m_1)},\dots, \vec{\boldsymbol{\tau}}_M^{(m_M)}\rangle 
\end{equation}
where
\begin{equation}\label{eq:moller_multiple_wg}
\braket{\hat{\Omega}_{-}^\dagger}_{\vec{\boldsymbol{\tau}}_1^{(m_1)},\vec{\boldsymbol{\tau}}_2^{(m_2)},\dots, \vec{\boldsymbol{\tau}}_M^{(m_M)}} = \bra{\textbf{0}_\text{S}}\bra{\vec{\boldsymbol{\tau}}_1^{(m_1)},\dots, \vec{\boldsymbol{\tau}}_M^{(m_M)}} U(\infty, 0) |\textbf{0}_B\rangle |\psi_\text{S}(0)\rangle.
\end{equation}
Following a procedure similar to that outlined in Sec.~\ref{sec:scatteringmatrices}, Eq.~\ref{eq:moller_multiple_wg} can be simplified to
\begin{equation}
\braket{\hat{\Omega}_{-}^\dagger}_{\vec{\boldsymbol{\tau}}_1^{(m_1)},\vec{\boldsymbol{\tau}}_2^{(m_2)},\dots, \vec{\boldsymbol{\tau}}_M^{(m_M)}} = \bra{\textbf{0}_S}\bra{\vec{\boldsymbol{\tau}}_1^{(m_1)},\dots, \vec{\boldsymbol{\tau}}_M^{(m_M)}} U(\tau_\text{max}, 0) |\textbf{0}_B\rangle |\psi_\text{S}(0)\rangle ,
\end{equation}
where now $\tau_\text{max} = \max(T_P, \vec{\boldsymbol{\tau}}_1^{(m_1)}, \vec{\boldsymbol{\tau}}_2^{(m_2)},\dots, \vec{\boldsymbol{\tau}}_M^{(m_M)})$. As in the case of a single waveguide, evaluating this expression requires discretizing the waveguide basis and the temporal evolution of the system. The discretized annihilation operator $\Delta B_{i,j}$ creating an excitation in the time bin $(j\Delta t, (j+1)\Delta t)$ in the $i^\text{th}$ waveguide can be defined by
\begin{equation}
\Delta B_{i,j} = \frac{1}{\sqrt{\Delta t}} \int_{j\Delta t}^{(j+1)\Delta t} \textrm{d}\tau\  b_{i,\tau}.
\end{equation}
The temporal evolution of the system can be discretized by approximating the propagator evolving the system from $k\Delta t$ to $(k+1)\Delta t$ with
\begin{subequations}
\begin{eqnarray}
U((k+1)\Delta t, k\Delta t) &=& U[k+1,k] \\
&\approx& U_\text{S}[k+1,k]\prod_{i=1}^M U_\text{swap}^{(i)}[k+1, k],
\end{eqnarray}
\end{subequations}
where we need to define the swap operator corresponding to each waveguide as
\begin{equation}
U_\text{swap}^{(i)}[k+1,k] = \exp\big(\sqrt{\gamma_i \Delta t}\big(\Delta B_{i,0}^\dagger[k] a_i-a_i^\dagger \Delta B_{i,0}[k] \big) \big)
\end{equation}
with
\begin{equation}
\Delta B_{i,0}^\dagger[k] = \frac{1}{\sqrt{\Delta t}} \int_{k\Delta t}^{(k+1)\Delta t} \textrm{d}\tau\  b_{i,0}(t).
\end{equation}
Again, the propagator $U_\text{S}[k+1,k]$ contains the evolution of the system Hamiltonian by itself (Eq. \ref{eq:USitself}). Notably, all operators for different waveguides commute and those in the same waveguide obey the commutation from Eq. \ref{eq:commuteBjBk}, so
\begin{equation}
\left[\Delta B_{i,j},\Delta B_{l,0}^\dagger[k]\right]=\delta_{il}\delta[j-k].
\end{equation}

With this definition of $U[k+1,k]$, and following the procedure outlined in Sec.~\ref{sec:gensol}, we obtain a generalization of Eq.~\ref{eq:scatter3} for multiple waveguides:
\begin{subequations}
\begin{eqnarray}
\braket{\hat{\Omega}_{-}^\dagger}_{\tilde{\bm{\tau}}^{(N)}}&\equiv&\braket{\hat{\Omega}_{-}^\dagger}_{\vec{\boldsymbol{\tau}}_1^{(m_1)},\vec{\boldsymbol{\tau}}_2^{(m_2)},\dots, \vec{\boldsymbol{\tau}}_M^{(m_M)}} \\
&=&\bra{\textbf{0}_\text{S}} U_\text{eff}(\tau_\text{max}, \tilde{\tau}_N) \overleftarrow{\prod_{q=1}^N} \sqrt{\gamma_{Q[q]}}a_{Q[q]} U_\text{eff}(\tilde{\tau}_q,\tilde{\tau}_{q-1}) \ket{\psi_\text{S}(0)}\label{eq:scat_multiple_wg}
\end{eqnarray}
\end{subequations}
as a projection onto $|\vec{\boldsymbol{\tau}}_1^{(m_1)},\vec{\boldsymbol{\tau}}_2^{(m_2)},\dots, \vec{\boldsymbol{\tau}}_M^{(m_M)}\rangle$. Here, we unpack several new definitions:
\begin{itemize}
\item The total number of photons scattered is $N = m_1+m_2+ \cdots+ m_M$.
\item The time indices $\tilde{\tau}_q\in\tilde{\bm{\tau}}^{(N)}$ with $0<\tilde{\tau}_1 \leq \tilde{\tau}_2 \leq\cdots \leq \tilde{\tau}_N$ and $\tau_\text{max}=\text{max}(T_P,\tilde{\tau}_N)$.
\item $\tilde{\bm{\tau}}^{(N)}$ is a chronologically sorted set of all time indices from the $\vec{\boldsymbol{\tau}}_i^{(m_i)}$'s
\begin{equation}
\tilde{\bm{\tau}}^{(N)}= \text{sort}\{\vec{\boldsymbol{\tau}}_1^{(m_1)}+ \vec{\boldsymbol{\tau}}_2^{(m_2)}+ \cdots+ \vec{\boldsymbol{\tau}}_M^{(m_M)}\}.
\end{equation}
\item $Q[q]$ is the index of the waveguide corresponding to the photon scattered at $\tilde{\tau}_q$.
\item $U_\text{eff}(\tau_q,\tau_{q-1})$ is the propagator generated by the effective Hamiltonian $H_\text{eff}(t)$ evolution, where
\begin{equation}
H_\text{eff}(t) = H_\text{S}(t)-\textrm{i}\sum_{i=1}^M \frac{\gamma_i}{2}a_i^\dagger a_i.
\end{equation}
\end{itemize}
Eq.~\ref{eq:scat_multiple_wg} can be recast in terms of Heisenberg-like operators $\tilde{a}_i(\tau) = U_\text{eff}(0,\tau) a_i U_\text{eff}(\tau,0)$ to obtain a result similar to Eq.~\ref{eq:scatter_heisenberg}
\begin{equation}\label{eq:scat_multiple_wg_heis}
\braket{\hat{\Omega}_{-}^\dagger}_{\tilde{\bm{\tau}}^{(N)}} = 
\bra{\textbf{0}_S} U_\text{eff}(T_P,0) \overleftarrow{\prod_{q=1}^N} \sqrt{\gamma_{Q[q]}}\,\tilde{a}_{Q[q]}(\tilde{\tau}_q) \ket{\psi_S(0)}.
\end{equation}

Finally, we note that we can obtain a visually compact synthesis equation for $|\psi_{\text{B,I}}(\infty)\rangle$ by writing the state vector as
\begin{subequations}
\begin{eqnarray}
\ket{\tilde{\bm{\tau}}^{(N)}}&\equiv&|\vec{\boldsymbol{\tau}}_1^{(m_1)},\vec{\boldsymbol{\tau}}_2^{(m_2)},\dots, \vec{\boldsymbol{\tau}}_M^{(m_M)}\rangle \\
&=& \prod_{q=1}^N b_{Q[q],\tilde{\tau}_q}^\dagger|\mathbf{0}\rangle
\end{eqnarray}
\end{subequations}
and defining
\begin{equation}
\sum_{\{N\}}\int \mathop{\textrm{d}\tilde{\boldsymbol{\tau}}^{(N)}}=\sum_{m_1 =0}^\infty  \cdots \sum_{m_M=0}^\infty \int \mathop{\textrm{d}\vec{\boldsymbol{\tau}}_1^{(m_1)}} \cdots \int \mathop{\textrm{d}\vec{\boldsymbol{\tau}}_M^{(m_M)}}.
\end{equation}
Notably, each unique $\tilde{\bm{\tau}}^{(N)}$ defines a unique final state. Then,
\begin{equation}
|\psi_{\text{B,I}}(\infty)\rangle = \sum_{\{N\}}\int \mathop{\textrm{d}\tilde{\boldsymbol{\tau}}^{(N)}}\braket{\hat{\Omega}_{-}^\dagger}_{\tilde{\boldsymbol{\tau}}^{(N)}} \ket{\tilde{\boldsymbol{\tau}}^{(N)}}.
\end{equation}


\subsection{Connection to quantum trajectories and measurement theory}

In this section, we make a connection between our derived scattering amplitudes and quantum measurement theory. For a traditional approach to understanding how a system $H_\text{S}$ interacts with its environmental baths, a stochastic Schr{\"o}dinger equation is arrived at for the system's wavefunction \cite{breuer2002theory}. This equation gives a pure-state evolution of the system under the influence of the $M$ baths, at the expense of turning the system's wavefunction into a stochastic process. A complete realization of the stochastic process (from $t=0$ to $t\rightarrow\infty$) is uniquely identified by a sequence of collapse events, called the `measurement record'. The wavefunction conditioned on the measurement record undergoes a series of discontinuous jumps at times $\tilde{\bm{\tau}}^{(N)}=\{\tilde{\tau}_1,\dots,\tilde{\tau}_N\}$. Each jump collapses the wavefunction at time $\tilde{\tau}_q\in\tilde{\bm{\tau}}^{(N)}$ according to the operator $a_{Q[q]}$, where $Q[q]$ determines which bath `caused' the jump. This type of evolution is also equivalent to modeling measurement of the system by $M$ detectors, each coupled to the system via $a_{Q[q]}$.

Then, for the wavefunction conditioned on a measurement record \cite{gardiner2004quantum}
\begin{eqnarray}\label{eq:SSE}
\text{d}\psi_\text{c}(t)&\equiv&\psi_\text{c}(t+\mathop{\text{d}t})-\psi_\text{c}(t) \\
&=&\left[\left(-\text{i}H_\text{eff}(t) + \sum_{q=1}^{N}\frac{\gamma_{Q[q]}}{2}\langle a_{Q[q]}^\dagger a_{Q[q]}\rangle_\text{c}\right)\mathop{\text{d}t} + \sum_{q=1}^{N}\left(\frac{a_{Q[q]}}{\sqrt{\langle a_{Q[q]}^\dagger a_{Q[q]}\rangle_\text{c}}}-1\right)\mathop{\text{d}N_q(t)}\right]\psi_\text{c}(t),\nonumber
\end{eqnarray}
where $\braket{\dots}_\text{c}\equiv\braket{\psi_\text{c}(t)|\dots|\psi_\text{c}(t)}$. The Poisson increment $\mathop{\text{d}N_q(t)}$ represents a counting process that increments for each collapse registered by any detector, up to $N$ for a trajectory parameterized by $\tilde{\bm{\tau}}^{(N)}$. Using techniques from stochastic calculus, one can arrive at the probability density for a given trajectory \cite{kiukas2015equivalence,breuer2002theory}. Critically, this probability density is equivalent to the modulus square of our results in Sec. \ref{sec:multiplewaveguides}
\begin{eqnarray}
\mathbb{P}(\tilde{\bm{\tau}}^{(N)})=\left| \braket{\hat{\Omega}_{-}^\dagger}_{\tilde{\bm{\tau}}^{(N)}} \right|^2.\label{eq:pnsme}
\end{eqnarray}
Hence, we have shown the intricate connection between photon emission probabilities in a stochastic Schr{\"o}dinger equation and our microscopic scattering theory based on temporal waveguide modes.

On a historical note, Eq. \ref{eq:SSE} was often postulated or `unravelled' from the unconditioned dynamics of the system's density matrix \cite{carmichael2009open}. As a consequence, these stochastic Schr{\"o}dinger methods were originally derived by tracing over all bath degrees of freedom. Hence, there still exists a misconception that quantum trajectories are only the conditional dynamics of the system based on measurement records. In other words, the sequence of collapse events is often interpreted to exist in an abstract `detection Fock space'. We have shown here, by deriving the photon scattering amplitudes based on a microscopic theory of the system-bath interaction, without any trace operation over the waveguide modes, that the emission amplitudes from measurement theory have a physical connection to the waveguides' photonic state. We of course note that modern understanding of Eq. \ref{eq:SSE} also yields a similar interpretation of photon emission probabilities \cite{wiseman1994quantum,baragiola2017quantum}.


\section{EXAMPLES FOR COHERENTLY DRIVEN SYSTEMS}

While Eqs. \ref{eq:scat_multiple_wg} or \ref{eq:scat_multiple_wg_heis} may be difficult to compute for large $N$, experimentally relevant systems often only emit a few photons, and hence $\{N\}$ may be truncated to a few small integers---we will provide several examples of this nature here. In fact, some of the most experimentally relevant systems scatter just a few photons, which may potentially be used as quantum resources in quantum communication and computing \cite{o2009photonic}. We will consider two types: single- and two-photon sources from quantum two-level systems and from spontaneous parametric downconversion or four-wave mixing. We solve the two-level system's emitted state analytically and the downconversion/mixing source's emitted state numerically.

We offer a brief note about the numerical applicability of our approach. Suppose that the waveguides are each discretized into $B$ bins. Then, the total computational complexity scales polynomially with the number of bins, given that we need to compute no more than $(M B)^N/{N!}$ scattering amplitudes, where $N$ is the number of photons in the state and $M$ is the number of waveguides. In most applications involving the generation of pulsed quantum light, $B\gg M,N$ and hence this problem is tractable. Further, the computation can be executed in a highly efficient manner: only $B^2$ propagators must be computed and then the state amplitudes can all be assembled in parallel on a GPU.


\subsection{Quantum two-level system}

In this section, we compute the output state of a single waveguide coupled to a two-level system driven by a short optical pulse (scenario (a) in Fig. \ref{figure:1}). The quantum two-level system is one of the most fundamental building blocks of quantum optics \cite{Cohen-Tannoudji1992-uo}, which models a single discrete atomic transition. This type of system has been behind fundamental discoveries such as photon anti-bunching \cite{Michler2000-dx,Kimble1977-rw}, Mollow triplets \cite{Flagg2009-zi,Mollow1969-dv}, and quantum interference of indistinguishable photons \cite{Santori2002-wi,Hong1987-lv}. After almost two decades of development in a solid-state environment, the quantum two-level system is now poised to serve the pivotal role of an on-demand single-photon source \cite{Michler2000-dx,Santori2002-wi,he2013demand,schneider2015single,yuan2002electrically,claudon2010highly,unsleber2016highly,schlehahn2016electrically,somaschi2016near,ding2016demand,michler,senellart2017high}---by converting laser pulses with Poissonian counting statistics to single photons---for quantum networks \cite{Loredo2016-jz,Aharonovich2016-pq,o2009photonic}.
More recently, multi-photon quantum state generators have found strong interest as replacements for the single-photon source in many quantum applications  \cite{Rundquist2014-kf,munoz2014emitters,afek2010high}. To this end, we recently discovered that two-level systems may also generate pulses containing two-photons \cite{fischer2017signatures,fischer2017pulsed}. Here, we provide analytic models for both of these phenomena, by computing scattered fields from the driven two-level system.

\begin{figure}
  \centering
  \includegraphics[scale=1]{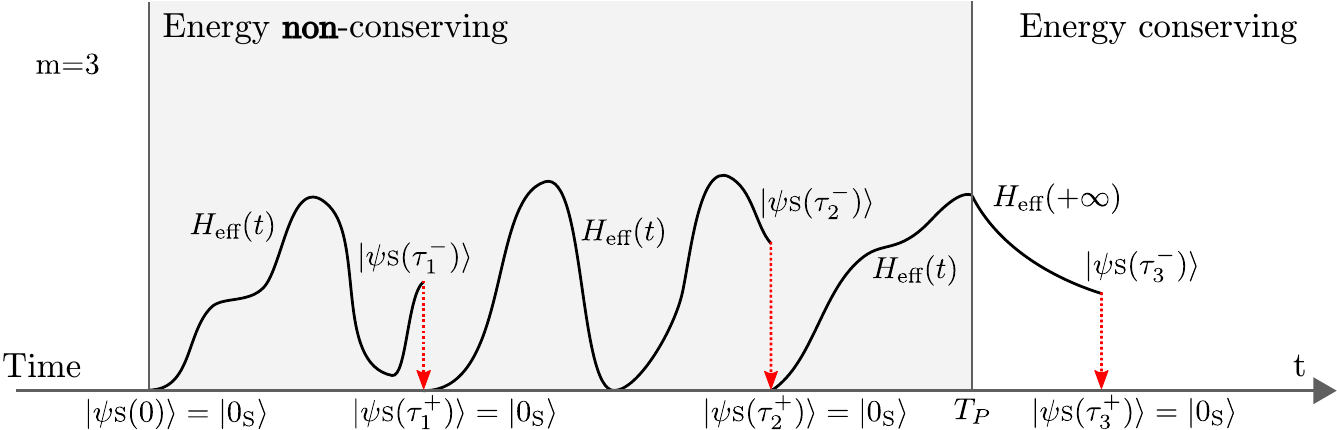}
  \caption{Sketch of a path for two-level system evolution, formalized by Eq. \ref{eq:TLSsimp}, and specifically for the path amplitude with a realization of the photon emission times $\vec{\bm{\tau}}^{(3)}=\{\tau_1,\tau_2,\tau_3\}$ and $\tau_2<T_P<\tau_3$. Compared to an arbitrary system, the two-level system has the interesting property that it can only store one excitation at a time. As a result, every photon emission collapses the system in its ground state $\ket{0_\textrm{S}}$. Similarly, a maximum of only one emission may occur in the energy conserving phase.}
  \label{figure:4}
\end{figure}

The two-level system has the interesting property that it can only store one excitation at a time (Fig. \ref{figure:4}).
As a result, the system operator $a$ is truncated to one excitation, $a\rightarrow\sigma$, where $\sigma$ is the atomic dipole operator. It is defined by its action on the system's Hilbert space, which is spanned by
\begin{equation}
\ket{0_\textrm{S}},\quad\sigma\ket{0_\textrm{S}}=0,\quad\textrm{and}\quad\ket{1_\textrm{S}}=\sigma^\dagger\ket{0_\textrm{S}}.
\end{equation}
The Hamiltonian for the bare two-level system has the simple form
\begin{equation}
H_{0\textrm{S}}=\omega_0 \sigma^\dagger \sigma.
\end{equation}


\subsubsection{\textit{Traditional theory as a single-photon source}}


We first consider the case where the two-level system conserves energy for all time, i.e. $T_P\rightarrow 0$. From Eq. \ref{eq:ham}, we then have the time-dependent part of $H_{\textrm{S}}$ as
\begin{equation}
H_{1\textrm{S}}(t)=0.
\end{equation}
In this scenario, we will prepare the system in its excited state
\begin{equation}
\ket{\psi_\textrm{S}(0)}=\ket{1_\textrm{S}}\label{eq:initstate0}
\end{equation}
and calculate the single-photon wavepacket \textit{spontaneously} emitted from the system. 

The effective non-Hermitian Hamiltonian that governs the evolution is given by
\begin{equation}
H_\textrm{eff}=H_{0\textrm{S}}-\textrm{i}\frac{\gamma}{2} \sigma^\dagger \sigma.
\end{equation}
Because energy is conserved, we know that the output field will be a superposition of field states with one excitation. Furthermore, due to the time-independence of $H_\textrm{eff}$, we can simplify the integral operator in $\braket{\hat{\Omega}_-^\dagger}_{\tau_1}$ to an exponential one
\begin{subequations}
\begin{eqnarray}
\braket{\hat{\Omega}_-^\dagger}_{\tau_1}&=&\sqrt{\gamma}\bra{1_\textrm{S}}\mathcal{T}e^{-\textrm{i}\int_{0}^{\tau_1}\mathop{\textrm{d}t} H_\textrm{eff}}\ket{1_\textrm{S}}\\
&=&\sqrt{\gamma}\bra{1_\textrm{S}}e^{-\textrm{i}H_\textrm{eff}\tau_1}\ket{1_\textrm{S}}\\
&=&\sqrt{\gamma}\,\textrm{e}^{-\textrm{i} \omega_0 \tau_1}\textrm{e}^{-\gamma \tau_1/2}.\label{eq:spemit1}
\end{eqnarray}
\end{subequations}
Hence, the final state of the waveguide after spontaneous emission can be written as
\begin{equation}
\ket{\psi_\textrm{B,I}(\infty)}=\sqrt{\gamma}\int_0^\infty \mathop{\textrm{d}\tau_1} \textrm{e}^{-\textrm{i} \omega_0 \tau_1}\textrm{e}^{-\gamma \tau_1/2} \ket{\tau_1}.
\end{equation}
This is a standard result and has been derived through many different means, including the Bethe ansatz, frequency-mode scattering theory, and direct integration \cite{liao2016photon,roy2016strongly}. This has been the traditional theory of using two-level systems as single-photon sources, ignoring the precise mechanism that excites the system at time $t=0$, as in Eq. \ref{eq:initstate0}.


\subsubsection{\textit{General theory of photon emission}}


However, to fully understand the two-level system as a photon source, it's important to consider the excitation dynamics of the system. The way to generate the highest purity single-photon source from a two-level system is to excite the system with a coherent laser pulse \cite{fischer2016dynamical}. Here, the coherent pulse drives Rabi oscillations in the two-level system, which are terminated when the system is maximally excited. Recently, we began investigating these dynamics for two-photon emission \cite{fischer2017signatures,fischer2017pulsed}---both to characterize the two-photon errors that spoil single-photon emission and the potential to use the two-photon state as a quantum resource.

In this scenario, the laser will inject energy into the system and hence we set
\begin{equation}
\ket{\psi_\textrm{S}(0)}=\ket{0_\textrm{S}}.
\end{equation}
The scattering matrix elements for the two-level system simplify significantly. To show this, suppose we insert the atomic identity operator $\mathbb{1}_\textrm{S}=\ket{0_s}\bra{0_s} + \ket{1_s}\bra{1_s}$ before the operator $\sigma$. Because
\begin{subequations}
\begin{eqnarray}
\mathbb{1}_\textrm{S}\sigma&=&\left(\ket{0_s}\bra{0_s} + \ket{1_s}\bra{1_s}\right)\sigma\\
&=&\ket{0_s}\bra{0_s}\sigma,
\end{eqnarray}
\end{subequations}
we then make the replacement $\sigma\rightarrow\ket{0_s}\bra{0_s}\sigma$ in Eq. \ref{eq:scatter3}
\begin{subequations}
\begin{eqnarray}
\braket{\hat{\Omega}_-^\dagger}&=&\bra{{0}_\textrm{S}} U_\text{eff}(\tau_\text{max}, \tau_m)\sqrt{\gamma}\sigma\, U_\text{eff}(\tau_m, \tau_{m-1}) \cdots\sqrt{\gamma}\sigma\,U_\text{eff}(\tau_2,\tau_1)\sqrt{\gamma}\sigma\,U_\text{eff}(\tau_1,0)\ket{0_\textrm{S}}   \qquad \\
&=&\bra{{0}_\textrm{S}} U_\text{eff}(\tau_\text{max}, \tau_m)\ket{0_\textrm{S}}\bra{0_\textrm{S}}\sqrt{\gamma}\sigma\, U_\text{eff}(\tau_m, \tau_{m-1})\ket{0_\textrm{S}}\cdots \nonumber\\
&& \hspace{100pt}\cdots\bra{0_\textrm{S}}\sqrt{\gamma}\sigma\,U_\text{eff}(\tau_2,\tau_1)\ket{0_\textrm{S}}\bra{0_\textrm{S}}\sqrt{\gamma}\sigma\,U_\text{eff}(\tau_1,0)\ket{0_\textrm{S}}.
\end{eqnarray}
\end{subequations}
Now, we see the result that each evolution between emissions, i.e. from each $\tau_{q-1}$ to $\tau_q$, is uncorrelated! Hence, the expectations may be evaluated independently. In this section, we will be evaluating $U_\text{eff}$ by inspection, so it is easier to use the definition from Eq. \ref{eq:Ueff} to write the solution in the form (defining $\tau_0=0$)
\begin{eqnarray}
\hspace{-10pt}&&\braket{\hat{\Omega}_-^\dagger}_{\vec{\bm{\tau}}^{(m)}}=\nonumber\\
\hspace{-10pt}&&\begin{cases}
(\sqrt{\gamma})^m\bra{{0}_\textrm{S}} \mathcal{T}\textrm{e}^{-\textrm{i}\int_{\tau_{m}}^{T_P}\mathop{\textrm{d}t} H_\textrm{eff}(t)}\ket{0_\textrm{S}}\prod_{q=1}^m\bra{{1}_\textrm{S}}\mathcal{T}\textrm{e}^{-\textrm{i}\int_{\tau_{q-1}}^{\tau_{q}}\mathop{\textrm{d}t} H_\textrm{eff}(t)}\ket{0_\textrm{S}} & \text{if } \tau_m<T_P\\
(\sqrt{\gamma})^m\prod_{q=1}^m\bra{{1}_\textrm{S}}\mathcal{T}\textrm{e}^{-\textrm{i}\int_{\tau_{q-1}}^{\tau_{q}}\mathop{\textrm{d}t} H_\textrm{eff}(t)}\ket{0_\textrm{S}}  & \text{if } \tau_{m-1}<T<\tau_m\\
0&\text{otherwise}
\end{cases},\qquad\quad\label{eq:TLSsimp}
\end{eqnarray}
where energy conservation dictates that only one emission can occur after $T_P$ due to the fact that the system holds a maximum of one excitation. Also, note: unlike in Eq. \ref{eq:scatter3}, the products are no longer ordered because of their statistical independence.

From a numerical implementation perspective, the expression $\mathcal{T}\textrm{e}^{-\textrm{i}\int_{\tau_{q-1}}^{\tau_q}\mathop{\textrm{d}t} H_\textrm{eff}(t)}\ket{0_\textrm{S}}$ means to integrate the Schr{\"o}dinger equation as $\textrm{i}\frac{\partial}{\partial t}\ket{\psi_\textrm{S}(t)}=H_\textrm{eff}(t)\ket{\psi_\textrm{S}(t)}$, subject to $\ket{\psi_\textrm{S}(\tau_{q-1})}=\ket{0_\textrm{S}}$. Then $\ket{\psi_\textrm{S}(\tau_{q})}=\mathcal{T}\textrm{e}^{-\textrm{i}\int_{\tau_{q-1}}^{\tau_{q}}\mathop{\textrm{d}t} H_\textrm{eff}(t)}\ket{0_\textrm{S}}$ and take either $\braket{0_\textrm{S}|\psi_\textrm{S}(\tau_q)}$ or $\braket{1_\textrm{S}|\psi_\textrm{S}(\tau_q)}$, which correspond either to a waiting period or an emission, respectively. This process is repeated for each $\tau_{q-1}\rightarrow\tau_q$ and then the expectations are assembled according to Eq. \ref{eq:TLSsimp}.

Now, consider the specific form of $H_{1\text{S}}(t)$. If we consider the frequency of the driving laser pulse to be resonant with the two-level system, then we have the time-dependent part of $H_\textrm{S}$ as
\begin{equation}
H_\textrm{1S}(t) = f(t)\textrm{e}^{-\textrm{i}\omega_0 t}\sigma^\dagger +f^*(t) \textrm{e}^{\textrm{i}\omega_0 t}\sigma,
\end{equation}
where $f(t)$ is an arbitrary temporal function that contains the pulse shape and the overlap of the system's dipole moment with the pulse's electric field (see Appendix C for a detailed derivation of this driving term via a Mollow transformation; we assume a semi-classical limit here). Then, the non-Hermitian Hamiltonian that governs the evolution is also time-dependent
\begin{equation}
H_\textrm{eff}(t)=H_\textrm{0S}+H_\textrm{1S}(t)-\textrm{i}\frac{\gamma}{2} \sigma^\dagger \sigma.
\end{equation}

We choose a simple square shape for the pulse $f(t)$, so that we may arrive at nice analytic expressions for the scattered fields (though one could easily numerically integrate Eq. \ref{eq:TLSsimp} for more complicated pulse shapes)
\begin{equation}
H_{1\textrm{S}}(t)=
\begin{cases}
\Omega\left(\textrm{i}\textrm{e}^{\textrm{-i}\omega_0 t}\sigma^\dagger - \textrm{i} \textrm{e}^{\textrm{i}\omega_0 t}\sigma\right) & \text{if } 0<t<T_P\\
0 & \text{otherwise},
\end{cases}
\end{equation}
where $\Omega$ is the Rabi frequency (and has no relation to the Moller scattering operators). Now, the expectations in Eq. \ref{eq:TLSsimp} that end with a photon emission simplify into two categories
\begin{eqnarray}
&&\bra{1_\textrm{S}}e^{-\textrm{i}\int_{\tau_{q-1}}^{\tau_q}\mathop{\textrm{d}t} H_\textrm{eff}(t)}\ket{0_\textrm{S}}=\\
&&\qquad\qquad\begin{cases}
\bra{1_\textrm{S}}e^{-\textrm{i}H_\textrm{eff}(0^+)(\tau_q-\tau_{q-1})}\ket{0_\textrm{S}} &\;\textrm{ if }\tau_q<T_P\\
\bra{1_\textrm{S}}e^{-\textrm{i}H_\textrm{eff}(\infty)(\tau_{q}-T_P)}\ket{1_\textrm{S}}\bra{1_\textrm{S}}e^{-\textrm{i}H_\textrm{eff}(0^+)(T_P-\tau_{q-1})}\ket{0_\textrm{S}}&\;\textrm{ if }\tau_{q-1}<T_P<\tau_{q}\nonumber
\end{cases},
\end{eqnarray}
where conservation of energy for $t>T_P$ allows us to insert the projector $\ket{1_\textrm{S}}\bra{1_\textrm{S}}$. Computing these expectations, we arrive at
\begin{eqnarray}
\bra{1_\textrm{S}}e^{-\textrm{i}H_\textrm{eff}(\infty)\tau}\ket{1_\textrm{S}}= \textrm{e}^{-\textrm{i} \omega_0 \tau} \textrm{e}^{-\gamma \tau/2},
\end{eqnarray}
like in Eq. \ref{eq:spemit1}, and
\begin{eqnarray}
\bra{1_\textrm{S}}e^{-\textrm{i}H_\textrm{eff}(0^+)\tau}\ket{0_\textrm{S}}
= \textrm{e}^{-\textrm{i} \omega_0 \tau}\textrm{e}^{-\gamma \tau/4} \frac{\Omega}{\Omega'}\sin{\left(\Omega' \tau\right)}
\end{eqnarray}
with the (potentially \textbf{complex}) Rabi frequency
\begin{equation}
\Omega'\equiv \sqrt{\Omega^2-\left(\frac{\gamma}{4}\right)^2}.
\end{equation}
Then, for emission paths that end during the energy-nonconserving phase, we also need the waiting integral
\begin{subequations}
\begin{eqnarray}
\bra{0_\textrm{S}}e^{-\textrm{i}\int_{\tau_m}^{T_P}\mathop{\textrm{d}t} H_\textrm{eff}(t)}\ket{0_\textrm{S}}&=&\bra{0_\textrm{S}}e^{-\textrm{i}H_\textrm{eff}(0^+)(T_P-\tau_m)}\ket{0_\textrm{S}}\\
&=&\textrm{e}^{-\gamma (T_P-\tau_m)/4} \left(\cos{\left(\Omega' (T_P-\tau_m)\right)}+\frac{\gamma}{4} \frac{\sin{\left(\Omega' (T_P-\tau_m)\right)}}{\Omega'} \right),\qquad\quad
\end{eqnarray}
\end{subequations}
which calculates the amplitude no photon emission occurs between $\tau_m$ and $T_P$.

Combining these together, we arrive at the general solution for the scattered field from a two-level system driven by a square pulse
\begin{eqnarray}
\braket{\hat{\Omega}_-^\dagger}_{\vec{\bm{\tau}}^{(m)}}=&\label{eq:TLScomplete}\\
&\hspace{-87pt}\begin{cases}
(\sqrt{\gamma})^m \textrm{e}^{-\textrm{i} \omega_0 \tau_m}\textrm{e}^{-\gamma T_P/4}\left(\cos{\left(\Omega' (T_P-\tau_m)\right)}+\frac{\gamma}{4} \frac{\sin{\left(\Omega' (T_P-\tau_m)\right)}}{\Omega'} \right)\prod_{q=1}^m\frac{\Omega}{\Omega'}\sin{\left(\Omega' (\tau_{q}-\tau_{q-1})\right)} & \text{if } \tau_m<T_P\\
(\sqrt{\gamma})^m\textrm{e}^{-\textrm{i} \omega_0 \tau_m}\textrm{e}^{-\gamma (\tau_m-T_P)/2}\textrm{e}^{-\gamma T_P/4}\frac{\Omega}{\Omega'}\sin{\left(\Omega' (T_P-\tau_{m-1})\right)}\prod_{q=1}^{m-1}\frac{\Omega}{\Omega'}\sin{\left(\Omega' (\tau_{q}-\tau_{q-1})\right)} & \text{if } \tau_{m-1}<T_P<\tau_m \\
0&\textrm{otherwise }\\
\end{cases}\nonumber.
\end{eqnarray}
In either the strong driving limit $\Omega\gg\gamma$ (previously identified by Mollow \cite{mollow1975pure}) or the short pulse limit $T_P \ll \frac{1}{\gamma}$, the scattering elements have a particularly simple form
\begin{eqnarray}
\braket{\hat{\Omega}_-^\dagger}_{\vec{\bm{\tau}}^{(m)}}\approx&\hspace{-30pt}\\
&\hspace{-68pt}\begin{cases}
(\sqrt{\gamma})^m \textrm{e}^{-\textrm{i} \omega_0 \tau_m}\textrm{e}^{-\gamma T_P/4} \cos{\left(\Omega (T_P-\tau_m)\right)}\prod_{q=1}^m\sin{\left(\Omega (\tau_{q}-\tau_{q-1})\right)} & \text{if } \tau_m<T_P\\
(\sqrt{\gamma})^m\textrm{e}^{-\textrm{i} \omega_0 \tau_m}\textrm{e}^{-\gamma (\tau_m-T_P)/2}\textrm{e}^{-\gamma T_P/4}\sin{\left(\Omega (T_P-\tau_{m-1})\right)}\prod_{q=1}^{m-1}\sin{\left(\Omega (\tau_{q}-\tau_{q-1})\right)} & \text{if } \tau_{m-1}<T_P<\tau_m \\
0&\text{otherwise}
\end{cases}\nonumber.
\end{eqnarray}
For strong driving $\Omega'\approx\Omega$, and for weak short pulses $\frac{\Omega}{\Omega'}\sin{(\Omega'\,\tau)}\approx\sin{(\Omega\,\tau)}$.

From these scattering elements, it is quite simple to identify the origin of photon antibunching (similarly did Mollow \cite{mollow1975pure} and Pletyukhov \cite{pletyukhov2012scattering}). Photon antibunching is a statement that the intensity correlation between two different points in space or time is zero. If we approximate $\sqrt{\omega}\approx \sqrt{\omega_0}$ in the intensity operator since $\gamma\ll\omega_0$, we may equally look at photon-number correlations. Then, consider the second-order coherence function of the field at position $\vec{r}=0^+$ \cite{loudon2000quantum}, which is a special case of Eq. \ref{eq:allcorr}. This is the correlation that would be measured were a photon-number-resolving detector placed \textit{right} in front of the two-level system and measuring the scattered field
\begin{eqnarray}
G^{(2)}(t_1, t_2)&=&\braket{\psi_\textrm{B,I}(\infty)|b_0^\dagger(t_1) b_0^\dagger(t_2)b_0(t_2)b_0(t_1)|\psi_\textrm{B,I}(\infty)}\label{eq:g2t1t2}.
\end{eqnarray}
(This is a slight abuse of notation, because the actual state has propagated out to infinity. This statement is anyways true since the interaction of the system with the waveguide is spatially and temporally localized.)

To understand how the form of $\braket{\hat{\Omega}_-^\dagger}_{\vec{\bm{\tau}}^{(m)}}$ requires antibunching, consider the operator $$b_0^\dagger(t_1) b_0^\dagger(t_2)b_0(t_2)b_0(t_1).$$ It removes two photons from both the initial and final states, so neither zero-photon nor single-photon basis states contribute. Further, the expectation is a sum of terms, each based on $\braket{\hat{\Omega}_-^\dagger}_{\vec{\bm{\tau}}^{(m)}}$, where terms with different numbers of photons $m$ do not interfere (see Appendices A and B).

Then, look at only $b_0(t_2)b_0(t_1)\ket{\psi_\textrm{B,I}(\infty)}$, and the interaction-picture state is most easily written in the temporal mode basis. We want to commute the two annihilation operators towards the right into the state. Each commutation of an annihilation operator with a temporal mode operator will yield a delta function (Eq. \ref{eq:commute1}). Thus, each nonzero term in the photon-subtracted initial state will contain a product of two delta functions like $\delta(t_1-\tau_q)\delta(t_2-\tau_p)$, where $q\neq p$ and $p,q\in\{1,\dots,m\}$ (see Appendix B). When $t_1\rightarrow t_2$, then at least two time indices in each term of $\braket{\hat{\Omega}_-^\dagger}_{\vec{\bm{\tau}}^{(m>1)}}$ will approach each other and be the same. When two time-indices approach each other, $\braket{\hat{\Omega}_-^\dagger}_{\vec{\bm{\tau}}^{(m>1)}}\rightarrow 0$ for the two-level system from Eq. \ref{eq:TLScomplete}, and there is perfect anti-bunching to all orders of photon number in the scattered field state!


\subsubsection{\textit{Short pulse regime}}

The short pulse regime $T_P \ll \frac{1}{\gamma}$, corresponds to the previously identified single-photon and two-photon emission regimes \cite{fischer2017signatures,fischer2017pulsed}. There, the system undergoes somewhat high-fidelity Rabi oscillations between its ground $\ket{0_\textrm{s}}$ and excited states $\ket{1_\textrm{s}}$ as a function of the interacted pulse area $A_\textrm{R}$, with the Rabi frequency defined as
\begin{equation}
\Omega=\frac{A_\textrm{R}}{2T_P}.
\end{equation}
In this section, we will keep all terms to first order in $\gamma T_P$ to arrive at nice analytic forms for various quantities of interest.

\begin{figure}[t]
  \centering
  \includegraphics[scale=0.6]{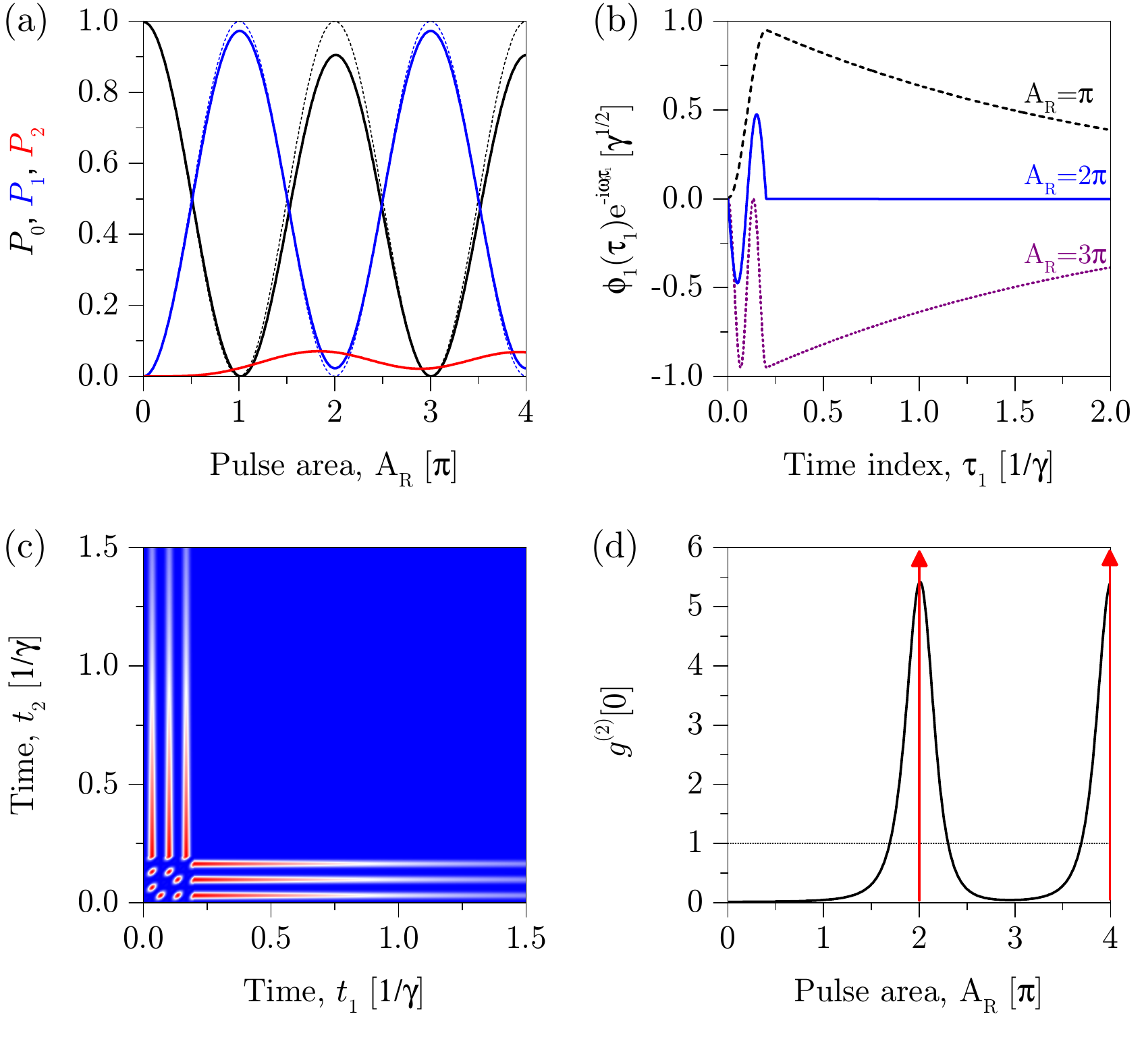}
  \caption{Statistical characterization of photon emission under Rabi oscillations---a two-level system is driven by a short optical pulse $T_P=0.2/\gamma$. (a) Signatures of Rabi oscillations in the emitted photon numbers versus pulse area, showing vacuum $P_0$, single-photon $P_1$, and two-photon $P_2$ contributions. Dotted lines show ideal values for arbitrarily short pulses. (b) Envelopes of single-photon amplitudes for different pulse areas $\braket{\tau_1|\phi_1}\text{e}^{\text{i}\omega_0\tau_1}$. (c) Second-order coherence $G^{(2)}(t_1,t_2)$ for $A_\textrm{R}=6\pi$. (d) Pulse-wise second-order coherence $g^{(2)}[0]$, black shows case of $T_P=0.2/\gamma$, while red arrows depict the singularities for even-$\pi$ pulses of arbitrarily short length.}
  \label{figure:5}
\end{figure}

First, consider the case where \textbf{no photons are scattered}. Then,
\begin{subequations}
\begin{eqnarray}
\braket{\mathbf{0}|\Psi(t\rightarrow\infty)} &=& \bra{0_\textrm{S}}\mathcal{T}e^{-\textrm{i}\int_{0}^{T_P}\mathop{\textrm{d}t} H_\textrm{eff}(t)}\ket{0_\textrm{S}}\\
&=&\textrm{e}^{-\gamma T_P/4}\left(\cos{\left(\Omega' \,T_P\right)}+\frac{\gamma}{4} \frac{\sin{\left(\Omega' \,T_P\right)}}{\Omega'} \right)\\
&\approx&\textrm{e}^{-\gamma T_P/4} \left(\cos{(A_\textrm{R}/2)}+\frac{\gamma T_P}{2A}\sin{(A_\textrm{R}/2)}\right).
\end{eqnarray}
\end{subequations}
Hence, the probability of scattering zero photons is given by
\begin{subequations}
\begin{eqnarray}
P_0 &\equiv& \left|\braket{\mathbf{0}|\Psi(t\rightarrow\infty)} \right|^2\label{eq:P0gen}\\
& \approx & \textrm{e}^{-\gamma T_P/2}\left(\cos{(A_\textrm{R}/2)}+\frac{\gamma T_P}{2A_\textrm{R}}\sin{(A_\textrm{R}/2)}\right)^2,
\end{eqnarray}
\end{subequations}
which is shown as the black curve in Fig. \ref{figure:5}a for a short pulse.

Second, consider the cases where \textbf{one photon is scattered}. Then,
\begin{subequations}
\begin{eqnarray}
&&\hspace{-20pt}\braket{0_\textrm{S},\tau_1|\Psi(t\rightarrow\infty)} \nonumber\\
&&\hspace{-20pt}\qquad=\sqrt{\gamma}\bra{0_\textrm{S}}\mathcal{T}e^{-\textrm{i}\int_{\tau_1}^{T_P}\mathop{\textrm{d}t} H_\textrm{eff}(t)}\ket{0_\textrm{S}}\bra{1_\textrm{S}}\mathcal{T}e^{-\textrm{i}\int_{0}^{\tau_1}\mathop{\textrm{d}t} H_\textrm{eff}(t)}\ket{0_\textrm{S}}\\
&&\hspace{-20pt}\qquad=
\begin{cases}
\sqrt{\gamma}\,\textrm{e}^{-\textrm{i} \omega_0 \tau_1}\textrm{e}^{-\gamma T_P /4}\left(\cos{\left(\Omega' (T_P-\tau_m)\right)}+\frac{\gamma}{4} \frac{\sin{\left(\Omega' (T_P-\tau_m)\right)}}{\Omega'} \right)\frac{\Omega}{\Omega'}\sin{\left(\Omega' \,\tau_1\right)} &\quad\textrm{ if }\tau_1<T_P\\
\sqrt{\gamma}\,\textrm{e}^{-\textrm{i} \omega_0 \tau_1}\textrm{e}^{-\gamma (\tau_1-T_P)/2}\textrm{e}^{-\gamma T_P /4}\frac{\Omega}{\Omega'}\sin{\left(\Omega' \,T_P\right)} & \quad\textrm{ if } T_P<\tau_1
\end{cases}\nonumber\\
&&\hspace{-20pt}\qquad\approx
\begin{cases}
\sqrt{\gamma}\,\textrm{e}^{-\textrm{i} \omega_0 \tau_1}\textrm{e}^{-\gamma T_P /4}\cos{\left(\frac{A_\textrm{R}}{2T_P}(T_P-\tau_1)\right)}\sin{\left(\frac{A_\textrm{R}}{2T_P}\tau_1 \right)} &\quad\textrm{ if }\tau_1<T_P\\
\sqrt{\gamma}\,\textrm{e}^{-\textrm{i} \omega_0 \tau_1}\textrm{e}^{-\gamma (\tau_1-T_P)/2}\textrm{e}^{-\gamma T_P /4}\sin{\left(A_\textrm{R}/2\right)} & \quad\textrm{ if } T_P<\tau_1
\end{cases}.
\end{eqnarray}
\end{subequations}
Hence, the single-photon part of the bath wavefunction is written as
\begin{subequations}
\begin{eqnarray}
\ket{\phi_1}&=&\int_0^\infty\mathop{\textrm{d}\tau_1}\braket{0_\textrm{S},\tau_1|\Psi(t\rightarrow\infty)} \ket{\tau_1}\\
&\approx& \sqrt{\gamma}\,\textrm{e}^{-\gamma T_P /4}\int_0^{T_P}\mathop{\textrm{d}\tau_1} \textrm{e}^{-\textrm{i} \omega_0 \tau_1}\cos{\left(\frac{A_\textrm{R}}{2T_P}(T_P-\tau_1)\right)}\sin{\left(\frac{A_\textrm{R}}{2T_P}\tau_1 \right)}\ket{\tau_1}+ \\
&&\qquad\qquad \sin{\left(A_\textrm{R}/2\right)}\textrm{e}^{-\gamma T_P /4}\int_{T_P}^\infty\mathop{\textrm{d}\tau_1} \sqrt{\gamma}\,\textrm{e}^{-\textrm{i} \omega_0 \tau_1}\textrm{e}^{-\gamma (\tau_1-T_P)/2}\ket{\tau_1}.
\end{eqnarray}
\end{subequations}
This amplitude shows Rabi oscillations during the pulse period $0<\tau_1<T_P$, followed by exponential decay (Fig. \ref{figure:5}b). Quite interesting, is that the single-photon amplitude is zero outside of the pulse period when $A_\textrm{R}=2\pi$ to second-order in $\gamma T_P$. The total probability of emitting one photon is then given by
\begin{subequations}
\begin{eqnarray}
P_1 &\equiv& \braket{\phi_1|\phi_1} \\
&\approx& \frac{1}{2}\textrm{e}^{-\gamma T_P /2} \left(1 -\cos{\left(A_\textrm{R}\right)} + \frac{\gamma T_P}{2}\left( 1-\cos{(A_\textrm{R})}/2 - \frac{\sin{(A_\textrm{R})}}{2A_\textrm{R}}\right)\right),
\end{eqnarray}
\end{subequations}
which is shown as the blue curve in Fig. \ref{figure:5}a for a short pulse.

Third, consider the cases where \textbf{two photons are scattered}. Then,
\begin{eqnarray}
&&\hspace{-20pt}\braket{0_\textrm{S},\{\tau_1 ,\tau_2\}|\Psi(t\rightarrow\infty)} \nonumber\\
&&\hspace{-20pt}=\begin{cases}
\bra{0_\textrm{S}}\mathcal{T}e^{-\textrm{i}\int_{\tau_2}^{T_P}\mathop{\textrm{d}t} H_\textrm{eff}(t)}\ket{0_\textrm{S}}\bra{1_\textrm{S}}\mathcal{T}e^{-\textrm{i}\int_{\tau_1}^{\tau_2}\mathop{\textrm{d}t} H_\textrm{eff}(t)}\ket{0_\textrm{S}}\bra{1_\textrm{S}}\mathcal{T}e^{-\textrm{i}\int_{0}^{\tau_1}\mathop{\textrm{d}t} H_\textrm{eff}(t)}\ket{0_\textrm{S}} &\textrm{if }\tau_2<T_P\nonumber\\
\bra{1_\textrm{S}}\mathcal{T}e^{-\textrm{i}\int_{\tau_1}^{\tau_2}\mathop{\textrm{d}t} H_\textrm{eff}(t)}\ket{0_\textrm{S}}\bra{1_\textrm{S}}\mathcal{T}e^{-\textrm{i}\int_{0}^{\tau_1}\mathop{\textrm{d}t} H_\textrm{eff}(t)}\ket{0_\textrm{S}} & \textrm{if } \tau_1<T_P<\tau_2
\end{cases}\\
&&\hspace{-20pt}\approx
\begin{cases}
(\sqrt{\gamma})^2\,\textrm{e}^{-\textrm{i} \omega_0 \tau_1}\textrm{e}^{-\gamma T_P /4}\cos{\left(\frac{A_\textrm{R}}{2T_P}(T_P-\tau_2)\right)}\sin{\left(\frac{A_\textrm{R}}{2T_P}(\tau_2-\tau_1) \right)}\sin{\left(\frac{A_\textrm{R}}{2T_P}\tau_1 \right)} &\textrm{if }\tau_2<T_P\\
(\sqrt{\gamma})^2\,\textrm{e}^{-\textrm{i} \omega_0 \tau_1}\textrm{e}^{-\gamma (\tau_2-T_P)/2}\textrm{e}^{-\gamma T_P /4}\sin{\left(\frac{A_\textrm{R}}{2T_P}(T_P-\tau_1) \right)}\sin{\left(\frac{A_\textrm{R}}{2T_P}\tau_1 \right)} & \textrm{if } \tau_1<T_P<\tau_2\\
0 & \textrm{otherwise } 
\end{cases}\nonumber.
\end{eqnarray}
Hence, the two-photon part of the bath wavefunction is written as
\begin{subequations}
\begin{eqnarray}
\ket{\phi_2}&\approx&\int_0^\infty\mathop{\textrm{d}\tau_1}\int_{\tau_1}^\infty\mathop{\textrm{d}\tau_2}\braket{0_\textrm{S},\{\tau_1 ,\tau_2\}|\Psi(t\rightarrow\infty)} \ket{\tau_1,\tau_2}\\
&=&(\sqrt{\gamma})^2\, \textrm{e}^{-\gamma T_P /4} \times\label{eq:phi2}\\
&&\hspace{10pt}\bigg[\int_0^{T_P}\mathop{\textrm{d}\tau_1}\int_{\tau_1}^{T_P}\mathop{\textrm{d}\tau_2} e^{-\textrm{i}\omega_0\tau_2}  \cos{\left(\tfrac{A_\textrm{R}}{2T_P}(T_P-\tau_2)\right)}\sin{\left(\tfrac{A_\textrm{R}}{2T_P}(\tau_2-\tau_1) \right)}\sin{\left(\tfrac{A_\textrm{R}}{2T_P}\tau_1 \right)}   \ket{\tau_1,\tau_2} \nonumber \\
&&\hspace{17pt}+\int_0^{T_P}\mathop{\textrm{d}\tau_1}\int_{\tau_1}^\infty\mathop{\textrm{d}\tau_2} e^{-\textrm{i}\omega_0\tau_2}\textrm{e}^{-\gamma (\tau_2-T_P)/2}  \sin{\left(\tfrac{A_\textrm{R}}{2T_P}(T_P-\tau_1) \right)}\sin{\left(\tfrac{A_\textrm{R}}{2T_P}\tau_1 \right)} \ket{\tau_1,\tau_2}\bigg].\nonumber
\end{eqnarray}
\end{subequations}
The total probability of emitting two photons is then given by
\begin{subequations}
\begin{eqnarray}
P_2 &\equiv& \braket{\phi_2|\phi_2} \label{eq:P2gen}\\
&\approx &\gamma\,\textrm{e}^{-\gamma T_P /2}\int_0^{T_P}\mathop{\textrm{d}\tau_1} \sin{\left(\frac{A_\textrm{R}}{2T_P}(T_P-\tau_1) \right)}\sin{\left(\frac{A_\textrm{R}}{2T_P}\tau_1 \right)} \\
&= &\frac{\gamma T_P}{8}\,\textrm{e}^{-\gamma T_P /2}\left(2 + \cos{(A_\textrm{R})} - 3 \frac{\sin{(A_\textrm{R})}}{A_\textrm{R}}\right),
\end{eqnarray}
\end{subequations}
where we used only the second integral of Eq. \ref{eq:phi2} because the first integral is $\mathcal{O}((\gamma T_P)^2)$. This probability is shown as the red curve in Fig. \ref{figure:5}a for a short pulse. Quite interesting are the points where $P_2>P_1$, which may not have been naively expected for the two-level system. We also note that one can easily use our solution in Eq. \ref{eq:TLScomplete} and compute its integrals numerically for higher values of $m$, to \textit{directly} calculate the oscillating $P_m$ extracted from photon correlations in Ref. \cite{lindkvist2014scattering}.

In experiment, the two-photon wavefunction is often characterized from the second-order coherence (Eq. \ref{eq:g2t1t2}). For short pulses, $\ket{\phi_3}$ is $\mathcal{O}((\gamma T_P)^2)$ and hence
\begin{eqnarray}
G^{(2)}(t_1, t_2) \approx\braket{\phi_2|b_0^\dagger(t_1)b_0^\dagger(t_2)b_0(t_2)b_0(t_1)|\phi_2},
\end{eqnarray}
which is
\begin{eqnarray}
&&G^{(2)}(t_1, t_2)^+\nonumber\\
&&=\gamma^2\,\textrm{e}^{-\gamma T_P /2} \left(\cos{\left(\tfrac{A_\textrm{R}}{2T_P}(T_P-t_2)\right)}\sin{\left(\tfrac{A_\textrm{R}}{2T_P}(t_2-t_1) \right)}\sin{\left(\tfrac{A_\textrm{R}}{2T_P}t_1 \right)}\right)^2\Theta(0<t_1<t_2<T) + \nonumber \\
&&\qquad\gamma^2\,\textrm{e}^{-\gamma T_P /2}\textrm{e}^{-\gamma (t_2-T_P)}  \left(\sin{\left(\tfrac{A_\textrm{R}}{2T_P}(T_P-t_1) \right)}\sin{\left(\tfrac{A_\textrm{R}}{2T_P}t_1 \right)} \right)^2\Theta(0<t_1<T<t_2),\nonumber\\
\end{eqnarray}
in the positive half plane where $t_1<t_2$. The second-order coherence is symmetric with respect to exchange of $t_1$ and $t_2$ (see Appendix B) so
\begin{eqnarray}
G^{(2)}(t_1, t_2)=G^{(2)}(t_1, t_2)^++G^{(2)}(t_2, t_1)^+,
\end{eqnarray}
which is shown in Fig. \ref{figure:5}c. Signatures of the Rabi oscillations are seen when $0<t_1<T_P$ or $0<t_2<T_P$, followed again by exponential decay to long times at a rate of $\gamma$: we have discussed the intuitive reasoning behind these multi-lobed correlations elsewhere \cite{fischer2017signatures} and will not repeat our discussion. Anyways, we comment that the separation of timescales---one photon order $T_P$ and one order $1/\gamma$---suggests the two-photon state is mostly separable and may be useful as a source of entangled photon pairs, though more investigation is required here. Such investigation would require computing the entanglement entropy between the two photons (after frequency filtering to route them into separate waveguides~\cite{law2004analysis,law2000continuous,blay2017effects}), as done for standard photon-pair sources. Our formalism should allow for such a calculation.

Experimentally accessing these temporal correlations or photocount distributions $P_m$ is quite challenging, and more typically a quantity called the pulse-wise second-order coherence is used \cite{fischer2016dynamical}
\begin{subequations}
\begin{eqnarray}
g^{(2)}[0]&=&\frac{\sum_m m(m-1) P_m }{(\sum_m m P_m)^2}\\
&\approx& \frac{2 P_2 }{(P_1 + 2P_2)^2}
\end{eqnarray}
\end{subequations}
for emission from a two-level system excited by a short pulse. This quantity has the property that $g^{(2)}[0]=1$ for a coherent laser pulse, $g^{(2)}[0]=0$ for a single-photon wavepacket, and $g^{(2)}[0]\gg 1$ for a two-photon wavepacket superposed with a strong $\ket{\mathbf{0}_\textrm{B}}$ component. Hence, it periodically oscillates, antibunches roughly when $P_1>P_2$, and bunches roughly when $P_1<P_2$. We plot this quantity for a short pulse in Fig. \ref{figure:5}d, which we used experimentally to verify the existence of these photon number oscillations \cite{fischer2017signatures}.

Notably, because $P_1$ and $P_2$ are roughly periodic, they have very simple expressions for areas that are multiples of~$\pi$, e.g.
\begin{center}
\begin{tabular}{| c |c |c |}
 \hline
  & & \\[-2ex]
  Area, A& $\;P_1(A)\textrm{e}^{\gamma T_P /2}\;$ & $\;P_2(A)\textrm{e}^{\gamma T_P /2}\;$ \\[0.8ex] \hline
  & & \\[-2ex]
  $\pi$ & $1+\frac{3}{8} \gamma T_P$ & $\frac{1}{8}\gamma T_P$ \\ [1.4ex]
  $2\pi$ & $\frac{1}{8}\gamma T_P$ & $\frac{3}{8}\gamma T_P$ \\  [1.ex]
 \hline
\end{tabular}
\end{center}
As a result
\begin{eqnarray}
\frac{P_2(A=2\pi)}{P_1(A=2\pi)}=3,
\end{eqnarray}
which is independent of the pulse length or system-waveguide coupling for short pulses! Hence the pulse-wise second-order coherence has the simple expression
\begin{eqnarray}
g^{(2)}[0](A=2\pi)\approx\frac{\textrm{e}^{+\gamma T_P /2}}{\gamma T_P}
\end{eqnarray}
that diverges at even areas (shown as the red arrows in Fig. \ref{figure:5}d) for arbitrarily short pulses. These results provide nice formalism and rigor to our previous studies on photon emission from two-level systems driven by short optical pulses \cite{fischer2016dynamical,fischer2017signatures,fischer2017pulsed}.


\subsection{Spontaneous parametric downconversion and four-wave mixing}

Photon pair generation from spontaneous parametric downconversion (SPDC) or four-wave mixing (SFWM) is a promising technology for use as state generators that source quantum light in various linear-optical quantum processors \cite{lu2016biphoton,mower2011efficient,silverstone2015qubit,pavesi2016silicon}. Recent progress has led to high-quality pair sources that are nanofabricated on-chip and integrated with linear-optical elements and high-efficiency photon counters. However, the existing models for photon pair emission are somewhat complicated, owing to expressing the coherent drive in the state vector of the input field \cite{yang2008spontaneous,liscidini2012asymptotic,helt2015spontaneous,dezfouli2014heisenberg}. Consequently, the equations of motion for every single field mode were derived and then perturbatively expanded. As a result, these models only apply when the probability of scattering photons into the output waveguides is very low. Further, these models carried around complex spatial and momentum integrals, which in our opinion are not strictly necessary to understand the basic physics behind pair emission. In contrast, our new formalism allows for inclusion of the coherent drive \textit{exactly} and also requires only the solutions of Heisenberg-like \textit{system} operators. Furthermore, given that nanofabricated optical resonators can support few-mode operation, all linear and nonlinear terms can easily be put in the system Hamiltonian.

The system of interest comprises two nonlinearly coupled cavity modes at $\omega_1$ and $\omega_2$ in the undepleted (classical) pumping regime. Then, the time-independent part of the system Hamiltonian is the sum of Hamiltonians of the individual cavity modes
\begin{equation}
H_{0\text{S}} = \omega_1 a_1^\dagger a_1 +\omega_2 a_2^\dagger a_2,
\end{equation}
where $a_1$ and $a_2$ annihilate photons at frequencies $\omega_1$ and $\omega_2$ respectively. For SPDC, these two cavity modes couple to a classical pump at frequency $\omega_\text{p} = \omega_1+\omega_2$ via a $\chi^{(2)}$ nonlinearity. The Hamiltonian describing such a coupling is given by
\begin{equation}
H_{1\text{S}} = g(t) \big(\textrm{e}^{\text{i}\omega_\text{p} t}a_1 a_2+\textrm{e}^{-\text{i}\omega_\text{p} t}a_1^\dagger a_2^\dagger \big),
\end{equation}
where $g(t)$ depends on the amplitude of the pump beam and the nonlinear susceptibility of the cavity. We note that an identical Hamiltonian can be used to model a four-wave mixing process that might occur in a $\chi^{(3)}$ nonlinear cavity with two cavity modes at $\omega_1$ and $\omega_2$ being driven by  classical pump beams at frequencies $\omega_{\text{p},1}$ and $\omega_{\text{p},2}$ with $\omega_1+\omega_2 = \omega_{\text{p},1}+\omega_{\text{p},2}$. The Hilbert space of this system is a tensor product of the Hilbert spaces of the two individual cavities. In particular, a convenient basis for describing the system state is given by the Fock states
\begin{equation}\label{eq:fock_state}
\ket{n_1, n_2} = \frac{(a_1^\dagger)^{n_1} (a_2^\dagger)^{n_2}}{\sqrt{n_1 ! n_2 !}} \ket{0,0},
\end{equation}
which have $n_1$ photons in the first cavity and $n_2$ photons in the second cavity. Each of the cavity modes then couples to a different waveguide, with linear couplings like in Fig. \ref{figure:1}b with $M=2$ and rates $\gamma_1$ and $\gamma_2$. Thus, we can use the results of Eq. \ref{eq:scat_multiple_wg_heis}.

Here, it is convenient to remove the high-frequency (i.e.~$\textrm{e}^{\pm\text{i}\omega_\text{p} t}$) terms in the above system Hamiltonian via a unitary transformation. Consider defining the interaction-picture state $\ket{\Psi_\textrm{I}}$ in terms of the actual state of the entire system $\ket{\Psi}$ via
\begin{equation}\label{eq:unitary_spdc}
\ket{\Psi_\textrm{I}(t)} = \exp\bigg[\text{i}\omega_1 t\bigg( a_1^\dagger a_1+\int \mathop{\textrm{d}\omega} b_{1,\omega}^\dagger b_{1,\omega}\bigg)\bigg] \exp\bigg[\text{i}\omega_2 t\bigg( a_2^\dagger a_2+\int \mathop{\textrm{d}\omega} b_{2,\omega}^\dagger b_{2,\omega}\bigg)\bigg] \ket{\Psi(t)}.
\end{equation}
A straightforward differentiation of this equation (with $\omega_\text{p} = \omega_1+\omega_2$) can show that the interaction-picture Hamiltonian for $\ket{\Psi_\textrm{I}}$ is now given by
\begin{equation}
H_\text{I}(t) = g(t) (a_1 a_2 + a_1^\dagger a_2^\dagger) + \text{i}\sum_{i=1}^2\sqrt{\gamma_i}\left(b_{i,\tau=0}^\dagger(t) a_i - b_{i,\tau=0}(t) a_i^\dagger\right).
\end{equation}
For the remainder of this section, we will work with the transformed state $|\Psi_\textrm{I}\rangle$ and can use Eq.~\ref{eq:unitary_spdc} to transform back to $|\Psi\rangle$. Thus, our central results of Eqs. \ref{eq:scat_multiple_wg} and \ref{eq:scat_multiple_wg_heis} still apply, but with the effective Hamiltonian
\begin{equation}
H_\text{eff}(t) = g(t)(a_1 a_2+a_1^\dagger a_2^\dagger)-\textrm{i}\frac{\gamma_1}{2}a_1^\dagger a_1-\textrm{i}\frac{\gamma_2}{2}a_2^\dagger a_2.
\end{equation}
Using this effective Hamiltonian, we can setup the differential equations governing the time-evolution of the Heisenberg-like operators $\tilde{a}_i(\tau) = U_\text{eff}(0,\tau) a_i U_\text{eff}(\tau,0)$:
\begin{equation}\label{eq:dyn_heis}
\frac{\textrm{d}}{\textrm{d}t} 
\begin{bmatrix}
\tilde{a}_1(\tau) \\
\tilde{a}_2(\tau) \\
\tilde{a^\dagger}_1(\tau) \\
\tilde{a^\dagger}_2(\tau)
\end{bmatrix}
= 
\begin{bmatrix}
-\gamma_1/2 & 0 & 0 & -\text{i}g(\tau) \\
0 & -\gamma_2/2 & -\text{i}g(\tau) & 0 \\
0 & \text{i}g(\tau) & \gamma_1/2 & 0 \\
\text{i}g(\tau) & 0  & 0 & \gamma_2/2 
\end{bmatrix}
\begin{bmatrix}
\tilde{a}_1(\tau) \\
\tilde{a}_2(\tau) \\
\tilde{a^\dagger}_1(\tau) \\
\tilde{a^\dagger}_2(\tau)
\end{bmatrix}.
\end{equation}
This system of equations can easily be integrated numerically, and $\tilde{a}_1(\tau)$ and $\tilde{a}_2(\tau)$ can be expressed in terms of the Schr{\"o}dinger operators $a_1$ and $a_2$ as
\begin{subequations}
\begin{eqnarray}
\tilde{a}_1(\tau) &=& \alpha_{1,1}(\tau) a_1 + \alpha_{1,2}(\tau) a_2^\dagger \label{eq:heis_schro1}\\
\tilde{a}_2(\tau) &=& \alpha_{2,1}(\tau) a_1^\dagger+\alpha_{2,2}(\tau) a_2.\label{eq:heis_schro2}
\end{eqnarray}
\end{subequations}

As preparation for computing the output state from the pair source, we next consider the computation of a state with the form $\ket{\phi_{n_1, n_2}(t)} = U_\text{eff}(t,0)|n_1, n_2\rangle$ where $|n_1, n_2\rangle$ is the Fock state defined in Eq.~\ref{eq:fock_state}. This is equivalent to solving the differential equation
\begin{equation}\label{eq:eff_schro}
\textrm{i}\frac{\textrm{d}\ket{\phi_{n_1, n_2}(t)}}{\textrm{d}t} = H_\text{eff}(t) \ket{\phi_{n_1, n_2}(t)}.
\end{equation}
Since the Fock states (Eq.~\ref{eq:fock_state}) form a complete basis in the Hilbert space of the two cavities, it is possible to expand $\ket{\phi_{n_1,n_2}(t)}$ onto this basis as
\begin{equation}\label{eq:expansion_explicit}
\ket{\phi_{n_1,n_2}(t)} = \sum_{m_1=0}^{\infty}\sum_{m_2=0}^\infty \xi_{m_1,m_2}^{(n_1,n_2)}(t) \ket{m_1,m_2}.
\end{equation}
However, this expansion can be considerably simplified if we note that the operator $a_1^\dagger a_1-a_2^\dagger a_2$ commutes with $H_\text{eff}(t)$---the difference in the photon numbers of the two cavities thus remains conserved. An immediate implication of this conservation law is that $\xi_{m_1,m_2}^{(n_1,n_2)} = 0$ if $m_1-m_2 \neq n_1-n_2$, which reduces the expansion in Eq.~\ref{eq:expansion_explicit} to
\begin{equation}
\ket{\phi_{n_1,n_2}(t)} = \sum_{p=-\min(n_1,n_2)}^{\infty} c_{p}^{(n_1,n_2)}(t)\ket{n_1+p,n_2+p}.
\end{equation}
Eq.~\ref{eq:eff_schro} can then be translated into an infinite system of equations for $c_{p}^{(n_1,n_2)}(t)$, which can be truncated and numerically integrated to compute $\ket{\phi_{n_1,n_2}(t)}$.

In what follows, we consider the cavities to be initially in the vacuum state, and the pump $g(t)$ being switched on at $t = 0$ till $t = T_P$. We first consider the case when \textbf{no photons are scattered}
\begin{equation}
\langle \textbf{0}_\text{B}|\psi_\text{B,I}(t\to \infty)\rangle = \langle 0, 0 | U_\text{eff}(T_P,0)|0, 0\rangle = c_0^{(0,0)}(T_P)
\end{equation}
and $P_0=\left|c_0^{(0,0)}(T_P)\right|^2$. 

Next, we consider the case where \textbf{two photons are scattered}---one photon in the first waveguide and another photon in the second waveguide. Specializing Eq.~\ref{eq:scat_multiple_wg_heis}, we obtain
\begin{align}
\langle \{\tau_{1,1}\},\{\tau_{2,1}\} | \psi_\text{B,I}(t\to \infty)\rangle=\sqrt{\gamma_1\gamma_2}
\begin{cases}
\langle 0,0 | U_\text{eff}(T_P,0) \tilde{a}_1(\tau_{1,1}) \tilde{a}_2(\tau_{2,1}) |0,0\rangle & \text{ if }  \tau_{2,1}\leq \tau_{1,1} \\
\langle 0,0 |U_\text{eff}(T_P,0) \tilde{a}_2(\tau_{2,1}) \tilde{a}_1(\tau_{1,1}) |0,0 \rangle & \text{ if } \tau_{1,1} \leq \tau_{2,1}  
\end{cases}.
\end{align}
Substituting for $\tilde{a}_1(\tau)$ and $\tilde{a}_2(\tau)$ from Eqs.~\ref{eq:heis_schro1} and Eq.~\ref{eq:heis_schro2}, we have the result
\begin{eqnarray}
&&\langle \{\tau_{1,1}\},\{\tau_{2,1}\} | \psi_\text{B,I}(t\to \infty)\rangle \\
&&\qquad\qquad= \sqrt{\gamma_1 \gamma_2}
\begin{cases}
\alpha_{1,2}(\tau_{1,1})[\alpha_{2,2}(\tau_{2,1}) c_0^{(0)}(T_P)+\alpha_{2,1}(\tau_{2,1}) c_{-1}^{(1,1)}(T_P)] & \text{ if }  \tau_{2,1}\leq \tau_{1,1} \\
\alpha_{2,1}(\tau_{2,1})[\alpha_{1,1}(\tau_{1,1}) c_0^{(0)}(T_P)+\alpha_{1,2}(\tau_{1,1}) c_{-1}^{(1,1)}(T_P)] & \text{ if }  \tau_{1,1} \leq \tau_{2,1} 
\end{cases}\nonumber
\end{eqnarray}
and
\begin{equation}
P_2=\int_0^\infty \mathop{\text{d}\tau_{1,1}}\int_0^\infty \mathop{\text{d}\tau_{2,1}}\left|\langle \{\tau_{1,1}\},\{\tau_{2,1}\} | \psi_\text{B,I}(t\to \infty)\rangle\right|^2.
\end{equation}
It also can be noted that since the difference in number of photons in the two cavities is conserved in the presence of the pump beam, it is not possible for an initially unexcited system to emit an unequal number of photons in the two waveguides. These results are exact, in contrast to all previous studies of photon pair generation, which could only provide perturbative estimates. 
\begin{figure}
  \centering
  \includegraphics[scale=0.6]{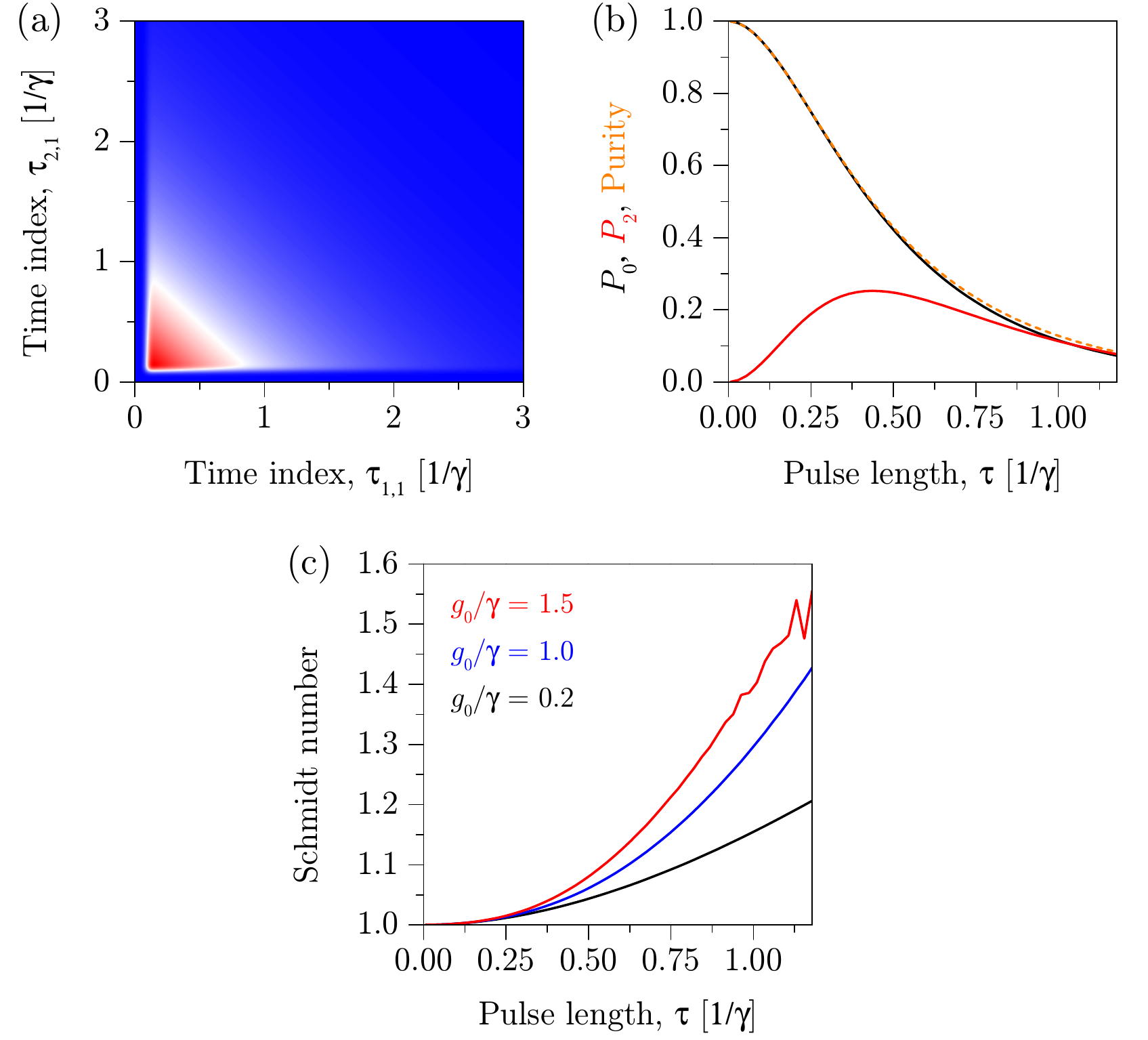}
  \caption{A photon pair source (SPDC or SFWM), in the classical driving limit and for two identical cavities (i.e.~$\gamma_1 = \gamma_2 = \gamma$). The solutions are numerically exact, in contrast to previous models. (a) Amplitude for two-photon part of the output state projected on the temporal basis, i.e. $\left|\langle \{\tau_{1,1}\},\{\tau_{2,1}\} | \psi_\text{B,I}(\infty)\rangle\right|^2$. (b) Variation in probability of single- and two-photon emission as a function of the pulse length $\tau$ (for driving strength $g_0 = \gamma$). (c) Variation of the Schmidt number as a function of the pulse length $\tau$ and the pump amplitude $g_0$.}
  \label{fig:spdc}
\end{figure}

As a specific numerical example, we consider driving the system with a Gaussian pulse
\begin{equation}
g(t) = g_0 \exp\bigg(-\frac{(t-t_0)^2}{2\tau^2} \bigg).
\end{equation}
(Physically this corresponds to the case where the pump drives extremely lossy cavity mode(s), which then couple to the modes $a_1$ and $a_2$ via the $\chi^{(2)}$ or $\chi^{(3)}$ nonlinearity.) Fig.~\ref{fig:spdc}a shows the modulus square of the two-photon wavefunction $|\braket{\{\tau_{1,1}\},\{\tau_{2,1}\}|\psi_\textrm{B,I}(\infty)}|^2$, as a function of $(\tau_{1,1}, \tau_{2,1})$. As one would intuitively expect, the photon wavefunction increases in magnitude while the system is driven by the pulse and decays in amplitude once the driving stops. The total photon emission probabilities $P_0$ and $P_2$ are shown in Fig.~\ref{fig:spdc}b as a function of pulse length---it can clearly be seen that at very short pulse lengths, the waveguides are almost entirely in the vacuum state, while at high pulse lengths the waveguides would almost entirely be in a superposition of higher-order Fock states. The probability of the output waveguides being in the two-photon states peaks at intermediate values of pulse lengths. 

For applications of SPDC or SFWM as a heralded single-photon source \cite{pavesi2016silicon}, the vacuum state is not important since the presence of a single photon in one waveguide is conditioned on the presence of a single photon in the other waveguide (which is detected with a photon counter). A suitable figure of merit for the `purity' of the output state of the pair source is thus the fraction of two photon state amongst the states excluding vacuum
\begin{equation}
\text{Purity = }\frac{P_2}{P_2+P_4+\cdots} = \frac{P_2}{1-P_0}.
\end{equation}
The purity of the output state as a function of pulse length is shown in Fig.~\ref{fig:spdc}b. The purity decreases with pulse length---this is a direct consequence of an increased emission of more than two photons at large pulse lengths.

The final figure of merit that we compute for the output state of this system is the Schmidt number---this measure quantifies the `extent' of entanglement between the output states of the two waveguides \cite{law2004analysis,law2000continuous}. The Schmidt decomposition of the two-photon output state $|\phi_{2}\rangle$ is equivalent to expressing it as a sum over a countably infinite set of unentangled two photon states:
\begin{equation}
|\phi_{2}\rangle = \sum_{n}\sqrt{\lambda_n}|\theta_{1,n}\rangle \ket{\theta_{2,n}},
\end{equation}
with the Schmidt number defined as
\begin{equation}
\text{Schmidt number} = \sum_n \frac{1}{\lambda_n^2}.
\end{equation}
A Schmidt number greater than 1 is a signature of an entangled system, while a Schmidt number equal to 1 corresponds to an unentangled system. One point to note is that since the output state of the waveguide is not completely a two-photon state, the norm of the two-photon component would be less than 1. To reconcile this with the Schmidt decomposition, which assumes the state to be entirely a two-photon state, we renormalize the two-photon component of the output state to have unity norm before computing the Schmidt number. Fig.~\ref{fig:spdc}c shows the Schmidt number as a function of pulse length for different values of the driving field $g_0$---it can clearly be seen that as the driving field or pulse length increases, the two waveguides become increasingly entangled to each other. This is intuitively expected, since the pump entangles the two cavity modes due to their nonlinear interaction, and the entanglement of the cavity modes is then transfered to the waveguides through photon emission. It is thus expected that this entanglement becomes stronger with an increase in the pump amplitude or the pulse length, since both of these lead to an increase in the strength of the interaction between the two cavity modes.

Since experimental realizations of photon pair generation typically rely on $\chi^{(2)}$ (three-wave mixing) or $\chi^{(3)}$ (four-wave mixing) nonlinearities, in addition to the mixing processes, there are often competing processes that might impact the response of these systems. Two particularly important processes are cross-phase modulation, where the number of photons in one cavity mode impact the resonant frequency of the second cavity mode, and self-phase modulation, where the number of photons in a cavity mode impacts its own resonant frequency. Both these processes can easily be modeled by the addition of more nonlinear terms to the system Hamiltonian \cite{quesada2017effects, vernon2015spontaneous}. The formalism outlined here can thus be easily extended to calculate the impact of such experimental non-idealities on the emission from such systems.


\section{CONCLUSIONS}

We have demonstrated a powerful technique to integrate the Schr{\"o}dinger equation for waveguide(s) coupled to a low-dimensional quantum system based on a coarse-graining of the temporal waveguide modes.  Our technique works even in the presence of singularities and time-dependent system Hamiltonians. The ease with which this method allows direct solutions to the Schr{\"o}dinger equation for problems with infinite dimensions suggests that  temporal modes may be an ideal basis to consider quantum-optical problems generally. This viewpoint seems to be gaining popularity in the field.

The theory described in this work applied only to low-dimensional systems with a single `ground' state, and the waveguides initially in the vacuum state.  An obvious extension of this work is to treat systems with multiple ground states which, for example, could lead to a more thorough understanding of spin-photon entanglement \cite{gao2012observation} or optically-controlled single-photon phase gates \cite{sun2016quantum}. Some of these scenarios might include cases in which a single photon impinges on an energy-nonconserving system \cite{shi2015multiphoton}, a situation we believe to be ideally suited for our formalism.

Finally, we note that while our theory only considered situations in which the system of interest underwent Hamiltonian evolution, an exciting avenue for further research would be the possible extension of this work to directly computing $N$-photon scattering super-operators under the addition of dissipation. For instance, phonon-induced dephasing is an important consideration in the solid state \cite{roy2011influence,muller2015ultrafast,gustin2017influence}, where a computation of the single-photon density matrix of the output field directly in the presence of this dephasing would be quite valuable.  We believe that our framework constitutes a compelling argument to reconsider the dynamics of photon emission in a temporal mode basis and provides a way to answer many interesting questions going forwards.


\section{ACKNOWLEDGEMENTS}

We gratefully acknowledge financial support from the National Science Foundation (Division of Materials Research---Grant No. 1503759) and the Air Force Office of Scientific Research (AFOSR) MURI Center for Quantum Metaphotonics and Metamaterials. KAF acknowledges support from the Lu Stanford Graduate Fellowship and the National Defense Science and Engineering Graduate Fellowship. RT acknowledges support from Kailath Stanford Graduate Fellowship.

The authors kindly thank Emily Davis, Leigh Martin, and Tatsuhiro Onodera for helpful discussions. The authors are also grateful to Joshua Combes for carefully reading the paper and providing thoughtful feedback.


\appendix
\addcontentsline{toc}{section}{\protect\numberline{}{APPENDICES}}
\section{Photon flux and interpretation of temporal modes}

We can now briefly comment on the reason for calling the modes `temporal' by examining the photon flux operator \cite{loudon2000quantum}. The expectation of the normally-ordered intensity operator at $\vec{r}=0$ in the Schr{\"o}dinger picture is
\begin{subequations}
\begin{eqnarray}
\braket{:I(\vec{r}=0):}&\propto& \braket{:E(\vec{r}=0)^2:}\\
&\propto&\braket{\int_0^\infty\mathop{\textrm{d}\omega_1}\int_0^\infty\mathop{\textrm{d}\omega_2}\sqrt{\omega_1\omega_2} \,b_{\omega_1}^\dagger b_{\omega_2}},
\end{eqnarray} 
\end{subequations}
which would be measured by an experimental photon detector ($:I:$ denotes the normal ordering of $I$). Making the same approximation in Sec. IIC that the frequency content of the state is narrowband around $\omega_0$
\begin{equation}
\braket{:I(0):}\propto \omega_0\braket{\int_{-\infty}^\infty\mathop{\textrm{d}\omega_1}\int_{-\infty}^\infty\mathop{\textrm{d}\omega_2}\,b_{\omega_1}^\dagger b_{\omega_2}}.
\end{equation}
We label this operator on the right and transform into the temporal mode basis
\begin{subequations}
\begin{eqnarray}
F_{\vec{r}=0}(0)&\equiv&\int_{-\infty}^\infty\mathop{\textrm{d}\omega_1}\int_{-\infty}^\infty\mathop{\textrm{d}\omega_2}\,b_{\omega_1}^\dagger b_{\omega_2}\\
&=&b_{\tau=0}^\dagger(0)b_{\tau=0}(0).
\end{eqnarray}
\end{subequations}
In the interaction picture, $F_0$ then acquires the time dependence
\begin{equation}
F_0(t)=b_{0}^\dagger(t)b_{0}(t),
\end{equation}
and hence $\braket{:I(0,t):}\propto \braket{F_0(t)}$ with the free evolution of the waveguide lumped in with the operator rather than the state. 

Let's evaluate the expectation of $F_0(t)$ for an arbitrary state in the temporal
mode basis, where we ignore any low-dimensional system's interaction with the waveguide $\gamma\rightarrow 0$
\begin{equation}
\braket{F_0(t)}=\braket{\psi_\textrm{B,I} | b_{0}^\dagger(t)b_{0}(t) | \psi_\textrm{B,I}}.
\end{equation}
(We note this correlation is also a special case of Eq. \ref{eq:allcorr}.) Expanding this expectation by inserting resolutions of the identity in the temporal mode basis
\begin{equation}
\braket{F_0(t)}=\sum_{m=0}^\infty\sum_{m'=0}^\infty\int \mathop{\textrm{d}\vec{\bm{\tau}}^{(m)}}\int \mathop{\textrm{d}\vec{\bm{\tau}}'^{(m')}}\braket{\psi_\textrm{B,I}|\vec{\bm{\tau}}^{(m)}}\braket{\vec{\bm{\tau}}^{(m)}|b_{0}^\dagger(t)b_{0}(t)|\vec{\bm{\tau}}'^{(m')}}\braket{\vec{\bm{\tau}}'^{(m')}|\psi_\textrm{B,I}}.
\end{equation}
Now, we evaluate
\begin{equation}
\braket{\vec{\bm{\tau}}^{(m)}|b_{0}^\dagger(t)b_{0}(t)|\vec{\bm{\tau}}'^{(m')}}=\delta_{mm'}\delta(\vec{\bm{\tau}}^{(m)}-\vec{\bm{\tau}}'^{(m)})\sum_{q=1}^m \delta(t-\tau_q)\,[m>0],
\end{equation}
given the explicit ordering of the indices $\tau_1<\cdots<\tau_m$. Then,
\begin{subequations}
\begin{eqnarray}
\hspace{-20pt}\braket{F_0(t)}&=&\sum_{m=1}^\infty\int \mathop{\textrm{d}\vec{\bm{\tau}}^{(m)}}\int \mathop{\textrm{d}\vec{\bm{\tau}}'^{(m)}}\braket{\psi_\textrm{B,I}|\vec{\bm{\tau}}^{(m)}}\braket{\vec{\bm{\tau}}'^{(m)}|\psi_\textrm{B,I}}\delta(\vec{\bm{\tau}}^{(m)}-\vec{\bm{\tau}}'^{(m)})\sum_{q=1}^m \delta(t-\tau_q)\\
\hspace{-20pt}&=&\sum_{m=1}^\infty\sum_{q=1}^m\int \mathop{\textrm{d}\vec{\bm{\tau}}^{(m)}}\left|\braket{\vec{\bm{\tau}}^{(m)}|\psi_\textrm{B,I}}\right|^2 \delta(t-\tau_q).\label{eq:f0}
\end{eqnarray}
\end{subequations}
Experimentally, an ideal photon detector is likely to detect a photon with probability proportional to $\braket{F_0(t)}$ \cite{loudon2000quantum}. This expression then has the interesting form that it represents a summation over all possible ways for the state to yield a single photon detection event at time $t$. Each $m$-photon wavepacket contributes independently. 

For instance, $\braket{F_0(t)}$ picks out just the occupation in the $t^\textrm{th}$ temporal mode from the single-photon wavepacket
\begin{equation}
\left|\braket{t|\psi_\textrm{B,I}}\right|^2.
\end{equation}
Then, for the two-photon wavepacket, there are two possible ways that a single-photon may be detected at time $t$: a photon was either detected before or after. Hence,
\begin{eqnarray}
&&\int_0^\infty\mathop{\textrm{d}\tau_1}\int_{\tau_1}^\infty\mathop{\textrm{d}\tau_2}\left|\braket{\tau_1,\tau_2|\psi_\textrm{B,I}}\right|^2\left(\delta(t-\tau_1)+\delta(t-\tau_2)\right)\nonumber\\
&&\hspace{50pt}=
\int_t^\infty\mathop{\textrm{d}\tau_2}\left|\braket{t,\tau_2|\psi_\textrm{B,I}}\right|^2+\int_0^t\mathop{\textrm{d}\tau_1}\left|\braket{\tau_1,t|\psi_\textrm{B,I}}\right|^2.
\end{eqnarray}
We also show the case for a three-photon wavepacket, where there are three possible ways that a single-photon may be detected at time $t$: two after, one before and one after, and two before
\begin{eqnarray}
&&\int_0^\infty\mathop{\textrm{d}\tau_1}\int_{\tau_1}^\infty\mathop{\textrm{d}\tau_2}\int_{\tau_2}^\infty\mathop{\textrm{d}\tau_3}\left|\braket{\tau_1,\tau_2,\tau_3|\psi_\textrm{B,I}}\right|^2\left(\delta(t-\tau_1)+\delta(t-\tau_2)+\delta(t-\tau_3)\right)=\\
&&\hspace{-20pt}\int_t^\infty\mathop{\textrm{d}\tau_2}\int_{\tau_2}^\infty\mathop{\textrm{d}\tau_3}\left|\braket{t,\tau_2,\tau_3|\psi_\textrm{B,I}}\right|^2 + \int_0^t\mathop{\textrm{d}\tau_1}\int_{t}^\infty\mathop{\textrm{d}\tau_3}\left|\braket{\tau_1,t,\tau_3|\psi_\textrm{B,I}}\right|^2 + \int_0^t\mathop{\textrm{d}\tau_1}\int_{\tau_1}^t\mathop{\textrm{d}\tau_2}\left|\braket{\tau_1,\tau_2,t|\psi_\textrm{B,I}}\right|^2.\quad\nonumber
\end{eqnarray}
The pattern continues rather intuitively, where Eq. \ref{eq:f0} is just a shorthand for counting all possible ways each $m$-photon wavepacket can contribute to a single excitation at time $t$. Notably, these expectations depend only on the magnitude of projections onto the $\tau$-modes, i.e. on $\left|\braket{\vec{\bm{\tau}}^{(m)}|\psi_\textrm{B,I}}\right|^2$. This corresponds to what we physically expect because intensity detectors do not respond to the quantum phase of wavefunctions \cite{loudon2000quantum}. Hence, we can construct classical probability density functions
\begin{equation}
\mathbb{P}(\vec{\bm{\tau}}^{(m)})\equiv p_m(\vec{\bm{\tau}}^{(m)})=\left|\braket{\vec{\bm{\tau}}^{(m)}|\psi_\textrm{B,I}}\right|^2
\end{equation}
that correspond to the density of detecting $m$ photons at times $t_1,\dots,t_m$ (like we used in \citet{fischer2017pulsed})---for a given scattered state, this is also a restatement of Eq. \ref{eq:pnsme}. Then, the probabilities to detect a given number of photons after the entire pulse has interacted with an ideal detector are given by the photocount distribution
\begin{equation}
P_m=\int\mathop{\textrm{d}\vec{\bm{\tau}}^{(m)}}p_m(\vec{\bm{\tau}}^{(m)}).
\end{equation}
From these results come the phrase `temporal mode', where an occupation in a given temporal mode indexed by $t$ can trigger a detection at time $t$ by a detector.


\section{Second-order coherence with temporal modes}

We also note that a similar process can be repeated for the second-order coherence
\begin{eqnarray}
G^{(2)}(t_1, t_2)&=&\braket{\psi_\text{B,I}|b_0^\dagger(t_1) b_0^\dagger(t_2)b_0(t_2)b_0(t_1)|\psi_\text{B,I}},
\end{eqnarray}
as in Appendix A. After inserting resolutions of the identity, we instead need to use
\begin{eqnarray}
\braket{\vec{\bm{\tau}}^{(m)}|b_{0}^\dagger(t_1)b_{0}^\dagger(t_2)b_{0}(t_2)b_{0}(t_1)|\vec{\bm{\tau}}'^{(m')}}=&\nonumber\\
&\hspace{-35ex}\delta_{mm'}\delta(\vec{\bm{\tau}}^{(m)}-\vec{\bm{\tau}}'^{(m)})\sum_{q_1=1}^m \sum_{q_2=1}^m\delta(t_1-\tau_{q_1})\delta(t_2-\tau_{q_2})\,[(m>1)\land (q_1\neq q_2)]
\end{eqnarray}
and then
\begin{eqnarray}
G^{(2)}(t_1, t_2)=\sum_{m=2}^\infty\sum_{q_1=1}^m \sum_{q_2=1}^m\int \mathop{\textrm{d}\vec{\bm{\tau}}^{(m)}}\left|\braket{\vec{\bm{\tau}}^{(m)}|\psi_\textrm{B,I}}\right|^2\delta(t_1-\tau_{q_1})\delta(t_2-\tau_{q_2})\,[q_1\neq q_2].
\end{eqnarray}
Intuitively, this expectation is used to count all possible ways to detect two-photons from an $m$-photon wavepacket. For instance, the three-photon wavepacket's contribution to $G^{(2)}(t_1, t_2)$ for $t_1<t_2$ is
\begin{eqnarray}
\int_{t_2}^{\infty}\mathop{\textrm{d}\tau_3}\left|\braket{t_1,t_2,\tau_3|\psi_\textrm{B,I}}\right|^2 + \int_{t_1}^{t_2}\mathop{\textrm{d}\tau_2}\left|\braket{t_1,\tau_2,t_2|\psi_\textrm{B,I}}\right|^2 + \int_{0}^{t_1}\mathop{\textrm{d}\tau_1}\left|\braket{\tau_1,t_1,t_2|\psi_\textrm{B,I}}\right|^2\label{eq:g23photon}
\end{eqnarray}
or
\begin{equation}
\int_{t_2}^{\infty}\mathop{\textrm{d}\tau_3}p_3(t_1,t_2,\tau_3) + \int_{t_1}^{t_2}\mathop{\textrm{d}\tau_2}p_3(t_1,\tau_2,t_2) + \int_{0}^{t_1}\mathop{\textrm{d}\tau_1}p_3(\tau_1,t_1,t_2).\label{eq:g23photon2}
\end{equation}
(If $t_1>t_2$ then exchange those indices in Eqs. \ref{eq:g23photon} or \ref{eq:g23photon2}.)

From this point, it should be clear that all $m$-photon wavepackets contribute to the coherence functions when $m$ is greater than the order of the coherence function. Thus, although every correlation from Eq. \ref{eq:allcorr} technically captures the entire state of the field, transforming between the correlations and a state vector is quite difficult, which of course also requires the assumption that the state is pure. Further, in the case of scattering from energy-nonconserving systems, it's often hard to know before doing a calculation how many correlation orders will contribute to a given $m$-photon wavepacket.

It is easy to extrapolate the pattern to the $R^\text{th}$ order correlation, where
\begin{eqnarray}
&&G^{(R)}(t_1, \dots,t_R)\\
&&=\sum_{m=R}^\infty\sum_{q_1=1}^m \cdots\sum_{q_R=1}^m\int \mathop{\textrm{d}\vec{\bm{\tau}}^{(m)}}\left|\braket{\vec{\bm{\tau}}^{(m)}|\psi_\textrm{B,I}}\right|^2\delta(t_1-\tau_{q_1})\delta(t_2-\tau_{q_2})\cdots\delta(t_R-\tau_{q_R})\,[q_l\neq q_k],\nonumber
\end{eqnarray}
where by $[q_l\neq q_k]$ we mean that we can only have one delta function at a given time.


\section{Mollow transformation}

\begin{figure}
  \begin{center}
  \centering
  \includegraphics[scale=0.9]{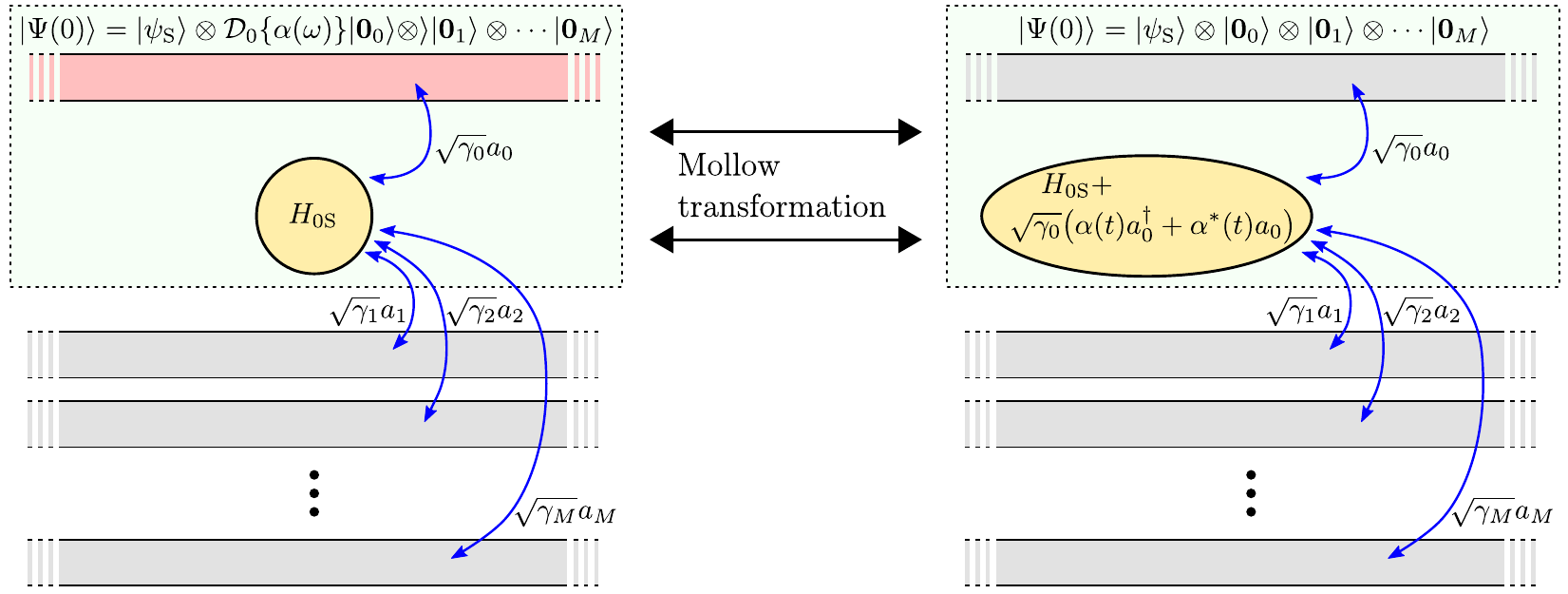}
  \end{center}
  \caption{The Mollow transformation is a unitary operation that removes a coherent state from a waveguide coupled to a system. It captures the effects of a coherent state drive in the system Hamiltonian itself. In the new frame, however, energy is not conserved.}
  \label{figure:6}
\end{figure}

Often, a classical laser pulse is incident on a quantum-mechanical system, causing it to undergo transitions between its various levels and then re-emit the absorbed energy. Typically the effect of the coherent pulse is lumped into the system Hamiltonian as a time-dependent operator (acting on $\mathcal{H}_\textrm{S}$) that has the time-dependence of the pulse's phase. Many often assume that this Hamiltonian is semi-classical, but Mollow identified it is actually exact, provided the loss channel where the coherent state originated is retained in the calculation \cite{mollow1975pure}. He did this by finding a unitary transformation that removes the coherent state from the waveguide (Fig. \ref{figure:6}). We now detail this calculation.

Consider a system with Hamiltonian $H_{0\text{S}}\{a_i\}$ coupled to a single chiral waveguide described by a continuum of modes $b_\omega$. The complete system can be modeled by
\begin{equation}\label{main_hamiltonian}
H = H_{0\text{S}}\{a_i\}+\int_{-\infty}^\infty \mathop{\textrm{d}\omega}\omega \,b_\omega^\dagger b_\omega +\xi \int_{-\infty}^\infty \mathop{\textrm{d}\omega}\left(a_0 b_\omega^\dagger+b_\omega a_0^\dagger\right),
\end{equation}
where $\xi = \sqrt{\gamma_0/2\pi}$ is the coupling constant between the waveguide and the local system, with $\kappa$ being the decay rate induced by the loss channel in the local system. The system is driven by a coherent state through the waveguide, which corresponds to the following state at $t=0$
\begin{equation}
|\psi(t=0)\rangle = \mathcal{D}\{\alpha(\omega)\}|0\rangle \otimes |\psi_\text{S}(0)\rangle,
\end{equation}
where $\mathcal{D}\{\alpha(\omega)\} = \exp\left(\int_{-\infty}^\infty \mathop{\textrm{d}\omega}\,(\alpha(\omega)b_\omega^\dagger-\alpha^*(\omega) b_\omega)\right)$ is the displacement operator creating a coherent state in the waveguide. We define a transformed state $|\tilde{\psi}(t)\rangle$ via
\begin{equation}\label{psi_tilde}
|\tilde{\psi}(t)\rangle = \mathcal{D}_t^\dagger \{\alpha(\omega)\}|\psi(t)\rangle,
\end{equation}
where
\begin{equation}
\mathcal{D}_t\{\alpha(\omega)\} = \exp\left(\int_{-\infty}^\infty \mathop{\textrm{d}\omega}\left(\alpha(\omega)\textrm{e}^{-\textrm{i}\omega t} b_\omega^\dagger- \alpha^*(\omega)\textrm{e}^{\textrm{i}\omega t} b_\omega\right)\right).
\end{equation}
The time evolution of the state $|\tilde{\psi}(t)\rangle$ can be computed by differentiating Eq.~\ref{psi_tilde}  to obtain an effective Hamiltonian
\begin{equation}\label{eff_hamil}
\tilde{H}(t) = \mathcal{D}_t^\dagger\{\alpha(\omega)\}H\mathcal{D}_t\{\alpha(\omega)\}+\text{i}\frac{\mathop{\textrm{d}}}{\mathop{\textrm{d}t}}\mathcal{D}^\dagger_t\{\alpha(\omega)\}\mathcal{D}_t\{\alpha(\omega)\}\quad\textrm{with}\quad \textrm{i}\frac{\mathop{\textrm{d}}}{\mathop{\textrm{d}t}}|\tilde{\psi}(t)\rangle = \tilde{H}(t)|\tilde{\psi}(t)\rangle.
\end{equation}
We next compute the effective Hamiltonian $\tilde{H}(t)$ using Eq.~\ref{main_hamiltonian}. In particular, it follows from the identities $\mathcal{D}^\dagger\{\alpha(\omega)\}b_\omega \mathcal{D}\{\alpha(\omega)\} = b_\omega+\alpha(\omega)$ and $\mathcal{D}^\dagger\{\alpha(\omega)\}b_\omega^\dagger\mathcal{D}\{\alpha(\omega)
\} = b_\omega^\dagger+\alpha^*(\omega)$ that
\begin{align}\label{term_1}
\mathcal{D}_t^\dagger\{\alpha(\omega)\}H\mathcal{D}_t\{\alpha(\omega)\} = H_{0\text{S}}\{a_i\}+\sqrt{\gamma_0} \left(\alpha(t)a_0^\dagger+\alpha^*(t)a_0\right)+\int_{-\infty}^\infty\mathop{\textrm{d}\omega} \omega b_\omega^\dagger b_\omega +\int_{-\infty}^\infty \mathop{\textrm{d}\omega}\omega|\alpha(\omega)|^2 +\nonumber \\ \int_{-\infty}^\infty\mathop{\textrm{d}\omega} \omega\left(b_\omega \alpha^*(\omega)\textrm{e}^{\textrm{i}\omega t}+b_\omega^\dagger \alpha(\omega) \textrm{e}^{-\textrm{i}\omega t}\right) +\xi \int_{-\infty}^\infty \mathop{\textrm{d}\omega}\left(a_0^\dagger b_\omega+b_\omega^\dagger a_0\right)  ,
\end{align}
where $\alpha(t) = \int_{-\infty}^\infty \frac{\mathop{\textrm{d}\omega}}{\sqrt{2\pi}}\textrm{e}^{-\textrm{i}\omega t}\alpha(\omega)$.

Now, we need to compute $\mathop{\textrm{d}\mathcal{D}_t^\dagger} \{\alpha(\omega)\}/\mathop{\textrm{d}t}$. Consider the problem of computing the derivative of an exponential of a time dependent operator $A(t) = \exp(\phi(t))$, with the property that $[\phi(t),\text{d}\phi(t)/\text{d}t] = \zeta(t)$, with $\zeta(t)$ being a scalar. The derivative of $A(t)$ can be computed as
\begin{equation}
\frac{\text{d}A(t)}{\text{d}t} = \lim_{\delta t \to 0} \frac{\exp(\phi(t+\delta t))-\exp(\phi(t))}{\delta t}.
\end{equation}
Using the Baker--Campbell--Hausdorff equality it follows that
\begin{equation}
\exp(\phi(t+\delta t)) \approx \exp \bigg(\phi(t)+\delta t \frac{d\phi(t)}{dt} \bigg) = \exp(\phi(t))\exp \bigg(\delta t \frac{\text{d}\phi(t)}{\text{d}t}\bigg) \exp(-\delta t\, \zeta (t)/2)
\end{equation}
resulting in
\begin{equation}
\frac{\text{d}A(t)}{\text{d}t} = \exp(\phi(t))\bigg(\frac{\text{d}\phi(t)}{\text{d}t}- \frac{\zeta(t)}{2}\bigg).
\end{equation}
Specializing this result to:
$A(t) = \mathcal{D}^\dagger_t\{\alpha(\omega)\}$ with $\phi(t) = \int_{-\infty}^\infty \mathop{\textrm{d}\omega}\left(\alpha^*(\omega)\textrm{e}^{\textrm{i}\omega t} b_\omega -\alpha(\omega) \textrm{e}^{-\textrm{i}\omega t} b_\omega^\dagger\right)$ and $\zeta(t) = 2\text{i} \int_{-\infty}^\infty \mathop{\textrm{d}\omega}\omega |\alpha(\omega)|^2$ then
\begin{equation}\label{BCH_formula_deriv}
\frac{\mathop{\textrm{d}\mathcal{D}_t^\dagger \{\alpha(\omega)\}}}{\mathop{\textrm{d}t}} = \textrm{i}\mathcal{D}_t^\dagger\{\alpha(\omega)\}\bigg(\int_{-\infty}^\infty \mathop{\textrm{d}\omega}\omega\left(\alpha^*(\omega) b_\omega \textrm{e}^{\textrm{i}\omega t} +\alpha(\omega) b_\omega^\dagger \textrm{e}^{-\textrm{i}\omega t} \right) -\int_{-\infty}^\infty \mathop{\textrm{d}\omega}\omega |\alpha(\omega)|^2\bigg),
\end{equation}
with which we obtain
\begin{equation}\label{deriv_disp}
\frac{\mathop{\textrm{d}\mathcal{D}_t^\dagger \{\alpha(\omega)\}}}{\mathop{\textrm{d}t}} \mathcal{D}_t\{\alpha(\omega)\} = \textrm{i}\int_{-\infty}^\infty \mathop{\textrm{d}\omega}\omega\left(\alpha^*(\omega) b_\omega \textrm{e}^{\textrm{i}\omega t}+\alpha(\omega) b_\omega^\dagger \textrm{e}^{-\textrm{i}\omega t}\right) +\int_{-\infty}^\infty\mathop{\textrm{d}\omega} \omega |\alpha(\omega)|^2.
\end{equation}
Putting together Eqs.~\ref{eff_hamil},~\ref{term_1} and \ref{deriv_disp}
\begin{equation}
\tilde{H}(t) = H_\text{S}(t)+\int_{-\infty}^\infty \mathop{\textrm{d}\omega} \omega\, b_\omega^\dagger b_\omega +\xi \int_{-\infty}^\infty \mathop{\textrm{d}\omega} \left(a_0^\dagger b_\omega +a_0 b_\omega^\dagger\right),
\end{equation}
where
\begin{equation}
H_\text{S}(t) =H_{0\text{S}}+H_{1\text{S}}(t)\quad\text{and}\quad H_{1\text{S}}(t)=\sqrt{\kappa}\left(\alpha(t)a_0^\dagger+\alpha^*(t) a_0\right)
\end{equation}
is the system Hamiltonian with an effective classical driving field added to it. (Here we drop the tilde in $H_\text{S}$ due to the appearance of the explicit time-dependence.) In this frame, energy is no longer conserved due to the time-dependent operator $H_{1\text{S}}(t)$ added to $H_{0\text{S}}$.

This transformation caries through with any arbitrary number of waveguides also coupled to the system Hamiltonian as depicted in Fig. \ref{figure:6}, though we did not want to clutter the calculation with the extra notation. In experiment, the coherent state sometimes has a very large photon number but is only weakly coupled to the system through $\gamma_0$, while other waveguides act as the primary decay channels for the system. In this case $\gamma_0\ll\gamma_1,\dots,\gamma_M$, and in the limit that $\gamma_0\rightarrow 0$ and $|\alpha(t)|\rightarrow\infty$ while keeping $\sqrt{\gamma_0}|\alpha(t)|$ fixed, the state of the system factorizes with the $0^\textrm{th}$ waveguide at all times while still pumping energy into $\mathcal{H}_\text{S}$ at a finite rate---this is the semi-classical coherent driving limit.


\addcontentsline{toc}{section}{\protect\numberline{}{REFERENCES}}
\bibliographystyle{unsrtnat}
\bibliography{bibliography}

\begin{thebibliography}{103}
\providecommand{\natexlab}[1]{#1}
\providecommand{\url}[1]{\texttt{#1}}
\expandafter\ifx\csname urlstyle\endcsname\relax
  \providecommand{\doi}[1]{doi: #1}\else
  \providecommand{\doi}{doi: \begingroup \urlstyle{rm}\Url}\fi

\bibitem[Johansson et~al.(2012)Johansson, Nation, and Nori]{johansson2012qutip}
JR~Johansson, PD~Nation, and Franco Nori.
\newblock Qutip: An open-source python framework for the dynamics of open
  quantum systems.
\newblock \emph{Computer Physics Communications}, 183\penalty0 (8):\penalty0
  1760--1772, 2012.
\newblock \doi{10.1016/j.cpc.2012.02.021}.

\bibitem[Carmichael(2009)]{carmichael2009open}
Howard Carmichael.
\newblock \emph{An open systems approach to quantum optics: lectures presented
  at the Universit{\'e} Libre de Bruxelles, October 28 to November 4, 1991},
  volume~18.
\newblock Springer Science \& Business Media, 2009.
\newblock \doi{10.1007/978-3-540-47620-7}.

\bibitem[Gao et~al.(2012)Gao, Fallahi, Togan, Miguel-S{\'a}nchez, and
  Imamoglu]{gao2012observation}
WB~Gao, Parisa Fallahi, Emre Togan, Javier Miguel-S{\'a}nchez, and Atac
  Imamoglu.
\newblock Observation of entanglement between a quantum dot spin and a single
  photon.
\newblock \emph{Nature}, 491\penalty0 (7424):\penalty0 426, 2012.
\newblock \doi{10.1038/nature11573}.

\bibitem[Yao et~al.(2005)Yao, Liu, and Sham]{yao2005theory}
Wang Yao, Ren-Bao Liu, and LJ~Sham.
\newblock Theory of control of the spin-photon interface for quantum networks.
\newblock \emph{Physical review letters}, 95\penalty0 (3):\penalty0 030504,
  2005.
\newblock \doi{10.1103/PhysRevLett.95.030504}.

\bibitem[Mollow(1975)]{mollow1975pure}
BR~Mollow.
\newblock Pure-state analysis of resonant light scattering: Radiative damping,
  saturation, and multiphoton effects.
\newblock \emph{Physical Review A}, 12\penalty0 (5):\penalty0 1919, 1975.
\newblock \doi{10.1103/PhysRevA.12.1919}.

\bibitem[Gardiner and Collett(1985)]{gardiner1985input}
CW~Gardiner and MJ~Collett.
\newblock Input and output in damped quantum systems: Quantum stochastic
  differential equations and the master equation.
\newblock \emph{Physical Review A}, 31\penalty0 (6):\penalty0 3761, 1985.
\newblock \doi{10.1103/PhysRevA.31.3761}.

\bibitem[Glauber(1963{\natexlab{a}})]{glauber1963quantum}
Roy~J Glauber.
\newblock The quantum theory of optical coherence.
\newblock \emph{Physical Review}, 130\penalty0 (6):\penalty0 2529,
  1963{\natexlab{a}}.
\newblock \doi{10.1103/PhysRev.130.2529}.

\bibitem[Glauber(1963{\natexlab{b}})]{glauber1963coherent}
Roy~J Glauber.
\newblock Coherent and incoherent states of the radiation field.
\newblock \emph{Physical Review}, 131\penalty0 (6):\penalty0 2766,
  1963{\natexlab{b}}.
\newblock \doi{10.1103/PhysRev.131.2766}.

\bibitem[Domokos et~al.(2002)Domokos, Horak, and Ritsch]{domokos2002quantum}
Peter Domokos, Peter Horak, and Helmut Ritsch.
\newblock Quantum description of light-pulse scattering on a single atom in
  waveguides.
\newblock \emph{Physical Review A}, 65\penalty0 (3):\penalty0 033832, 2002.
\newblock \doi{10.1103/PhysRevA.65.033832}.

\bibitem[Roulet and Scarani(2016)]{roulet2016solving}
Alexandre Roulet and Valerio Scarani.
\newblock Solving the scattering of {N} photons on a two-level atom without
  computation.
\newblock \emph{New Journal of Physics}, 18\penalty0 (9):\penalty0 093035,
  2016.
\newblock \doi{10.1088/1367-2630/18/9/093035}.

\bibitem[Pletyukhov and Gritsev(2015)]{pletyukhov2015quantum}
Mikhail Pletyukhov and Vladimir Gritsev.
\newblock Quantum theory of light scattering in a one-dimensional channel:
  Interaction effect on photon statistics and entanglement entropy.
\newblock \emph{Physical Review A}, 91\penalty0 (6):\penalty0 063841, 2015.
\newblock \doi{10.1103/PhysRevA.91.063841}.

\bibitem[Pletyukhov and Gritsev(2012)]{pletyukhov2012scattering}
Mikhail Pletyukhov and Vladimir Gritsev.
\newblock Scattering of massless particles in one-dimensional chiral channel.
\newblock \emph{New Journal of Physics}, 14\penalty0 (9):\penalty0 095028,
  2012.
\newblock \doi{10.1088/1367-2630/14/9/095028}.

\bibitem[Chang et~al.(2016)Chang, Gonz{\'a}lez-Tudela, Mu{\~n}oz,
  Navarrete-Benlloch, and Shi]{chang2016deterministic}
Yue Chang, Alejandro Gonz{\'a}lez-Tudela, Carlos~S{\'a}nchez Mu{\~n}oz, Carlos
  Navarrete-Benlloch, and Tao Shi.
\newblock Deterministic down-converter and continuous photon-pair source within
  the bad-cavity limit.
\newblock \emph{Physical review letters}, 117\penalty0 (20):\penalty0 203602,
  2016.
\newblock \doi{10.1103/PhysRevLett.117.203602}.

\bibitem[Shi and Sun(2009)]{shi2009lehmann}
T~Shi and CP~Sun.
\newblock Lehmann-{S}ymanzik-{Z}immermann reduction approach to multiphoton
  scattering in coupled-resonator arrays.
\newblock \emph{Physical Review B}, 79\penalty0 (20):\penalty0 205111, 2009.
\newblock \doi{10.1103/PhysRevB.79.205111}.

\bibitem[See et~al.(2017)See, Noh, and Angelakis]{see2017diagrammatic}
Tian~Feng See, Changsuk Noh, and Dimitris~G Angelakis.
\newblock Diagrammatic approach to multiphoton scattering.
\newblock \emph{Physical Review A}, 95\penalty0 (5):\penalty0 053845, 2017.
\newblock \doi{10.1103/PhysRevA.95.053845}.

\bibitem[Liao et~al.(2016)Liao, Zeng, Nha, and Zubairy]{liao2016photon}
Zeyang Liao, Xiaodong Zeng, Hyunchul Nha, and M~Suhail Zubairy.
\newblock Photon transport in a one-dimensional nanophotonic waveguide {QED}
  system.
\newblock \emph{Physica Scripta}, 91\penalty0 (6):\penalty0 063004, 2016.
\newblock \doi{10.1088/0031-8949/91/6/063004}.

\bibitem[Roy et~al.(2017)Roy, Wilson, and Firstenberg]{roy2016strongly}
Dibyendu Roy, Christopher~M Wilson, and Ofer Firstenberg.
\newblock Colloquium: Strongly interacting photons in one-dimensional
  continuum.
\newblock \emph{Reviews of Modern Physics}, 89\penalty0 (2):\penalty0 021001,
  2017.
\newblock \doi{10.1103/RevModPhys.89.021001}.

\bibitem[Zheng et~al.(2010)Zheng, Gauthier, and Baranger]{zheng2010waveguide}
Huaixiu Zheng, Daniel~J Gauthier, and Harold~U Baranger.
\newblock Waveguide {QED}: Many-body bound-state effects in coherent and
  fock-state scattering from a two-level system.
\newblock \emph{Physical Review A}, 82\penalty0 (6):\penalty0 063816, 2010.
\newblock \doi{10.1103/PhysRevA.82.063816}.

\bibitem[Xu and Fan(2017)]{xu2017input}
Shanshan Xu and Shanhui Fan.
\newblock Input-output formalism for few-photon transport.
\newblock In \emph{Quantum Plasmonics}, pages 1--23. Springer, 2017.
\newblock \doi{10.1007/978-3-319-45820-5_1}.

\bibitem[Xu and Fan(2015)]{xu2015input2}
Shanshan Xu and Shanhui Fan.
\newblock Input-output formalism for few-photon transport: A systematic
  treatment beyond two photons.
\newblock \emph{Physical Review A}, 91\penalty0 (4):\penalty0 043845, 2015.
\newblock \doi{10.1103/PhysRevA.91.043845}.

\bibitem[Fan et~al.(2010)Fan, Kocaba{\c{s}}, and Shen]{fan2010input}
Shanhui Fan, {\c{S}}{\"u}kr{\"u}~Ekin Kocaba{\c{s}}, and Jung-Tsung Shen.
\newblock Input-output formalism for few-photon transport in one-dimensional
  nanophotonic waveguides coupled to a qubit.
\newblock \emph{Physical Review A}, 82\penalty0 (6):\penalty0 063821, 2010.
\newblock \doi{10.1103/PhysRevA.82.063821}.

\bibitem[Caneva et~al.(2015)Caneva, Manzoni, Shi, Douglas, Cirac, and
  Chang]{caneva2015quantum}
Tommaso Caneva, Marco~T Manzoni, Tao Shi, James~S Douglas, J~Ignacio Cirac, and
  Darrick~E Chang.
\newblock Quantum dynamics of propagating photons with strong interactions: a
  generalized input--output formalism.
\newblock \emph{New Journal of Physics}, 17\penalty0 (11):\penalty0 113001,
  2015.
\newblock \doi{10.1088/1367-2630/17/11/113001}.

\bibitem[Schneider et~al.(2016)Schneider, Sproll, Stawiarski, Schmitteckert,
  and Busch]{schneider2016green}
Michael~P Schneider, Tobias Sproll, Christina Stawiarski, Peter Schmitteckert,
  and Kurt Busch.
\newblock Green's-function formalism for waveguide {QED} applications.
\newblock \emph{Physical Review A}, 93\penalty0 (1):\penalty0 013828, 2016.
\newblock \doi{10.1103/PhysRevA.93.013828}.

\bibitem[Hurst and Kok(2018)]{hurst2017analytic}
David~L Hurst and Pieter Kok.
\newblock Analytic few-photon scattering in waveguide {QED}.
\newblock \emph{Physical Review A}, 97\penalty0 (4):\penalty0 043850, 2018.
\newblock \doi{10.1103/PhysRevA.97.043850}.

\bibitem[Trivedi et~al.()Trivedi, Fischer, Fan, and Vu\v{c}kovi\'c]{rahul}
Rahul Trivedi, Kevin Fischer, Shanhui Fan, and Jelena Vu\v{c}kovi\'c.
\newblock Few-photon scattering and emission from open quantum systems.
\newblock \emph{Manuscript in preparation}.

\bibitem[Bombardelli(2016)]{bombardelli2016s}
Diego Bombardelli.
\newblock S-matrices and integrability.
\newblock \emph{Journal of Physics A: Mathematical and Theoretical},
  49\penalty0 (32):\penalty0 323003, 2016.
\newblock \doi{10.1088/1751-8113/49/32/323003}.

\bibitem[Yudson and Reineker(2008)]{yudson2008multiphoton}
VI~Yudson and P~Reineker.
\newblock Multiphoton scattering in a one-dimensional waveguide with resonant
  atoms.
\newblock \emph{Physical Review A}, 78\penalty0 (5):\penalty0 052713, 2008.
\newblock \doi{10.1103/PhysRevA.78.052713}.

\bibitem[Yudson(1985)]{yudson1985dynamics}
VI~Yudson.
\newblock Dynamics of integrable quantum systems.
\newblock \emph{Zh. Eksp. Teor. Fiz}, 88:\penalty0 1757--1770, 1985.

\bibitem[Yudson(1988)]{yudson1988dynamics}
VI~Yudson.
\newblock Dynamics of the integrable one-dimensional system “photons+
  two-level atoms”.
\newblock \emph{Physics Letters A}, 129\penalty0 (1):\penalty0 17--20, 1988.
\newblock \doi{10.1016/0375-9601(88)90465-3}.

\bibitem[Rupasov(1982)]{rupasov1982complete}
VI~Rupasov.
\newblock Complete integrability of the quasi-one-dimensional quantum model of
  dicke superradiance.
\newblock \emph{JETP Letters}, 36:\penalty0 142--146, 1982.

\bibitem[Bassi and LeClair(1999)]{bassi1999one}
Zorawar~S Bassi and Andr{\'e} LeClair.
\newblock A one-dimensional model for n-level atoms coupled to an
  electromagnetic field.
\newblock \emph{Journal of Mathematical Physics}, 40\penalty0 (8):\penalty0
  3723--3731, 1999.
\newblock \doi{10.1063/1.532922}.

\bibitem[Rupasov and Yudson(1984)]{rupasov1984exact}
VI~Rupasov and VI~Yudson.
\newblock Exact dicke superradiance theory: Bethe wavefunctions in the discrete
  atom model.
\newblock \emph{Zh. Eksp. Teor. Fiz}, 86:\penalty0 825, 1984.

\bibitem[LeClair(1997)]{leclair1997qed}
Andr{\'e} LeClair.
\newblock {QED} for a fibrillar medium of two-level atoms.
\newblock \emph{Physical Review A}, 56\penalty0 (1):\penalty0 782, 1997.
\newblock \doi{10.1103/PhysRevA.56.782}.

\bibitem[Konik and LeClair(1998)]{konik1998scattering}
Robert Konik and Andr{\'e} LeClair.
\newblock Scattering theory of oscillator defects in an optical fiber.
\newblock \emph{Physical Review B}, 58\penalty0 (4):\penalty0 1872, 1998.
\newblock \doi{10.1103/PhysRevB.58.1872}.

\bibitem[LeClair et~al.(1997)LeClair, Lesage, Lukyanov, and
  Saleur]{leclair1997maxwell}
A~LeClair, F~Lesage, S~Lukyanov, and H~Saleur.
\newblock The {M}axwell-{B}loch theory in quantum optics and the kondo model.
\newblock \emph{Physics Letters A}, 235\penalty0 (3):\penalty0 203--208, 1997.
\newblock \doi{10.1016/S0375-9601(97)00602-6}.

\bibitem[Leclair(1999)]{leclair1999eigenstates}
Andr{\'e} Leclair.
\newblock Eigenstates of the atom--field interaction and the binding of light
  in photonic crystals.
\newblock \emph{Annals of Physics}, 271\penalty0 (2):\penalty0 268--293, 1999.
\newblock \doi{10.1006/aphy.1998.5874}.

\bibitem[Konyk and Gea-Banacloche(2016)]{konyk2016quantum}
William Konyk and Julio Gea-Banacloche.
\newblock Quantum multimode treatment of light scattering by an atom in a
  waveguide.
\newblock \emph{Physical Review A}, 93\penalty0 (6):\penalty0 063807, 2016.
\newblock \doi{10.1103/PhysRevA.93.063807}.

\bibitem[Nysteen et~al.(2015)Nysteen, Kristensen, McCutcheon, Kaer, and
  M{\o}rk]{nysteen2015scattering}
Anders Nysteen, Philip~Tr{\o}st Kristensen, Dara~PS McCutcheon, Per Kaer, and
  Jesper M{\o}rk.
\newblock Scattering of two photons on a quantum emitter in a one-dimensional
  waveguide: exact dynamics and induced correlations.
\newblock \emph{New Journal of Physics}, 17\penalty0 (2):\penalty0 023030,
  2015.
\newblock \doi{10.1088/1367-2630/17/2/023030}.

\bibitem[Baragiola and Combes(2017)]{baragiola2017quantum}
Ben~Q. Baragiola and Joshua Combes.
\newblock Quantum trajectories for propagating fock states.
\newblock \emph{Physical Review A}, 96:\penalty0 023819, Aug 2017.
\newblock \doi{10.1103/PhysRevA.96.023819}.

\bibitem[Pan et~al.(2016)Pan, Dong, and Zhang]{pan2016exact}
Yu~Pan, Daoyi Dong, and Guofeng Zhang.
\newblock Exact analysis of the response of quantum systems to two-photons
  using a {QSDE} approach.
\newblock \emph{New Journal of Physics}, 18\penalty0 (3):\penalty0 033004,
  2016.
\newblock \doi{10.1088/1367-2630/18/3/033004}.

\bibitem[Shi et~al.(2015)Shi, Chang, and Cirac]{shi2015multiphoton}
Tao Shi, Darrick~E Chang, and J~Ignacio Cirac.
\newblock Multiphoton-scattering theory and generalized master equations.
\newblock \emph{Physical Review A}, 92\penalty0 (5):\penalty0 053834, 2015.
\newblock \doi{10.1103/PhysRevA.92.053834}.

\bibitem[Baragiola et~al.(2012)Baragiola, Cook, Bra{\'n}czyk, and
  Combes]{baragiola2012n}
Ben~Q Baragiola, Robert~L Cook, Agata~M Bra{\'n}czyk, and Joshua Combes.
\newblock N-photon wave packets interacting with an arbitrary quantum system.
\newblock \emph{Physical Review A}, 86\penalty0 (1):\penalty0 013811, 2012.
\newblock \doi{10.1103/PhysRevA.86.013811}.

\bibitem[Verstraete and Cirac(2010)]{verstraete2010continuous}
Frank Verstraete and J~Ignacio Cirac.
\newblock Continuous matrix product states for quantum fields.
\newblock \emph{Physical review letters}, 104\penalty0 (19):\penalty0 190405,
  2010.
\newblock \doi{10.1103/PhysRevLett.104.190405}.

\bibitem[Osborne et~al.(2010)Osborne, Eisert, and
  Verstraete]{osborne2010holographic}
Tobias~J Osborne, Jens Eisert, and Frank Verstraete.
\newblock Holographic quantum states.
\newblock \emph{Physical review letters}, 105\penalty0 (26):\penalty0 260401,
  2010.
\newblock \doi{10.1103/PhysRevLett.105.260401}.

\bibitem[Guimond et~al.(2017)Guimond, Pletyukhov, Pichler, and
  Zoller]{guimond2017delayed}
Pierre-Olivier Guimond, Mikhail Pletyukhov, Hannes Pichler, and Peter Zoller.
\newblock Delayed coherent quantum feedback from a scattering theory and a
  matrix product state perspective.
\newblock \emph{Quantum Science and Technology}, 2017.
\newblock \doi{10.1088/2058-9565/aa7f03}.

\bibitem[Pichler and Zoller(2016)]{pichler2016photonic}
Hannes Pichler and Peter Zoller.
\newblock Photonic circuits with time delays and quantum feedback.
\newblock \emph{Physical review letters}, 116\penalty0 (9):\penalty0 093601,
  2016.
\newblock \doi{10.1103/PhysRevLett.116.093601}.

\bibitem[Cuevas et~al.(2017)Cuevas, Schuch, Perez-Garcia, and
  Cirac]{cuevas2017continuum}
Gemma De~las Cuevas, Norbert Schuch, David Perez-Garcia, and J~Ignacio Cirac.
\newblock Continuum limits of matrix product states.
\newblock \emph{arXiv preprint arXiv:1708.00880}, 2017.

\bibitem[Whalen et~al.(2017)Whalen, Grimsmo, and Carmichael]{whalen2017open}
SJ~Whalen, AL~Grimsmo, and HJ~Carmichael.
\newblock Open quantum systems with delayed coherent feedback.
\newblock \emph{Quantum Science and Technology}, 2:\penalty0 044008, 2017.
\newblock \doi{10.1088/2058-9565/aa8331}.

\bibitem[O'Brien et~al.(2009)O'Brien, Furusawa, and
  Vu{\v{c}}kovi{\'c}]{o2009photonic}
Jeremy~L O'Brien, Akira Furusawa, and Jelena Vu{\v{c}}kovi{\'c}.
\newblock Photonic quantum technologies.
\newblock \emph{Nature Photonics}, 3\penalty0 (12):\penalty0 687--695, 2009.
\newblock \doi{10.1038/nphoton.2009.229}.

\bibitem[Lee et~al.(2018)Lee, Noh, and Kim]{lee2017effective}
Chang-Woo Lee, Changsuk Noh, and Jaewan Kim.
\newblock Effective formalism for open-quantum-system dynamics:
  Time-coarse-graining approach.
\newblock \emph{Physical Review A}, 97\penalty0 (1):\penalty0 012102, 2018.
\newblock \doi{10.1103/PhysRevA.97.012102}.

\bibitem[Yang et~al.(2008)Yang, Liscidini, and Sipe]{yang2008spontaneous}
Zhenshan Yang, Marco Liscidini, and JE~Sipe.
\newblock Spontaneous parametric down-conversion in waveguides: a backward
  {H}eisenberg picture approach.
\newblock \emph{Physical Review A}, 77\penalty0 (3):\penalty0 033808, 2008.
\newblock \doi{10.1103/PhysRevA.77.033808}.

\bibitem[Liscidini et~al.(2012)Liscidini, Helt, and
  Sipe]{liscidini2012asymptotic}
M~Liscidini, LG~Helt, and JE~Sipe.
\newblock Asymptotic fields for a {H}amiltonian treatment of nonlinear
  electromagnetic phenomena.
\newblock \emph{Physical Review A}, 85\penalty0 (1):\penalty0 013833, 2012.
\newblock \doi{10.1103/PhysRevA.85.013833}.

\bibitem[Helt et~al.(2015)Helt, Steel, and Sipe]{helt2015spontaneous}
LG~Helt, MJ~Steel, and JE~Sipe.
\newblock Spontaneous parametric downconversion in waveguides: what's loss got
  to do with it?
\newblock \emph{New Journal of Physics}, 17\penalty0 (1):\penalty0 013055,
  2015.
\newblock \doi{10.1088/1367-2630/17/1/013055}.

\bibitem[Dezfouli et~al.(2014)Dezfouli, Dignam, Steel, and
  Sipe]{dezfouli2014heisenberg}
M~Kamandar Dezfouli, MM~Dignam, MJ~Steel, and JE~Sipe.
\newblock Heisenberg treatment of pair generation in lossy coupled-cavity
  systems.
\newblock \emph{Physical Review A}, 90\penalty0 (4):\penalty0 043832, 2014.
\newblock \doi{10.1103/PhysRevA.90.043832}.

\bibitem[Gardiner and Zoller(2004)]{gardiner2004quantum}
Crispin Gardiner and Peter Zoller.
\newblock \emph{Quantum noise: a handbook of Markovian and non-Markovian
  quantum stochastic methods with applications to quantum optics}, volume~56.
\newblock Springer Science \& Business Media, 2004.

\bibitem[Wiseman(1994)]{wiseman1994quantum}
Howard~Mark Wiseman.
\newblock \emph{Quantum trajectories and feedback}.
\newblock PhD thesis, University of Queensland, 1994.

\bibitem[Loudon(2000)]{loudon2000quantum}
Rodney Loudon.
\newblock \emph{The quantum theory of light}.
\newblock OUP Oxford, 2000.

\bibitem[Brecht et~al.(2015)Brecht, Reddy, Silberhorn, and
  Raymer]{brecht2015photon}
B~Brecht, Dileep~V Reddy, Ch~Silberhorn, and MG~Raymer.
\newblock Photon temporal modes: a complete framework for quantum information
  science.
\newblock \emph{Physical Review X}, 5\penalty0 (4):\penalty0 041017, 2015.
\newblock \doi{10.1103/PhysRevX.5.041017}.

\bibitem[Scully and Zubairy(1999)]{scully1999quantum}
Marlan~O Scully and M~Suhail Zubairy.
\newblock Quantum optics, 1999.

\bibitem[Lancaster and Blundell(2014)]{lancaster2014quantum}
Tom Lancaster and Stephen~J Blundell.
\newblock \emph{Quantum field theory for the gifted amateur}.
\newblock OUP Oxford, 2014.

\bibitem[Garc{\'\i}a-Ripoll(2006)]{garcia2006time}
Juan~Jos{\'e} Garc{\'\i}a-Ripoll.
\newblock Time evolution of matrix product states.
\newblock \emph{New Journal of Physics}, 8\penalty0 (12):\penalty0 305, 2006.
\newblock \doi{10.1088/1367-2630/8/12/305}.

\bibitem[Fischer(2017)]{fischer2017derivation}
Kevin Fischer.
\newblock Derivation of the quantum-optical master equation based on
  coarse-graining of time.
\newblock \emph{arXiv preprint arXiv:1712.00144}, 2017.

\bibitem[Breuer and Petruccione(2002)]{breuer2002theory}
Heinz-Peter Breuer and Francesco Petruccione.
\newblock \emph{The theory of open quantum systems}.
\newblock Oxford University Press on Demand, 2002.

\bibitem[Kiukas et~al.(2015)Kiukas, Gu{\c{t}}{\u{a}}, Lesanovsky, and
  Garrahan]{kiukas2015equivalence}
Jukka Kiukas, M{\u{a}}d{\u{a}}lin Gu{\c{t}}{\u{a}}, Igor Lesanovsky, and Juan~P
  Garrahan.
\newblock Equivalence of matrix product ensembles of trajectories in open
  quantum systems.
\newblock \emph{Physical Review E}, 92\penalty0 (1):\penalty0 012132, 2015.
\newblock \doi{10.1103/PhysRevE.92.012132}.

\bibitem[Cohen-Tannoudji et~al.(1992)Cohen-Tannoudji, Dupont-Roc, Grynberg, and
  Thickstun]{Cohen-Tannoudji1992-uo}
Claude Cohen-Tannoudji, Jacques Dupont-Roc, Gilbert Grynberg, and Patricia
  Thickstun.
\newblock \emph{Atom-photon interactions: basic processes and applications}.
\newblock Wiley Online Library, 1992.

\bibitem[Michler et~al.(2000)Michler, Kiraz, Becher, Schoenfeld, Petroff,
  Zhang, Hu, and Imamoglu]{Michler2000-dx}
P~Michler, A~Kiraz, C~Becher, W~V Schoenfeld, PM~Petroff, L~Zhang, E~Hu, and
  A~Imamoglu.
\newblock A quantum dot single-photon turnstile device.
\newblock \emph{Science}, 290\penalty0 (5500):\penalty0 2282--2285, 22~December
  2000.
\newblock \doi{10.1126/science.290.5500.2282}.

\bibitem[Kimble et~al.(1977)Kimble, Dagenais, and Mandel]{Kimble1977-rw}
HJ~Kimble, M~Dagenais, and L~Mandel.
\newblock Photon antibunching in resonance fluorescence.
\newblock \emph{Physical Review Letters}, 39\penalty0 (11):\penalty0 691--695,
  12~September 1977.
\newblock \doi{10.1103/PhysRevLett.39.691}.

\bibitem[Flagg et~al.(2009)Flagg, Muller, Robertson, Founta, Deppe, Xiao, Ma,
  Salamo, and Shih]{Flagg2009-zi}
EB~Flagg, A~Muller, JW~Robertson, S~Founta, DG~Deppe, M~Xiao, W~Ma, GJ~Salamo,
  and C~K Shih.
\newblock Resonantly driven coherent oscillations in a solid-state quantum
  emitter.
\newblock \emph{Nature Physics}, 5\penalty0 (3):\penalty0 203--207, 25~January
  2009.
\newblock \doi{10.1038/nphys1184}.

\bibitem[Mollow(1969)]{Mollow1969-dv}
B~R Mollow.
\newblock Power spectrum of light scattered by {Two-Level} systems.
\newblock \emph{Physical Review}, 188\penalty0 (5):\penalty0 1969--1975,
  25~December 1969.
\newblock \doi{10.1103/PhysRev.188.1969}.

\bibitem[Santori et~al.(2002)Santori, Fattal, Vu\v{c}kovi\'{c}, Solomon, and
  Yamamoto]{Santori2002-wi}
Charles Santori, David Fattal, Jelena Vu\v{c}kovi\'{c}, Glenn~S Solomon, and
  Yoshihisa Yamamoto.
\newblock Indistinguishable photons from a single-photon device.
\newblock \emph{Nature}, 419\penalty0 (6907):\penalty0 594--597, 10~October
  2002.

\bibitem[Hong et~al.(1987)Hong, Ou, and Mandel]{Hong1987-lv}
CK~Hong, ZY~Ou, and L~Mandel.
\newblock Measurement of subpicosecond time intervals between two photons by
  interference.
\newblock \emph{Physical Review Letters}, 59\penalty0 (18):\penalty0
  2044--2046, 2~November 1987.
\newblock \doi{10.1103/PhysRevLett.59.2044}.

\bibitem[He et~al.(2013)He, He, Wei, Wu, Atat{\"u}re, Schneider, H{\"o}fling,
  Kamp, Lu, and Pan]{he2013demand}
Yu-Ming He, Yu~He, Yu-Jia Wei, Dian Wu, Mete Atat{\"u}re, Christian Schneider,
  Sven H{\"o}fling, Martin Kamp, Chao-Yang Lu, and Jian-Wei Pan.
\newblock On-demand semiconductor single-photon source with near-unity
  indistinguishability.
\newblock \emph{Nature Nanotechnology}, 8\penalty0 (3):\penalty0 213--217,
  2013.
\newblock \doi{10.1038/nnano.2012.262}.

\bibitem[Schneider et~al.(2015)Schneider, Gold, Lu, H{\"o}fling, Pan, and
  Kamp]{schneider2015single}
C~Schneider, P~Gold, C-Y Lu, S~H{\"o}fling, J-W Pan, and M~Kamp.
\newblock Single semiconductor quantum dots in microcavities: bright sources of
  indistinguishable photons.
\newblock In \emph{Engineering the Atom-Photon Interaction}, pages 343--361.
  Springer, 2015.
\newblock \doi{10.1007/978-3-319-19231-4_13}.

\bibitem[Yuan et~al.(2002)Yuan, Kardynal, Stevenson, Shields, Lobo, Cooper,
  Beattie, Ritchie, and Pepper]{yuan2002electrically}
Zhiliang Yuan, Beata~E Kardynal, R~Mark Stevenson, Andrew~J Shields, Charlene~J
  Lobo, Ken Cooper, Neil~S Beattie, David~A Ritchie, and Michael Pepper.
\newblock Electrically driven single-photon source.
\newblock \emph{Science}, 295\penalty0 (5552):\penalty0 102--105, 2002.
\newblock \doi{10.1126/science.1066790}.

\bibitem[Claudon et~al.(2010)Claudon, Bleuse, Malik, Bazin, Jaffrennou,
  Gregersen, Sauvan, Lalanne, and G{\'e}rard]{claudon2010highly}
Julien Claudon, Jo{\"e}l Bleuse, Nitin~Singh Malik, Maela Bazin, P{\'e}rine
  Jaffrennou, Niels Gregersen, Christophe Sauvan, Philippe Lalanne, and
  Jean-Michel G{\'e}rard.
\newblock A highly efficient single-photon source based on a quantum dot in a
  photonic nanowire.
\newblock \emph{Nature Photonics}, 4\penalty0 (3):\penalty0 174--177, 2010.

\bibitem[Unsleber et~al.(2016)Unsleber, He, Gerhardt, Maier, Lu, Pan,
  Gregersen, Kamp, Schneider, and H{\"o}fling]{unsleber2016highly}
Sebastian Unsleber, Yu-Ming He, Stefan Gerhardt, Sebastian Maier, Chao-Yang Lu,
  Jian-Wei Pan, Niels Gregersen, Martin Kamp, Christian Schneider, and Sven
  H{\"o}fling.
\newblock Highly indistinguishable on-demand resonance fluorescence photons
  from a deterministic quantum dot micropillar device with 74\% extraction
  efficiency.
\newblock \emph{Optics Express}, 24\penalty0 (8):\penalty0 8539--8546, 2016.
\newblock \doi{10.1364/OE.24.008539}.

\bibitem[Schlehahn et~al.(2016)Schlehahn, Thoma, Munnelly, Kamp, H{\"o}fling,
  Heindel, Schneider, and Reitzenstein]{schlehahn2016electrically}
A~Schlehahn, A~Thoma, P~Munnelly, M~Kamp, Sven H{\"o}fling, T~Heindel,
  C~Schneider, and S~Reitzenstein.
\newblock An electrically driven cavity-enhanced source of indistinguishable
  photons with 61\% overall efficiency.
\newblock \emph{APL Photonics}, 1\penalty0 (1):\penalty0 011301, 2016.
\newblock \doi{10.1063/1.4939831}.

\bibitem[Somaschi et~al.(2016)Somaschi, Giesz, De~Santis, Loredo, Almeida,
  Hornecker, Portalupi, Grange, Ant{\'o}n, Demory, et~al.]{somaschi2016near}
N~Somaschi, V~Giesz, L~De~Santis, JC~Loredo, MP~Almeida, G~Hornecker,
  SL~Portalupi, T~Grange, C~Ant{\'o}n, J~Demory, et~al.
\newblock Near-optimal single-photon sources in the solid state.
\newblock \emph{Nature Photonics}, 10\penalty0 (5):\penalty0 340--345, 2016.
\newblock \doi{10.1038/nphoton.2016.23}.

\bibitem[Ding et~al.(2016)Ding, He, Duan, Gregersen, Chen, Unsleber, Maier,
  Schneider, Kamp, H{\"o}fling, et~al.]{ding2016demand}
Xing Ding, Yu~He, Z-C Duan, Niels Gregersen, M-C Chen, S~Unsleber, Sebastian
  Maier, Christian Schneider, Martin Kamp, Sven H{\"o}fling, et~al.
\newblock On-demand single photons with high extraction efficiency and
  near-unity indistinguishability from a resonantly driven quantum dot in a
  micropillar.
\newblock \emph{Physical Review Letters}, 116\penalty0 (2):\penalty0 020401,
  2016.
\newblock \doi{10.1103/PhysRevLett.116.020401}.

\bibitem[Michler(2017)]{michler}
Peter Michler, editor.
\newblock \emph{Quantum Dots for Quantum Information Technologies}.
\newblock Springer, 2017.
\newblock \doi{10.1007/978-3-319-56378-7}.

\bibitem[Senellart et~al.(2017)Senellart, Solomon, and
  White]{senellart2017high}
Pascale Senellart, Glenn Solomon, and Andrew White.
\newblock High-performance semiconductor quantum-dot single-photon sources.
\newblock \emph{Nature Nanotechnology}, 12\penalty0 (11):\penalty0 1026, 2017.
\newblock \doi{10.1038/nnano.2017.218}.

\bibitem[Loredo et~al.(2017)Loredo, Broome, Hilaire, Gazzano, Sagnes, Lemaitre,
  Almeida, Senellart, and White]{Loredo2016-jz}
JC~Loredo, MA~Broome, P~Hilaire, O.~Gazzano, I~Sagnes, A~Lemaitre, MP~Almeida,
  P.~Senellart, and AG~White.
\newblock Boson sampling with single-photon fock states from a bright
  solid-state source.
\newblock \emph{Physical Review Letters}, 118:\penalty0 130503, Mar 2017.
\newblock \doi{10.1103/PhysRevLett.118.130503}.

\bibitem[Aharonovich et~al.(2016)Aharonovich, Englund, and
  Toth]{Aharonovich2016-pq}
Igor Aharonovich, Dirk Englund, and Milos Toth.
\newblock Solid-state single-photon emitters.
\newblock \emph{Nature Photonics}, 10\penalty0 (10):\penalty0 631--641,
  29~September 2016.
\newblock \doi{10.1038/nphoton.2016.186}.

\bibitem[Rundquist et~al.(2014)Rundquist, Bajcsy, Majumdar, Sarmiento, Fischer,
  Lagoudakis, Buckley, Piggott, and Vu\v{c}kovi\'{c}]{Rundquist2014-kf}
Armand Rundquist, Michal Bajcsy, Arka Majumdar, Tomas Sarmiento, Kevin Fischer,
  Konstantinos~G Lagoudakis, Sonia Buckley, Alexander~Y Piggott, and Jelena
  Vu\v{c}kovi\'{c}.
\newblock Nonclassical higher-order photon correlations with a quantum dot
  strongly coupled to a photonic-crystal nanocavity.
\newblock \emph{Physical Review A}, 90\penalty0 (2):\penalty0 023846, 25~August
  2014.
\newblock \doi{10.1103/PhysRevA.90.023846}.

\bibitem[Mu{\~n}oz et~al.(2014)Mu{\~n}oz, Del~Valle, Tudela, M{\"u}ller,
  Lichtmannecker, Kaniber, Tejedor, Finley, and Laussy]{munoz2014emitters}
C~S{\'a}nchez Mu{\~n}oz, E~Del~Valle, A~Gonz{\'a}lez Tudela, K~M{\"u}ller,
  S~Lichtmannecker, M~Kaniber, C~Tejedor, JJ~Finley, and FP~Laussy.
\newblock Emitters of {N}-photon bundles.
\newblock \emph{Nature Photonics}, 8\penalty0 (7):\penalty0 550--555, 2014.
\newblock \doi{10.1038/nphoton.2014.114}.

\bibitem[Afek et~al.(2010)Afek, Ambar, and Silberberg]{afek2010high}
Itai Afek, Oron Ambar, and Yaron Silberberg.
\newblock High-{NOON} states by mixing quantum and classical light.
\newblock \emph{Science}, 328\penalty0 (5980):\penalty0 879--881, 2010.
\newblock \doi{10.1126/science.1188172}.

\bibitem[Fischer et~al.(2017{\natexlab{a}})Fischer, Hanschke, Wierzbowski,
  Simmet, Dory, Finley, Vu{\v{c}}kovi{\'c}, and
  M{\"u}ller]{fischer2017signatures}
Kevin~A Fischer, Lukas Hanschke, Jakob Wierzbowski, Tobias Simmet, Constantin
  Dory, Jonathan~J Finley, Jelena Vu{\v{c}}kovi{\'c}, and Kai M{\"u}ller.
\newblock Signatures of two-photon pulses from a quantum two-level system.
\newblock \emph{Nature Physics}, 13\penalty0 (7):\penalty0 649,
  2017{\natexlab{a}}.
\newblock \doi{10.1038/nphys4052}.

\bibitem[Fischer et~al.(2017{\natexlab{b}})Fischer, Hanschke, Kremser, Finley,
  M{\"u}ller, and Vu{\v{c}}kovi{\'c}]{fischer2017pulsed}
Kevin~A Fischer, Lukas Hanschke, Malte Kremser, Jonathan~J Finley, Kai
  M{\"u}ller, and Jelena Vu{\v{c}}kovi{\'c}.
\newblock Pulsed rabi oscillations in quantum two-level systems: beyond the
  area theorem.
\newblock \emph{Quantum Science and Technology}, 3\penalty0 (1):\penalty0
  014006, 2017{\natexlab{b}}.
\newblock \doi{10.1088/2058-9565/aa9269}.

\bibitem[Fischer et~al.(2016)Fischer, M{\"u}ller, Lagoudakis, and
  Vu{\v{c}}kovi{\'c}]{fischer2016dynamical}
Kevin~A Fischer, Kai M{\"u}ller, Konstantinos~G Lagoudakis, and Jelena
  Vu{\v{c}}kovi{\'c}.
\newblock Dynamical modeling of pulsed two-photon interference.
\newblock \emph{New Journal of Physics}, 18\penalty0 (11):\penalty0 113053,
  2016.
\newblock \doi{10.1088/1367-2630/18/11/113053}.

\bibitem[Lindkvist and Johansson(2014)]{lindkvist2014scattering}
Joel Lindkvist and G{\"o}ran Johansson.
\newblock Scattering of coherent pulses on a two-level system-single-photon
  generation.
\newblock \emph{New Journal of Physics}, 16\penalty0 (5):\penalty0 055018,
  2014.
\newblock \doi{10.1088/1367-2630/16/5/055018}.

\bibitem[Law and Eberly(2004)]{law2004analysis}
CK~Law and JH~Eberly.
\newblock Analysis and interpretation of high transverse entanglement in
  optical parametric down conversion.
\newblock \emph{Physical review letters}, 92\penalty0 (12):\penalty0 127903,
  2004.
\newblock \doi{10.1103/PhysRevLett.92.127903}.

\bibitem[Law et~al.(2000)Law, Walmsley, and Eberly]{law2000continuous}
CK~Law, IA~Walmsley, and JH~Eberly.
\newblock Continuous frequency entanglement: effective finite hilbert space and
  entropy control.
\newblock \emph{Physical Review Letters}, 84\penalty0 (23):\penalty0 5304,
  2000.
\newblock \doi{10.1103/PhysRevLett.84.5304}.

\bibitem[Blay et~al.(2017)Blay, Steel, and Helt]{blay2017effects}
Daniel~R Blay, MJ~Steel, and LG~Helt.
\newblock Effects of filtering on the purity of heralded single photons from
  parametric sources.
\newblock \emph{Physical Review A}, 96\penalty0 (5):\penalty0 053842, 2017.
\newblock \doi{10.1103/PhysRevA.96.053842}.

\bibitem[Lu et~al.(2016)Lu, Jiang, Zhang, and Lin]{lu2016biphoton}
Xiyuan Lu, Wei~C Jiang, Jidong Zhang, and Qiang Lin.
\newblock Biphoton statistics of quantum light generated on a silicon chip.
\newblock \emph{ACS Photonics}, 3\penalty0 (9):\penalty0 1626--1636, 2016.
\newblock \doi{10.1021/acsphotonics.6b00204}.

\bibitem[Mower and Englund(2011)]{mower2011efficient}
Jacob Mower and Dirk Englund.
\newblock Efficient generation of single and entangled photons on a silicon
  photonic integrated chip.
\newblock \emph{Physical Review A}, 84\penalty0 (5):\penalty0 052326, 2011.
\newblock \doi{10.1103/PhysRevA.84.052326}.

\bibitem[Silverstone et~al.(2015)Silverstone, Santagati, Bonneau, Strain,
  Sorel, O’Brien, and Thompson]{silverstone2015qubit}
Joshua~W Silverstone, Raffaele Santagati, Damien Bonneau, Michael~J Strain,
  Marc Sorel, Jeremy~L O’Brien, and Mark~G Thompson.
\newblock Qubit entanglement between ring-resonator photon-pair sources on a
  silicon chip.
\newblock \emph{Nature communications}, 6, 2015.
\newblock \doi{10.1038/ncomms8948}.

\bibitem[Pavesi and Lockwood(2016)]{pavesi2016silicon}
Lorenzo Pavesi and David~J Lockwood.
\newblock \emph{Silicon Photonics III: Systems and Applications}, volume 122.
\newblock Springer Science \& Business Media, 2016.

\bibitem[Quesada and Sipe(2017)]{quesada2017effects}
Nicolas Quesada and John Sipe.
\newblock The effects of self-and cross-phase modulation in the generation of
  bright twin beams using spdc.
\newblock In \emph{CLEO: Science and Innovations}, pages JW2A--23. Optical
  Society of America, 2017.
\newblock \doi{10.1364/CLEO_AT.2017.JW2A.23}.

\bibitem[Vernon and Sipe(2015)]{vernon2015spontaneous}
Z~Vernon and JE~Sipe.
\newblock Spontaneous four-wave mixing in lossy microring resonators.
\newblock \emph{Physical Review A}, 91\penalty0 (5):\penalty0 053802, 2015.
\newblock \doi{10.1103/PhysRevA.91.053802}.

\bibitem[Sun et~al.(2016)Sun, Kim, Solomon, and Waks]{sun2016quantum}
Shuo Sun, Hyochul Kim, Glenn~S Solomon, and Edo Waks.
\newblock A quantum phase switch between a single solid-state spin and a
  photon.
\newblock \emph{Nature Nanotechnology}, 11\penalty0 (6):\penalty0 539--544,
  2016.
\newblock \doi{10.1038/nnano.2015.334}.

\bibitem[Roy and Hughes(2011)]{roy2011influence}
Chiranjeeb Roy and Stephen Hughes.
\newblock Influence of electron--acoustic-phonon scattering on intensity power
  broadening in a coherently driven quantum-dot--cavity system.
\newblock \emph{Physical Review X}, 1\penalty0 (2):\penalty0 021009, 2011.
\newblock \doi{10.1103/PhysRevX.1.021009}.

\bibitem[M{\"u}ller et~al.(2015)M{\"u}ller, Fischer, Rundquist, Dory,
  Lagoudakis, Sarmiento, Kelaita, Borish, and
  Vu{\v{c}}kovi{\'c}]{muller2015ultrafast}
Kai M{\"u}ller, Kevin~A Fischer, Armand Rundquist, Constantin Dory,
  Konstantinos~G Lagoudakis, Tomas Sarmiento, Yousif~A Kelaita, Victoria
  Borish, and Jelena Vu{\v{c}}kovi{\'c}.
\newblock Ultrafast polariton-phonon dynamics of strongly coupled quantum
  dot-nanocavity systems.
\newblock \emph{Physical Review X}, 5\penalty0 (3):\penalty0 031006, 2015.
\newblock \doi{10.1103/PhysRevX.5.031006}.

\bibitem[Gustin and Hughes(2017)]{gustin2017influence}
Chris Gustin and Stephen Hughes.
\newblock Influence of electron-phonon scattering for an on-demand quantum dot
  single-photon source using cavity-assisted adiabatic passage.
\newblock \emph{Physical Review B}, 96\penalty0 (8):\penalty0 085305, 2017.
\newblock \doi{10.1103/PhysRevB.96.085305}.

\end{thebibliography}

\end{document}